\documentclass{aa}

\RequirePackage[normalem]{ulem} 
\RequirePackage{color}\definecolor{RED}{rgb}{1,0,0}\definecolor{BLUE}{rgb}{0,0,1} 

\usepackage[utf8]{inputenc}
\usepackage{graphicx}
\usepackage[colorlinks=true, allcolors=blue]{hyperref}
\usepackage{txfonts,textcomp}
\makeatletter
\renewcommand*\aa@pageof{, page \thepage{} of \pageref*{LastPage}}
\makeatother

\usepackage{subcaption}
\usepackage{multirow}
\usepackage{amsmath}

\begin{document}

\date{}

\title{Statistical study of a large and cleaned sample of ultraluminous and hyperluminous  X-ray sources\thanks{Tables A.1 and A.2 are only available in electronic form
at the CDS via anonymous ftp to cdsarc.cds.unistra.fr (130.79.128.5)
or via https://cdsarc.cds.unistra.fr/cgi-bin/qcat?J/A+A/}}
\author{Hugo Tranin\inst{\ref{irap},\ref{iccub}}, Natalie Webb\inst{\ref{irap}}, Olivier Godet\inst{\ref{irap}} and Erwan Quintin\inst{\ref{irap}}}
\institute{IRAP, Université de Toulouse, CNRS, CNES, 9 avenue du Colonel Roche, 31028 Toulouse, France
\label{irap}
\and
Institut d’Estudis Espacials de Catalunya, Universitat de Barcelona
(ICC-UB), Martí i Franquès 1, 08028 Barcelona, Spain\\
e-mail: \texttt{hugo.tranin@irap.omp.eu} \label{iccub}
}

\abstract
    {Ultraluminous and hyperluminous X-ray (ULX and HLX) sources could constitute interesting laboratories to further improve our understanding of the supermassive black hole growth through super-Eddington accretion episodes and successive mergers of lighter holes. ULXs are thought to be powered by super-Eddington accretion onto stellar-mass compact objects, while HLXs are of an unknown nature, but they could be good candidates for accreting intermediate mass black holes (IMBHs). However, a significant portion of the sample of ULX and HLX candidates derived from catalogue searches are in fact background active galactic nuclei (AGN).}
    {Here we build samples of ULXs and HLXs from the three largest X-ray catalogues available, compiled from \textit{XMM-Newton}, \textit{Swift}-XRT, and \textit{Chandra} detections, and the GLADE catalogue containing 1.7 million galaxies at D<1000\,Mpc. We aim to characterise the frequency, environment, hardness, and variability of ULXs and HLXs to better assess their differences and understand their populations.}
    {After a thorough classification of these X-ray sources, we were able to remove $\sim$42\% of  sources with a signal-to-noise ratio (S/N) $>3$ which were shown to be contaminants, to obtain the cleanest sample of ULXs and HLXs to date. From a sample of 1342 ULXs and 191 HLXs detected with a S/N $>3\sigma$, we study the occupation fraction, hardness, variability, radial distribution, and preferred environment of the sources. We built their Malmquist-corrected X-ray luminosity functions (XLFs) and compared them with previous studies. Thanks to the unprecedented size of the sample, we were able to statistically compare ULXs and HLXs and assess the differences in their nature. The interpretation of HLXs as IMBHs is investigated.}
    {A significant break is seen in the XLF at$~\sim 10^{40}$ erg~s$^{-1}$. With our ULX sample, containing $\lesssim 2$\% of contaminants, we are able to confirm that ULXs are located preferentially in spiral galaxies and galaxies with higher star formation rates. While X-ray binaries (XRBs), ULXs, and most HLXs share common hardness and variability distributions, a fraction of HLXs appear significantly softer. Unlike ULXs, HLXs seem to reside equally in spiral as well as lenticular and elliptical galaxies. We note that 35\% of the HLX candidates have an optical counterpart, and we estimate the mass of 120 of them to be in the range of $2\times 10^3 - 10^5 \mathrm{M}_\odot$. Most of the HLX population is found to be consistent with an accreting massive black hole in a dwarf galaxy satellite of the primary host.  This diverse nature needs to be confirmed with deeper optical and infrared observations, as well as upcoming X-ray facilities.}
    {}
    {}

\keywords{      Catalogs --
                        X-rays: general --
                        X-rays: binaries --
                        X-rays: galaxies --
            Methods: statistical
            }

\titlerunning{Statistical study of ULXs and HLX}
\authorrunning{Tranin, Webb, Godet \& Quintin}

\maketitle

\section{Introduction}
\label{sec:1} 

Supermassive black holes (SMBHs, in the mass range $10^6-10^{10} \mathrm{M}_\odot$) have been found in the centres of the most massive galaxies observed so far and they are thought to play a major role in galaxy evolution, notably during their violent episodes of accretion and ejection (\citealt{Fabian2012,Padovani2017}). However, the formation mechanisms of these objects have yet to be fully understood. Some SMBHs observed at $z>6$, which is less than 1\,Gyr after the Big Bang, are more massive than $10^9\,\mathrm{M}_\odot$ (e.g. \citealt{Mortlock2011, Wang2021}), implying rapid and efficient growth mechanisms that may be a mixture of intermediate-mass black hole (IMBH, $10^2-10^5 \mathrm{M}_\odot$) mergers and sustained super-Eddington accretion onto these seeds \citep{Volonteri2008,Haiman2013,Pacucci2022}. However, compelling evidence for the existence of IMBHs is still missing, with the exception of some recent discoveries through gravitational waves and X-ray observations (e.g. \citealt{Farrell2009, Mezcua2017, GW190521}). 

The search and study of X-ray sources in nearby galaxies, especially in the luminosity range $10^{39}-10^{42}$ erg~s$^{-1}$, could be a promising way to gain insight into the SMBH growth mechanisms.
Indeed, for an object spherically accreting hydrogen gas,
the luminosity must be lower than the Eddington limit -- $L_{Edd} = 1.26\cdot 10^{38} (\frac{M}{\mathrm{M}_\odot})$~erg~s$^{-1}$  --  for the accretion flow to be sustained. It is interesting to note, however, that in rare cases there appears to be physical mechanisms that help stabilise the accretion flow in the super-Eddington regime, as suggested by both recent models and observations (e.g. \citealt{Inayoshi2016, Massonneau2022, Lin2017, Belfiore2020}). 
However, using $L<L_{Edd}$ is one way to estimate a lower limit on the accretor mass. Extra-galactic and off-nuclear ultraluminous X-ray (ULX) sources with X-ray luminosities $L_X>10^{39}$~erg~s$^{-1}$  were thus first thought to be good IMBH candidates (e.g. \citealt{Colbert1999,LiuBreg2005}). However, the spectral curvature seen in ULXs, hinting at an accretion regime different than for X-ray binaries (e.g. \citealt{Bachetti2013}), and the discovery of pulsating ULXs (e.g. \citealt{Bachetti2014,Fuerst2016,Israel2017,Carpano2018,Quintin2021}) later revealed that some of these sources are instead neutron stars accreting above the Eddington limit. In ULX spectra, the inner temperature of the disc is cooler than usually found in black hole X-ray binaries ($kT= 0.1-0.3$~keV), while the power law is steeper ($\Gamma = 2-4.5$) and cut off at much lower energies ($2-7$ keV compared to $\gtrsim 60$ keV in black hole X-ray binaries). The absorption is intermediate, with column densities generally in the range $1-3\times 10^{21}$~cm$^{-2}$. This led the community to imagine a different accretion state than in X-ray binaries (XRBs), namely the ultraluminous state \citep{Gladstone2009, Kaaret2017}. ULXs are also likely to cause important feedback on their environment, with optical and radio signatures of a `bubble' surrounding them, inflated by powerful winds and intense photo-ionisation (e.g. \citealt{Abolmasov2008,Berghea2020,Gurpide2022}), which may seem inconsistent with a narrow beaming of X-ray emission that could explain a super-Eddington luminosity \citep{Pakull2002,Berghea2010}.

On the other hand, the more luminous and much rarer hyperluminous X-ray (HLX) sources ($L_X>10^{41}$~erg~s$^{-1}$) may still remain excellent candidates to look for IMBHs, as evidenced with ESO\,243--49 HLX--1 \citep{Farrell2009}, which is one of the best IMBH candidates known, reaching an X-ray luminosity of $10^{42}$ erg s$^{-1}$ in the 0.3--10 keV band. It shows spectral evolution similar to that observed in Galactic black holes \citep{Godet2009, Servillat2011, Godet2012}. The black hole in ESO\,243--49 HLX--1 is thought to be fed by episodes of mass transfer, induced by repetitive partial stripping of a white dwarf-like star when passing at periapsis \citep{godet2014}. This black hole may be embedded in a stellar cluster or be the central black hole in a dwarf galaxy stripped by or in interaction with ESO 243--49 \citep{Webb2010,Webb2017,Farrell2012}. Very few other HLX candidates are known so far, and they are most often located at a high distance \citep{Swartz2011,Zolotukhin2016}, where confusion with a nuclear or background source is more likely. In a recent study, \cite{Barrows2019} identified a large sample of 169 HLX candidates thanks to the resolving power and high sensitivity of two surveys, namely \textit{Chandra} and a  galaxy catalogue based on \textit{Sloan Digital Sky Survey} (SDSS) images. However, most of their candidates are at $D>1000$~Mpc, where a background source is difficult to detect in the deepest large optical surveys. Their median unabsorbed luminosity of $4.6 \times 10^{42}$~erg~s$^{-1}$  in the band $2-10$~keV, as well as the significant fraction of HLXs with an optical counterpart (28\%), could also be explained by a significant number of objects being background sources, although they interpret them as active galactic nuclei (AGN) in satellites of the host galaxy and report a background contamination rate of $7-8$\% estimated from the cosmic X-ray background curves of \cite{Moretti2003}. In contrast, NGC 5907 ULX-1 occasionally exceeded $10^{41}$~erg~s$^{-1}$ in luminosity while hosting a neutron star, appearing as an exception in this class of objects 
\citep{Israel2017}
. In the following, HLX refers to objects whose average broad-band luminosity over all X-ray detections by a given instrument exceeds $10^{41}$~erg~s$^{-1}$, which is not the case for this one.

Studying ULX samples has led to a statistical picture of the ULX population, using many catalogues of ULX candidates from old (\citealt{Swartz2004,LiuBreg2005,LiuMir2005,Swartz2011,Walton2011,Wang2016}) and recent (\citealt{Earnshaw2019,Kovlakas2020,Inoue2021,Bernadich2021,Walton2022}) releases of large X-ray catalogues. Focussing on the hardness (e.g. \citealt{Earnshaw2019}) and variability of ULXs (e.g. \citealt{Sutton2013,Bernadich2021}), their environment and counterparts (e.g. \citealt{Kovlakas2020}), or on their X-ray luminosity function (XLF, \citealt{Mineo2012,Wang2016}), these studies showed several features for the ULX population. ULXs are more prominent in spiral, star-forming galaxies. Their frequency increases with galaxy mass and their luminosity function may be explained as a mere power-law extension of the high-mass X-ray binary luminosity function, for which a break is noted at a few $10^{38}$~erg~s$^{-1}$ in elliptical galaxies (e.g. \citealt{KimFabbiano2004}, \citealt{Wang2016}). A break or cutoff at $1-2\times 10^{40}$~erg~s$^{-1}$  may also be present \citep{Swartz2011,Mineo2012} in the XLF of spiral galaxies; however, this is still under debate (see e.g. \citealt{Walton2011,Wang2016}). Moreover, no break is systematically observed in these galaxies at the Eddington luminosity of a $2\mathrm{M}_\odot$ neutron star ($L_{Edd}\sim 2\times 10^{38}$ erg~s$^{-1}$) or a $100 \mathrm{M}_\odot$ stellar black hole ($L_{Edd}\sim 10^{41}$ erg~s$^{-1}$), as would be expected if the Eddington luminosity was a hard limit \citep{Fabbiano2006,Kaaret2017}. ULXs in spiral  and elliptical galaxies are found to be consistent with a dominant population of high-mass and low-mass X-ray binaries (HMXBs and LMXBs), respectively, which is in agreement with the expected environment for these two classes. HMXBs and LMXBs consist of a black hole or a neutron star accreting matter from a close and more massive (respectively less massive) star, either because of stellar winds (Bondi-Hoyle accretion, \citealt{BondiHoyle1944, Bondi1952}) or by the star overflowing its Roche lobe and forming an accretion disc (e.g. \citealt{vanDenHeuvel1973}). In HMXBs, because mass transfer by Roche lobe overflow is unstable when the binary has a large mass ratio, accretion is generally driven by stellar winds (e.g. \citealt{IbenLivio1993,Nelemans2000}; however see \citealt{Pavlovskii2017} where this point is discussed). LMXBs reside near the galactic bulge or in globular clusters, while HMXBs, for which the compact object is most often identified as a neutron star (e.g. \citealt{Sidoli2018}), are located close to the galactic plane and at larger separations from the bulge \citep{Grimm2002, Repetto2017}, where massive stars are more frequent due to more recent star formation episodes.

Some differences were still reported between XRB and ULX populations. However, only a select number of the brightest (and often closest) ULXs have been studied in detail. Many of the conclusions on the population of ULXs are often drawn from simple spectral model fitting, which does not give a physical description of the data (e.g. \citealt{Swartz2004}), or from easily computed quantities such as the hardness ratio between two bands. A large sample approach was carried out by \cite{Walton2011} who reported a small offset of ULXs towards harder accretion states when residing in spiral galaxies. \cite{Earnshaw2019} found no significant difference between the hardness ratio distributions of ULXs and XRBs. Likewise, the larger dataset of \cite{Bernadich2021} showed similar locations for these two populations in the hardness-hardness diagrams. In contrast, ULXs seem less prone to exhibit high variability on the timescale of years compared to XRBs, and the most variable ones are hosted in spiral galaxies \citep{Bernadich2021}. Confirmed neutron star ULXs were reported to be the hardest ULXs, with highly variable hard radiation \citep{Gurpide2021}. Some ULXs are also variable on the timescale of a few kiloseconds (within a single observation), as found by \cite{Earnshaw2019} and studies of individual ULXs suggest the presence of winds, outflows, or pulsations (e.g \citealt{Sutton2013,Koliopanos2019}).

Nevertheless, ULX population studies suffer from a moderately high contamination fraction due to foreground stars and (mainly) background AGN, often estimated to be $\sim$20\% from the $\log N- \log S$ diagram of cosmic X-ray background sources (e.g. \citealt{Walton2011}), increasing with X-ray luminosity (e.g. \citealt{Bernadich2021}). Despite evidence for ULXs being super-Eddington stellar-mass accretors, the physical mechanism behind the apparent super-Eddington luminosities is still poorly constrained. Modelling over the last twenty years has suggested different possible scenarios such as anisotropic (beamed) emission (e.g. \citealt{King2009,Wiktorowicz2019}); supercritical accretion discs around stellar-mass black holes with radiation-driven winds (e.g. \citealt{Poutanen2007,Middleton2015}) or around neutron stars (e.g. \citealt{Erkut2019,Kuranov2020}); thermal-timescale mass transfer,  in particular from a Helium-burning secondary (e.g. \citealt{Wiktorowicz2015,Pavlovskii2017}); accretion flows around pulsating highly magnetised neutron stars (e.g. \citealt{Bachetti2014,Mushtukov2015,Israel2017, Mushtukov2017, Koliopanos2017}); and accretion of clumpy stellar wind enhanced by X-ray ionisation \citep{Krticka2022}.

For this work, we developed a novel approach to build cleaner ULX and HLX samples, using a general-purpose classification of X-ray sources developed in \cite{Tranin2022}. We assessed the added value of such a clean sample, and we conducted a statistical study of our sample -- the largest to date -- to offer a more complete view of ULX and HLX properties.

In Section \ref{sec:2} we explain our ULX selection method, involving the use of a classification of X-ray sources, and the way we validated the sample and built a complete sub-sample. The sample of contaminants and the selection of HLX candidates are also described. In Section \ref{sec:3} we present a statistical study of our clean sample, in terms of the X-ray luminosity function, ULX rate evolution with the environment, and hardness and variability properties. Samples of XRBs, ULXs, and HLXs are compared. We discuss these results in Section \ref{sec:4} and compare them to catalogues and results obtained in previous works. The nature of HLXs is also investigated. We summarise our study in Section \ref{sec:5}. Unless stated otherwise, errors are quoted at the 1-$\sigma$ level.

\section{The sample}
\label{sec:2}
\subsection{X-ray catalogues}

To obtain the largest possible ULX sample, we consider the three largest X-ray catalogues to date, generated from observations with \textit{XMM-Newton}, the \textit{Chandra} X-ray observatory and the \textit{Swift} X-ray telescope (XRT). From the largest of these catalogues, 4XMM-DR11 \citep{Webb2020}, we select the 496645 point-like sources that have a detection with a reasonable detection flag $\texttt{SUM\_FLAG}\leq 1$ (source parameters may be affected, but spurious detections are unlikely). A few extended sources are considered as point-like by the \textit{XMM-Newton} pipeline ($\texttt{SC\_EXTENT}=0$) and remain in this selection. They have typically $\texttt{SC\_EXT\_ML}>100$ and $\texttt{SC\_SUM\_FLAG}>1$ or $\texttt{SC\_EXT\_ML}>10^4$, so we remove the 561 corresponding sources.  We also remove sources below the detection likelihood $\texttt{SC\_DET\_ML}=10$ threshold, to limit the false source rate while still keeping a high fraction of the catalogue sources (83\%). 

The \textit{Chandra} CSC2 catalogue \citep{EvansCSC2010,EvansCSC2019} is treated in a similar way: removing sources flagged as extended or ambiguous (as given in the source name), saturated or overlapping a bright streak. Unlike in the \textit{XMM-Newton} pipeline, extended sources seem to be all flagged as extended in the \textit{Chandra} pipeline, but some genuinely point-like sources may also be flagged likewise: from visual inspection, we choose to keep 1217 sources flagged as extended but not being in a confused area and whose median major axis across energy bands (computed in the \textit{Chandra} pipeline) is at most 0.5 arcsec (i.e. the pixel scale of ACIS cameras). We keep only sources in the likelihood class \texttt{TRUE} 
and with a non-zero flux, having detections with a \texttt{conf\_code} lower than 256 (to remove sources with unreliable parameters due to an overlapping extended source); 214757 sources remain after this filtering process. As detailed in Section \ref{sec:3.1}, 199 sources not included at this stage (because flagged as extended or confused by the Chandra pipeline) were added a posteriori to the ULX catalogue, after visual inspection.  

Likewise, from the \textit{Swift}-XRT catalogue 2SXPS \citep{Evans2020} we select sources having $\texttt{detflag}=0$ and $\texttt{fieldflag}\leq 1$ to remove possibly spurious sources and sources in polluted fields, as well as sources with a zero flux.
Since \textit{Swift}-XRT has the largest PSF among the three X-ray telescopes, some supposedly point-like sources may actually be extended at this stage, and no extent estimate is made available. This was notably confirmed by a visual inspection of some fields containing galaxy clusters. We thus flagged 2SXPS sources matching the extent of an \textit{XMM-Newton} or \textit{Chandra} extended source, unless they also match a reliable source at less than 10 arcsec. In galaxy clusters, because the cluster is extended, its \textit{Simbad} and XRT position are sometimes separated by $\sim$20 arcsec: 188 sources at less than 20 arcsec from a \textit{Simbad} galaxy cluster were conservatively flagged as well, as possibly due to the hot gas extended emission. This leaves a sample of  130162 clean sources.

\subsection{X-ray matches}

To assess the number of unique ULXs left in our final sample, and to be able to probe long-term variability of these sources, we perform a crossmatch of the three catalogues with each other using the TOPCAT software (Tool for Operations on Catalogues and Tables, \citealt{Taylor2005}). The `sky with errors' algorithm is used with the 3--$\sigma$ position error of each catalogue. Grouped matches are found, meaning the ambiguous association of a source to at least two sources from the other catalogue, either in crowded fields (like galaxies) or because of spatial resolution issues (it is not unusual that two close \textit{Chandra} sources are confused into a single \textit{XMM-Newton} or \textit{Swift} one, since these instruments cannot resolve them). We choose to remove these associations and flag the corresponding source, unless they become ungrouped when a crossmatch using the 1--$\sigma$ position error is performed. In this way, ambiguous associations in which one of the potential associations is clearly favoured are well retrieved, while more ambiguous associations are flagged and removed.

\begin{figure}
    \centering
    \includegraphics[width=8.5cm]{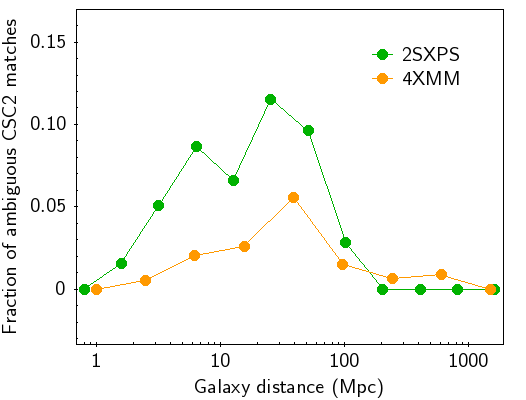}
    \caption{Fraction of 2SXPS-CSC2 and 4XMM-CSC2 ambiguous matches as a function of the host distance given by GLADE.}
    \label{fig:frac_ambiguous}
\end{figure}

From the 412242 \textit{XMM-Newton}, 214757 \textit{Chandra} and 130162 \textit{Swift} unique sources, the cross-correlation retrieves 687291 unique sources including 65366 with detections in at least two facilities and 4229 with a grouped match flag. In particular, 1.7\% of 2SXPS sources matching a CSC2 source are in ambiguous association. The median separation between the two counterparts of an ambiguous 2SXPS-CSC2 (4XMM-CSC2) association is 8 arcsec. Consequently, for sources associated with a galaxy, this fraction of ambiguous associations increases with galaxy distance and becomes $\sim$10\% in galaxies at 40~Mpc (Figure \ref{fig:frac_ambiguous}). Similarly, 0.7\% of 4XMM-CSC2 associations are ambiguous, their median separation is 5.5 arcsec, and $\sim$5\% of 4XMM-CSC2 associations are ambiguous in galaxies at 40~Mpc. At greater distances, some \textit{Chandra} sources must be confused as well. This source confusion issue is important in ULX studies because tight groups of XRB (resp. ULX) can be mistaken for ULXs (resp. HLXs) (e.g. \citealt{Wolter2015}).

\subsection{The galaxy sample}

The first published ULX catalogues were constructed by cross-correlating X-ray catalogues with rather small catalogues of bright galaxies (in particular RC3, \citealt{RC31991}). More recently, to  achieve more complete ULX catalogues, larger galaxy catalogues were considered, such as HyperLEDA \citep{Paturel2003}, the Catalogue of Neighbouring Galaxies \citep{Karachentsev2004} or HECATE (Heraklion Extragalactic Catalogue, \citealt{HECATE}). Here we use a recent compilation of galaxy catalogues, GLADE (Galaxy List for the Advanced Detector Era, \citealt{Glade2018}), intended to help locate the origin of gravitational wave events. It is 100\% complete in bright galaxies (defined as those accounting for half of the integrated Schechter luminosity function, \citealt{Glade2018}) up to 91\,Mpc and 90\% complete at 200\,Mpc. Unlike more recent versions, its 2016 version, which is based on HyperLEDA, GWGC (the Gravitational Wave Galaxy Catalogue, \citealt{GWGC}) and 2MASX (the 2MASS extended source catalogue, \citealt{Skrutskie2006}), contains information on the extent of each galaxy. It contains more than 270000 entries at D<200\,Mpc, $\sim$30\% more than HECATE for the same distance range. We found that some extended infrared sources from 2MASX, considered as galaxies in GLADE, are actually  young stellar objects or diffuse emission from nebular regions in our Galaxy: from a visual inspection of the ones matching an X-ray source, we found that their B magnitude, inferred in GLADE, is generally brighter than 14, or that their \textit{Gaia} colours are distinct from actual galaxies (whose $G$ and $BP$ \textit{Gaia} magnitudes follow $G-BP>0$). We therefore removed the corresponding 2MASX entries. We also remove entries matching a stellar object in \textit{Simbad} within 10 arcsec (typical angular size of the closest young stellar objects), unless they are also categorised as galaxy. Since the major axis of 2MASX entries is missing in the catalogue, we retrieve the \texttt{r\_fe} column native of 2MASX. Last but not least, because GLADE is a compilation of catalogues, some duplicate entries are found. We retrieve 6725 such galaxies from a \textit{Sky} internal crossmatch of 10 arcsec. Although their distances in each catalogue are generally close to each other, large differences can occur when the distance is photometrically estimated. In these cases, we favour the distance of HyperLEDA or GWGC over the one of 2MASX, as they are more consistent with each other \citep{Glade2018}. To remove Galactic globular clusters present in GLADE, where we do not expect the presence of ULXs, we limit the sample to distances above 1~Mpc. The resulting galaxy sample is composed of $\sim$1.7 million galaxies essentially at $B_{mag}$ brighter than 19 (median $B_{mag}=17.3$). 

Unlike HECATE, GLADE does not contain any information on the galaxy morphology, star formation rate (SFR) or stellar mass. The latter is however estimated from infrared integrated luminosity in the latest release of GLADE, GLADE+ \citep{Glade2021}, so we retrieve the corresponding column from this catalogue. The Hubble type \texttt{t} of galaxies is given in HyperLEDA, so this column is retrieved as well and provides morphology information for $\sim$460000 galaxies. Such information is necessary to distinguish the study of ULX populations in spiral and elliptical galaxies, which are significantly different as mentioned in Section \ref{sec:1}. Similar to the definition used by 
\cite{Earnshaw2019}
and 
\cite{Bernadich2021}
, we define spiral and elliptical galaxies as those with $\texttt{t}\geq 0$ and $\texttt{t}<0$, respectively. Irregular galaxies are thus included in the spiral sample in the same way as 
\cite{Walton2011} 
and 
\cite{Walton2022}
. To get a more complete census of spiral and elliptical galaxies, we cross-correlate GLADE with the catalogue of galaxy morphologies inferred by machine learning on PanSTARRS images \citep{PS1morph}. As a result, we obtain stellar masses and morphologies for $\sim$1.5 million and $\sim$1 million galaxies, respectively. Last but not least, as done in HECATE and introduced by \cite{Cluver2017} (see also \citealt{Kennicutt2012}), we cross-correlate GLADE with the WISE (Wide-field Infrared Survey Explorer) catalogue of infrared sources \citep{Cutri2012} to estimate the SFR from the W3-band (12~$\mu$m) magnitude and the galaxy distance. More than 99\% of our GLADE subset has a WISE counterpart. This method provides consistent values of SFR for late-type galaxies, while the heated dust content generally dominates in early-type galaxies, often leading to an SFR overestimation (e.g. \citealt{Galliano2018, HECATE}). 

\subsection{ULX candidates}
\label{sec:ulx_offnuc}

To form the ULX sample, the first step is to identify X-ray sources matching a galaxy, but outside its nuclear region to avoid AGN contaminants. We used the matching tool of TOPCAT to this end, with the `sky ellipses' algorithm to retrieve all sources whose X-ray error circle overlap the ellipse representing the galaxy area. The major axis of the galaxy is chosen to be the Holmberg diameter $D_{Holm}=1.26\times D_{25}$ as in \citealt{Walton2022}, where $D_{25}$ is the isophotal diameter at surface brightness 25~mag~arcsec$^{-2}$, in order to retrieve more candidates and study the spatial distribution of ULXs out to large radii. The resulting samples of X-ray matches are used later as input for source classification. They contain 18506, 13055 and 7243 sources for \textit{Chandra}, \textit{XMM-Newton} and \textit{Swift}, respectively.

To exclude the central region, we select sources satisfying the criterion 
\begin{equation}
{\label{eq:0}
~~~~~~~~~~~~~~~~~~~~~~~~d>3(\texttt{POSERR}+0.5)    
}\end{equation}

where $d$ is the angular separation to the galaxy centre and $\texttt{POSERR}$ is the X-ray position error at $1\sigma$. A minimal offset of 3 arcsec is required. The 0.5 arcsec term is applied to correct for astrometric errors on the galaxy centre reported in GLADE -- it is the typical offset between GLADE and SDSS positions. The mean observed X-ray luminosity over the instrument broad band (0.2-12 keV for \textit{XMM-Newton}, 0.5-7~keV for \textit{Chandra} and 0.3-10 keV for \textit{Swift}) is computed from the mean X-ray flux given in the X-ray catalogues and the galaxy distance, and must strictly exceed $L_X>10^{39}$~erg~s$^{-1}$ for the source to be considered as a ULX candidate. Sources brighter than $10^{41}$~erg~s$^{-1}$ are kept in the same sample at this stage. 

\begin{figure}
    \centering
    \includegraphics[width=9cm]{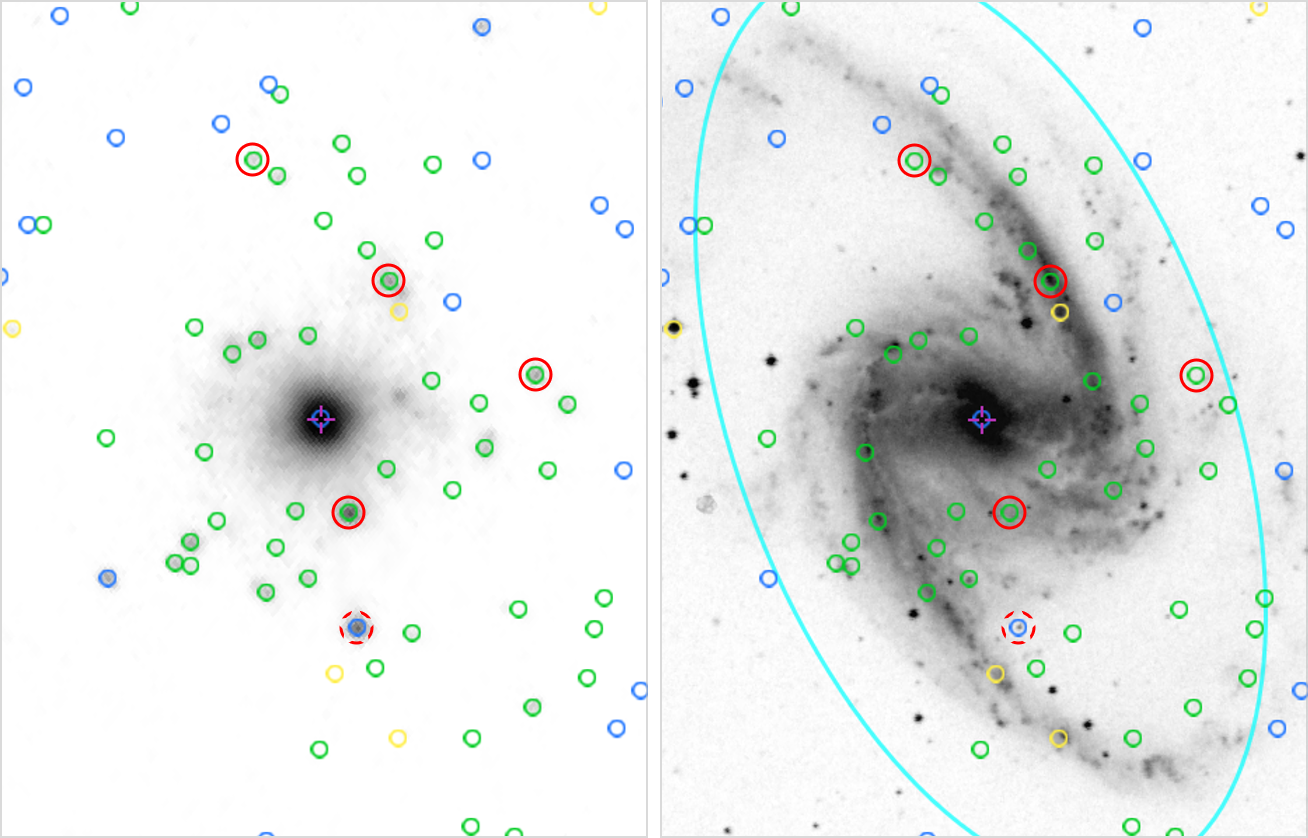}
    \caption{Illustration of ULX selection on the galaxy NGC~1365. (Left) \textit{XMM-Newton} X-ray image (Right) DSS Optical image. The $D_{25}$ ellipse of the galaxy is shown in grey. Blue, orange, green and red circles represent sources classified as AGN, soft sources, XRB and initial ULX candidates, respectively. The dashed red circle corresponds to a candidate classified as a contaminant, and is indeed a background AGN. An interactive view of this figure is available at \url{https://xmm-ssc.irap.omp.eu/claxson/xray_analyzer.php?srcquery=53.4019\%20-36.1406}}
    \label{fig:ngc1365}
\end{figure}

\subsection{Source classification to filter contaminants}

One main issue in ULX studies is the high rate of contaminants in ULX samples, due to the presence of foreground (essentially stars in our Galaxy) and background (AGN) sources. Some central AGN may also pass the criterion cited above because of bad astrometry. Previous attempts to remove these contaminants mainly focussed on removing known AGN and stars (e.g. \citealt{Kovlakas2020, Bernadich2021}) using the large existing datasets for these types. However, this is not sufficient to remove all contaminants, since many of them remain uncatalogued. Another option, applied in \cite{Bernadich2021}, is to remove all sources that are too bright in optical to be a ULX. To this end, they use \textit{Gaia} and PanSTARRS magnitudes obtained by a positional crossmatch with X-ray sources to compute the X-ray to optical flux ratio: any source with $\log(F_X/F_{Opt}) < -2.2$ is considered as a star, and other sources with $\log(F_X/F_{Opt}) < 0$ were visually inspected to flag likely background sources while keeping the bright HII star-forming regions where ULXs are preferentially found.

While efficient, this process is tedious and also misses all possible background AGN having $F_X>F_{Opt}$. To overcome this issue, we make use of the automated probabilistic classification of X-ray sources we recently developed \citep{Tranin2022}. In a nutshell, the dataset is first prepared by enriching the X-ray catalogue by completing X-ray sources with the following information: their optical and infrared counterparts from large ground-based surveys, identified with the Bayesian cross-matching tool \textsc{Nway} \citep{Salvato2018}; galaxies hosting the source, as explained in Section \ref{sec:3.3}; X-ray variability ratio between multi-instrument observations; and source identification given by external catalogues, to form the training sample. The classification scheme starts with a naive Bayes classifier based on the densities of the training sample for all source properties obtained after this enrichment. The scheme is then fine-tuned to maximise the classification performance of a chosen class (here X-ray binaries), by increasing the weight of the most discriminating properties. The probabilities of each class as well as the final class (giving maximum probability) are computed. Here, the X-ray sample of sources matching a GLADE galaxy is divided into a training sample of known sources and a test sample to classify. The classification is based on three classes: AGN (corresponding to background contaminants and most sources located in galaxy centres), X-ray binaries (the sample of in-situ genuine sources) and soft sources (a miscellaneous class containing stars, supernovae and supernova remnants, which are in-situ and foreground contaminants). The properties used in the classification process and the list of catalogues used to retrieve AGN, stars and XRB are detailed in Tables 1 and 2 of \cite{Tranin2022}. Supernovae and supernova remnants are retrieved through a match with \textit{Simbad}. The parameters of the classifier for each catalogue are summarised in Table \ref{tab:classif_param}, while the sample size of the training samples of \textit{XMM-Newton}, \textit{Chandra} and \textit{Swift} catalogues are detailed in Table \ref{tab:confmatrix}, as well as the classification results in the training samples. Four known ULXs are contained in the training samples of XRBs. The classifier is very efficient at retrieving AGN, and  retrieves more than 80\% of X-ray binaries. False positive rates are also low ($\lesssim$15\%) and can be further decreased by applying a probability threshold on the class under consideration.

\begin{figure}
    \centering
    \includegraphics[width=8.5cm]{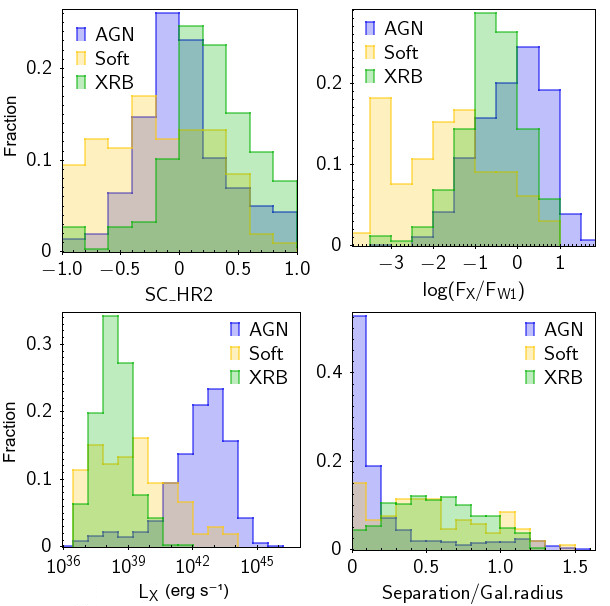}
    \caption{Densities of some source properties for each class of the 4XMM-DR11 training sample. \texttt{SC\_HR2}: second hardness ratio of 4XMM-DR11 (between bands 0.5-1 and 1-2 keV), from EPIC-pn and MOS cameras. $\log(F_X/F_{W1})$: logarithm of the X-ray to infrared (W1-band, 3.4~$\mu$m) flux ratio. $L_X$: mean 0.2--12~keV observed luminosity. Separation/Gal.~radius: galactocentric distance, i.e. the source separation to the host centre, in units of the host radius at the source position angle.}
    \label{fig:histo_classes}
\end{figure}

As a sanity check, we obtain that 99\% of sources at $L_X>10^{42}$~erg~s$^{-1}$  and 94\% of sources matching the galaxy centre (see  Equation (\ref{eq:0})) are classified as AGN. The vast majority of other central sources are classified as soft sources, mostly being absorbed (Seyfert II) AGN or hot gas.

\begin{table}
\caption{Classification parameters for 4XMM, CSC2 and 2SXPS.}
    \centering
    \begin{tabular}{cc}
         \multicolumn{2}{c}{4XMM} \\\hline
         Priors $\mathcal{P}_{AGN},~\mathcal{P}_{Soft},~\mathcal{P}_{XRB}$ & 67\%, 8\%, 25\%\\
         Weights $\alpha_{loc},~ \alpha_{spe},~ \alpha_{ctp},~ \alpha_{var}$ & 4.8, 9.1, 1.0, 5.5\\\\

         \multicolumn{2}{c}{CSC2} \\\hline
         Priors $\mathcal{P}_{AGN},~\mathcal{P}_{soft},~\mathcal{P}_{XRB}$ & 50\%, 10\%, 40\%\\
         Weights $\alpha_{loc},~ \alpha_{spe},~ \alpha_{ctp},~ \alpha_{var}$ & 6.9, 5.3, 3.6, 4.2\\\\

         \multicolumn{2}{c}{2SXPS}\\\hline
         Priors $\mathcal{P}_{AGN},~\mathcal{P}_{soft},~\mathcal{P}_{XRB}$ & 67\%, 8\%, 25\%\\
         Weights $\alpha_{loc},~ \alpha_{spe},~ \alpha_{ctp},~ \alpha_{var}$ & 9.6, 9.1, 1.3, 4.5\\\\
    \end{tabular}
    \tablefoot{For each catalogue are shown (first line) the prior proportions applied to each class and (second line) the weights for properties related to location, spectrum, multiwavelength counterparts and variability, once optimised to maximise the XRB classification performance. See Table~\ref{tab:classif_param} and Equation (3) of \cite{Tranin2022} for further details.}
    \label{tab:classif_param}
\end{table}

\begin{table}
    \caption{Confusion matrixes resulting from the classification of 4XMM-DR11, CSC2 and 2SXPS training samples.}
    \centering
    \begin{tabular}{ccccc}
    \multicolumn{5}{c}{4XMM}\\\hline
         &  AGN & Soft & XRB & Precision\\
        $\rightarrow$AGN & 1482 & 26 & 10 & 98\%\\
        $\rightarrow$Soft & 15 & 33 & 18 & 50\%\\
        $\rightarrow$XRB & 29 & 20 & 320 & 87\%\\
        Recall & 97\%& 42\%& 92\%& \\\\

    \multicolumn{5}{c}{CSC2}\\\hline
         &  AGN & Soft & XRB & Precision\\
        $\rightarrow$AGN & 996 & 27 & 58 & 92\%\\
        $\rightarrow$Soft & 25 & 140 & 27 & 73\%\\
        $\rightarrow$XRB & 81 & 80 & 830 & 84\%\\
        Recall & 90\% & 57\% & 91\% & \\\\

    \multicolumn{5}{c}{2SXPS}\\\hline
         &  AGN & Soft & XRB & Precision\\
        $\rightarrow$AGN & 1568 & 31 & 14 & 97\%\\
        $\rightarrow$Soft & 8 & 43 & 13 & 67\%\\
        $\rightarrow$XRB & 65 & 36 & 363 & 78\%\\
        Recall & 96\% & 39\% & 93\% & \\\\

    \end{tabular}
    \label{tab:confmatrix}
\end{table}

To remove contaminants, the probability that the source is an XRB is simply the probability that it is not a contaminant (indeed $P_{XRB}=1-P_{AGN}-P_{Soft}$). It is then compared to the fraction of contaminants in the galaxy area within the source separation. This fraction is obtained with the formula:

\begin{equation}\label{eq:1}
~~~~~~~~~~~~~~~~~~~~~~~~f_{cont}= 2\frac{\pi~a~b~n_{cont}~sep^2}{N_{ULX}}    
\end{equation}

where $a$ and $b$ are the galaxy semi-major and semi-minor axes, $n_{cont}$ is the density of contaminants computed with the analytical formula of the log(N)-log(S) relation in \cite{Moretti2003}, for the hard cosmic X-ray background, $sep$ is the galactocentric distance (i.e. the source separation in units of the galaxy radius at its position angle), and $N_{ULX}$ is the number of ULX candidates in the considered galaxy. Regardless of this quantity, sources with a probability to be a contaminant higher than $\sim$95\% are also classified as contaminant: it is notably the case of many foreground sources, and many of the few spurious sources remaining in the sample. Selected candidates are thus those following $P_{XRB}>\max(0.05, f_{cont})$. Here, $P_{XRB}$ should be understood as the probability that the ULX candidate is not a background or foreground contaminant, rather than the probability that the ULX is actually an XRB, which requires further study to be affirmed. 
From visual inspection of 150 sources, we estimate that more than 90\% of reliable ULXs are retrieved (as detailed in Section \ref{sec:2.8}) and that at most 15\% of selected candidates are compatible with background contaminants. Removed candidates represent $\sim$42\% of the initial candidates. This high fraction is the result of three factors: first, the selection out to $1.26 D_{25}$ increases the galaxy area by 60\% and the number of background contaminants by the same amount. Second, foreground contaminants and spurious sources (resulting from the X-ray detection pipelines) are also removed in this process. Third, a few valid ULX candidates are removed as well, because their properties used in this work are also compatible with an AGN (e.g. NGC 3921 ULX X-2, \citealt{Jonker2012}).

\subsection{Filtering remaining contaminants}

We visually verified a large number of our selected ULX candidates, which greatly helped to  develop and assess the filtering pipeline described above. To this end, we used the virtual observatory tools Aladin Lite\footnote{\url{https://aladin.u-strasbg.fr/AladinLite/}} \citep{Bonnarel2000,Boch2014}, the \textit{Simbad} database\footnote{\url{https://simbad.cds.unistra.fr/simbad/}} and the VizieR catalogue access tool\footnote{\url{https://vizier.cds.unistra.fr/viz-bin/VizieR}} \citep{Ochsenbein2000}, developed at CDS, Strasbourg Observatory, France.

Some contaminants remain after the filtering process described above: to remove them, we match our sample with \textit{Simbad} (3 arcsec) and exclude objects of types AGN or stellar objects (75\% of these objects were already successfully identified by the classification). We also visually inspect all sources with an optical counterpart and selected as ULX, having optical colour $b-r>0.5$, due to the redder nature of background AGN (about 700 sources). 
There are 135 sources from \textit{XMM-Newton}, 139 from \textit{Chandra} and 53 from \textit{Swift} that are discarded in this process, bringing the expected contamination rate to about 2\% (Section \ref{sec:nbcontam}). In the following, we refer to the resulting cleaned sample as "selected ULXs".

In the same way as the contamination rate, the false negative rate is expected to increase with luminosity, because the likelihood of the AGN class is enhanced (Figure \ref{fig:histo_classes}, lower left panel).

\subsection{HLX sample}
\label{sec:hlx_def}

Hyperluminous candidates were selected as described in Section \ref{sec:ulx_offnuc}, as off-nuclear sources with a mean observed X-ray luminosity $>10^{41}$~erg~s$^{-1}$ in the broad energy band. To limit the fraction of spurious sources, only candidates having a $S/N>3$ were kept. Respectively 195, 360 and 110 sources from CSC2, 4XMM and 2SXPS satisfy these criteria. Each of these HLX candidates was visually inspected to remove spurious sources (notably due to source confusion issues) and contaminants such as stars, hot gas overdensities and jet hotspots. This step led to the removal of 157 sources. Likewise, 33 sources were removed as being mistakenly associated to GLADE galaxies, whose extent was manifestly overestimated.

Unlike ULXs, which have well-constrained properties that are essentially similar to XRBs for the classifier, selecting HLXs from the classification results can induce important biases. In particular, bright, reliable HLX candidates having a Gaia \citep{gaiaedr3}, PanSTARRS \citep{Chambers2016} or DES (Dark Energy Survey, \citealt{DESDR1}) optical counterpart are misclassified as background AGN. A more robust approach to eliminate background AGN is to use the redshift (spectroscopic or photometric) of the optical counterpart (as done by \citealt{Barrows2019}). Recently, in addition to spectroscopic measurements of the redshift for a few million sources (e.g. in the SDSS-BOSS survey, \citealt{Bolton2012}), large and deep surveys led to the release of several billion photometric redshifts. In SDSS, this work was notably performed by \cite{Beck2016}, inferring the redshift of 208 million sources up to $z\sim 0.6$ from their (u,g,r,i,z) magnitudes. \cite{Tarrio2020} computed the photometric redshift of 1.1 billion sources using the (g,r,i,z,y) PanSTARRS bands, extending up to redshift $z\sim 1$. \cite{Zou2022} performed a similar work using the (g,r,i,z,Y) and (g,r,z,W1,W2) bands from the Dark Energy Survey and the DESI Legacy Survey, respectively. These three studies provide highly reliable photometric redshifts, with a typical accuracy better than $\sigma(\Delta z_{norm})=0.03$. Besides, most recently, identification as quasars or galaxies and redshifts were provided for 7.8 million \textit{Gaia} DR3 sources \citep{GaiaExtra2022}, using the low-resolution optical \textit{Gaia} spectra. We use this set of redshift catalogues to identify background and foreground contaminants: X-ray sources at $<3$~arcsec of an optical source with a redshift measurement inconsistent with the distance of the assumed host (i.e. $|z_{host}-z|>z_{err}$, with $z$ and $z_{err}$ the redshift value and error in the external catalogue) were discarded. This led to the removal of 283 background and 5 foreground sources, respectively. In contrast, sources having a distance match ($|z_{host}-z|<z_{err}$) have more chance to be indeed located in their associated galaxy: this is the case of 37 sources. At the end of this selection process, 115, 75 and 13 HLX candidates are found in 4XMM-DR11, CSC2 and 2SXPS respectively, totalling 191 unique sources including 11 observed by multiple X-ray instruments at the HLX level according to their mean luminosity (for X-ray sources observed with several instruments, the entry of smallest position error is kept).  From them, 63\% have no counterpart in \textit{Gaia}, PanSTARRS and DES catalogues, 24\% have a counterpart but no redshift measurement and 13\% have a redshift consistent with their host.

\subsection{Complete sub-sample}
\label{sec:2.7}

As introduced for example in \cite{Walton2011}, ULX properties and the luminosity function have to be assessed on a complete sample, so that the contribution of brighter ULXs is not overestimated. Indeed, the limiting sensitivity of the instrument defines the minimum flux that a source should have to be detected; for extragalactic sources, this translates into a limiting luminosity for each distance bin. Usually, ULX surveys keep only galaxies with limiting luminosity below $10^{39}$~erg~s$^{-1}$  to ensure that all ULXs are well-detected. For instance, \cite{Kovlakas2020} (resp. \citealt{Bernadich2021}) keep galaxies closer than 40~Mpc (resp. 29~Mpc). This leads to the removal of most bright ULXs located in more distant galaxies. In this study, we keep all sources with luminosities above the galaxy limiting luminosity $L_{lim}=4\pi F_{lim}D^2$. As further detailed in Section \ref{sec:3.1}, the contribution of ULXs in each luminosity bin is later weighted by the number of galaxies having $L_{lim}$ below this luminosity. 

\begin{figure}
    \centering
    \includegraphics[width=8.5cm]{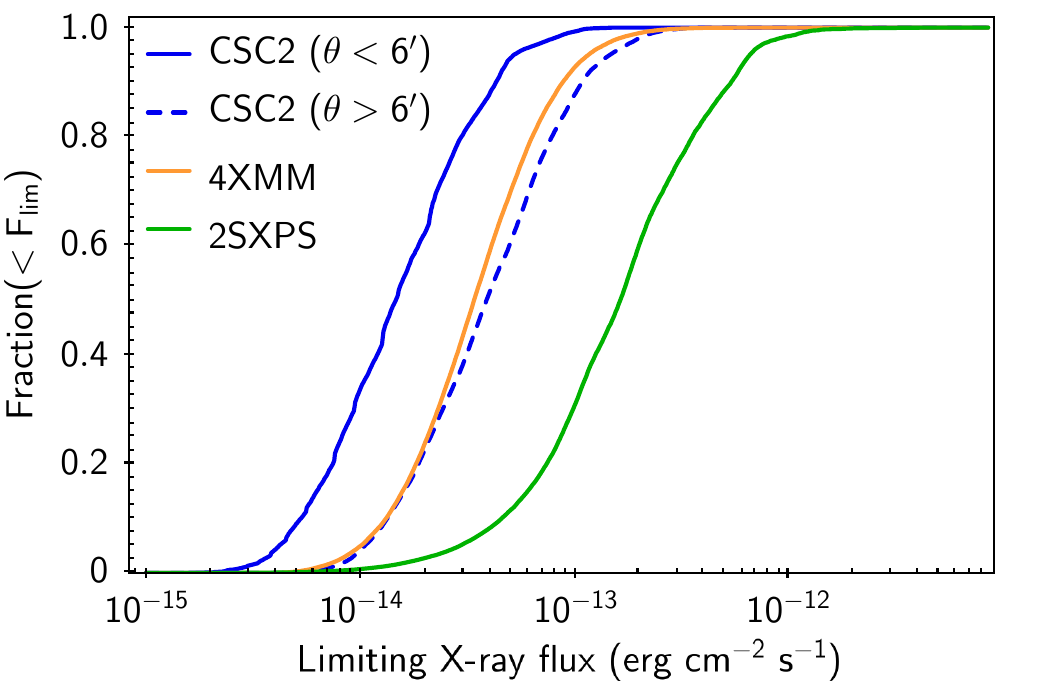}
    \caption{Cumulative distribution of GLADE galaxies X-ray 3$\sigma$ sensitivities, computed from Equation (\ref{eq:2}) as a function of flux. $\theta$ refers to the \textit{Chandra} off-axis angle.}
    \label{fig:flim_inst}
\end{figure}

The only prerequisite is thus to compute the sensitivity $F_{lim}$ for each galaxy in the sky coverage of each instrument. We use here a data-driven approach, for simplicity and because sensitivity maps are not accessible for all 3 instruments. For each X-ray catalogue, we infer the flux -- effective exposure time relation from subsets of $\sim 10^5$ detections with signal-to-noise close to 3. \textit{Chandra} sources show the highest deviation from a single power-law model, because a third parameter, the off-axis angle $\theta$, is determinant in the sensitivity value. We empirically find that once we consider a \textit{Chandra} effective exposure time $t'=2t/\max(\theta,2\text{ arcmin})$, the three X-ray subsets show a $\pm$0.5 dex deviation from a single power-law model. The resulting $F_{lim}$ relations are as follow:

 \begin{equation}\label{eq:2}
~~~~~~~~~~~~~~~~~~~~~~~~F_{lim} =
\begin{cases}
\big(\frac{1.2\times 10^{-15}}{t'}\big)^{0.75} & \text{for CSC2}\\
 \big(\frac{1\times 10^{-13}}{t}\big)^{0.8} & \text{for 4XMM}\\
 \big(\frac{5\times 10^{-13}}{t}\big)^{0.8} & \text{for 2SXPS}
\end{cases}
 \end{equation}

. The limiting X-ray sensitivities of \textit{Chandra}, \textit{XMM-Newton} and \textit{Swift} are illustrated in Figure \ref{fig:flim_inst}, showing the cumulative distribution of the limiting X-ray flux for GLADE galaxies in their sky coverage. The difference between on-axis and off-axis \textit{Chandra} galaxies is to be noted. Once applied to X-ray sources matching a GLADE galaxy, the $F>F_{lim}$ cut keeps 70 to 95\% of sources with $S/N>3$ and removes 80 to 90\% of sources with $S/N<3$. This empirical sensitivity can thus be considered as a proxy for the $3\sigma$ sensitivity.

\subsection{Sample of contaminants}
\label{sec:2.8}

To validate the automatic filtering method of the test sample, we compiled a sample of objects classified as contaminants from the three catalogues to compare it with the compilation of selected ULX. They are ULX candidates following $d > 3(\texttt{POSERR} + 0.5)$ and $P_{XRB}<\max(0.05,f_{cont})$, and having no identification as one of the three classes AGN, soft source or XRB in \textit{Simbad} or other catalogues, in order not to bias the classification assessment. This results in a sample of 1431 sources (1331 unique) with 494, 715 and 222 sources from CSC2, 4XMM and 2SXPS, respectively.

We visually inspected 100 sources having S/N>3 from this sample: 83\% were indeed contaminants (mostly background AGN), 7\% may be reliable ULXs missed by our classification, and 10\% were ambiguous cases, either lying far from the galaxy optical extent but having no optical counterpart, or having source confusion issues leading to unreliable properties.

\begin{figure}
    \centering
    \includegraphics[width=8.5cm]{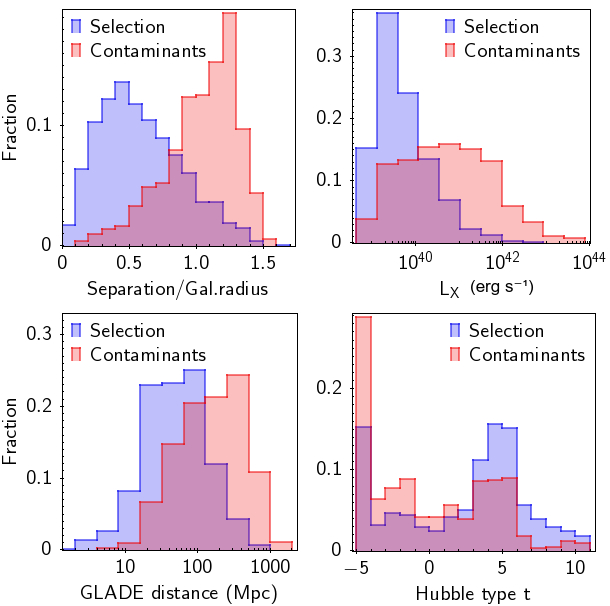}
    \caption{Normalised distributions of some properties of selected ULX candidates and candidates classified as contaminants, from the compilation of CSC2, 4XMM-DR11 and 2SXPS.}
    \label{fig:histo_contam}
\end{figure}

The distributions of galactocentric distance, mean X-ray luminosities, GLADE distance and Hubble type are shown in Figure \ref{fig:histo_contam} for both selected ULXs and contaminant samples. These results agree with what is reported in previous works: sources further from the galaxy centre, of higher luminosities or in elliptical galaxies are more prone to be contaminants. In particular, before the selection of the HLX candidates, all but six sources above $10^{42}$~erg~s$^{-1}$  were classified as contaminants. One of the 6 sources is 2CXO~J115324.3+493104, an AGN jet hotspot \citep{Sambruna2006}, proving the presence of unexpected types of contaminants in the ULX sample. We thus removed it from the HLX sample, as well as 2CXO~J003703.9-010904, another jet hotspot in the HLX luminosity range \citep{Martel1998,Kataoka2003}. 

\begin{figure}
    \centering
    \includegraphics[width=8.5cm]{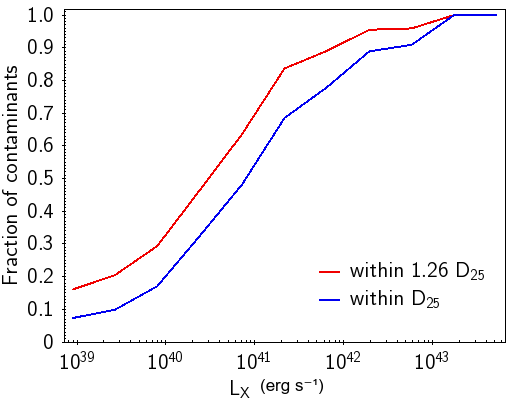}
    \caption{Fraction of contaminants per luminosity bin, for candidates in the $D_{25}$ and 1.26$\times D_{25}$ ellipses, among the initial ULX candidates (before filtering).}
    \label{fig:cont_frac}
\end{figure}

Figure \ref{fig:cont_frac} shows the fraction of sources classified as contaminants among initial ULX candidates as a function of the mean X-ray luminosity: in particular, for a search within the Holmberg diameter 1.26$\times D_{25}$, more than 30\% of candidates above $10^{40}$~erg~s$^{-1}$  are contaminants, even after removing known contaminants. This evolution is quantitatively consistent with the 70\% contamination rate observed for HLX candidates \citep{Zolotukhin2016,Kaaret2017}, although this is dependent on the apparent size of selected galaxies. Consequently, previous analyses of the bright end of the X-ray luminosity function of ULXs (e.g. \citealt{Swartz2011,Mineo2012,Wang2016}) may be severely affected by contamination.

\section{Results}
\label{sec:3}

The final ULX samples are made up of 1234, 667 and 304 sources from \textit{Chandra}, \textit{XMM-Newton}, and \textit{Swift}, totalling a compiled sample of 1901 unique sources (i.e. after removing duplicate entries for sources observed by several instruments; this sample is hereafter called the compiled sample). The rates of HLX candidates in each sample are 6\%, 16\% and 1.9\%, respectively. This difference is expected because of the different capabilities of these X-ray facilities, as detailed in Section \ref{sec:3.1}. Table \ref{tab:ulx_counts} summarises the number counts of ULXs, complete ULXs (i.e. with $L_X>L_{lim}$), HLXs, and galaxies of different morphologies in each ULX sample. To our knowledge, our work provides both the cleanest (about 2\% of contaminants, Section \ref{sec:nbcontam}) and the largest census of ULXs, just above the recently published multi-mission catalogue of \cite{Walton2022} (1843 sources). Only a handful of well-known ULXs are absent from our sample. For instance, out of the 17 well-studied ULXs compared in \cite{Gurpide2021}, only three ULXs are missed: Holmberg~IX~X-1 and Circinus~X-5, because the $1.26\times D_{25}$ ellipse given in GLADE are somewhat smaller than the actual galaxy area ; and NGC~55~X-1, because its mean X-ray luminosity is just below $10^{39}$~erg~s$^{-1}$ for each instrument. All the pulsating ULXs known so far, including the recent pulsating ULX (PULX) candidate of \cite{Quintin2021}, are retrieved in our sample. ESO 243-49 HLX-1 (along with other HLXs, Section \ref{sec:hlxsample}) is retrieved as well. Conversely, supersoft ULXs studied by \cite{Urquhart2016} are not in our sample because of having $\langle L_X \rangle <10^{39}$~erg~s$^{-1}$.

\begin{table}
\caption{Samples of ULXs, HLXs and galaxies under study}
    \centering
    \begin{tabular}{c|cccc}
         &  CSC2 & 4XMM & 2SXPS & Total (unique)\\\hline\\ [-1.5 ex]
        ULX & 1234 & 667 & 304 & 2205 (1901)\\
        cULX & 925 & 470 & 266 & 1661 (1448)\\
        HLX & 75 & 115 & 13 & 203 (191) \\
        cHLX & 63 & 92 & 13 & 168 (158)\\\\ [-1 ex]
        Galaxies & 741 & 599 & 207 & 1547 (1303)\\
        Spiral & 377 & 320 & 171 & 868 (713)\\
        Elliptical & 206 & 112 & 19 & 337 (325)\\
    \end{tabular}
    \tablefoot{Number counts of ULXs, complete ULXs (i.e. with $L_X>L_{lim}$, Section \ref{sec:2.7}), HLXs, complete HLXs and number of galaxies that host them (all types, spiral and elliptical), for CSC2,  4XMM-DR11, 2SXPS and their sum (in parenthesis: minus duplicates). In the compilation of unique sources, for X-ray sources observed with several instruments, the entry of smallest position error is kept. By construction of the HLX sample, the sources all have $S/N>3$ and thus a higher fraction of them are complete.}
    \label{tab:ulx_counts}
\end{table}

\subsection{Malmquist-corrected XLF}
\label{sec:3.1}

Previous works on the X-ray luminosity function of ULXs only kept galaxies with limiting X-ray luminosities $L_{lim}<10^{39}$~erg~s$^{-1}$  . While this method gives the intrinsic shape of the XLF, by probing the same volume in all bins, it also cuts out most of the detected bright sources, located in further galaxies. Here, we keep all sources brighter than the limiting luminosity of their galaxy $L_X>L_{lim, host}$. Thus, the distribution of $L_X$ is the XLF convolved with the survey volume, each luminosity bin $[L_X,L_X+dL_X]$ comes from a different survey volume (Malmquist bias, also known as the Eddington bias), and a 1/$V_{max}$ correction has to be applied to each bin. This volume $V_{max}$ is here given by the number of galaxies complete to the luminosity of the bin, that is to say having $L_{lim}<L_{X,bin}$. The deconvolved differential XLF is computed by dividing the luminosity distribution of ULXs by the cumulative limiting luminosity distribution of galaxies, $N(L_{lim}<L_{X,bin})$. Although this approach has been extensively used to constrain the luminosity functions of stars (e.g. \citealt{Leggett1988,Stobie1989,Kroupa1995}), galaxies (e.g. \citealt{Binggeli1988,Loveday1992,Bouwens2011}) and quasars (e.g. \citealt{LyndenBell1971,Ueda2003,Aird2010}), it seems absent from the literature of XRB and ULX statistical studies at the time of writing. The underlying assumption is that a universal XLF exists and is the same in the different survey volumes, which seems reasonable since we study the very local Universe $z<0.16$ (however see Section \ref{sec:4.1.2} for further discussion).

\begin{figure}[h!]
    \centering
    \includegraphics[width=8.5cm]{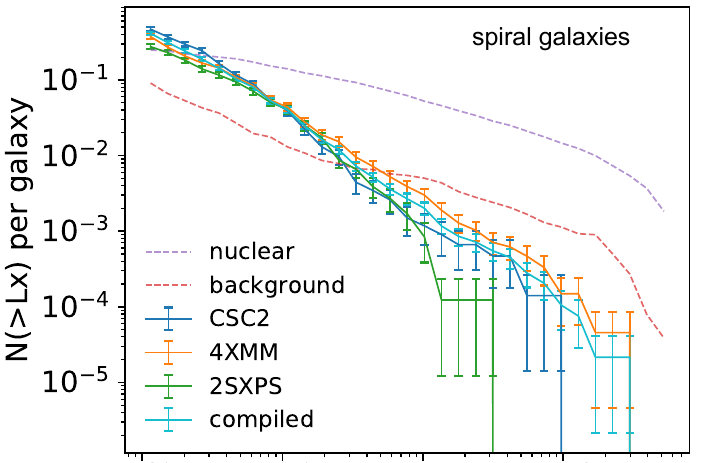}
    \includegraphics[width=8.4cm]{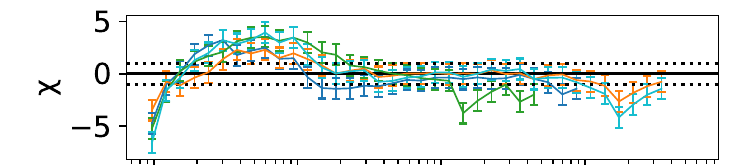}
    \includegraphics[width=8.5cm]{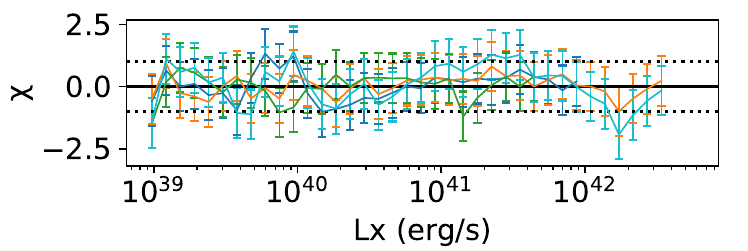}\vspace{0.5cm}
    \includegraphics[width=8.5cm]{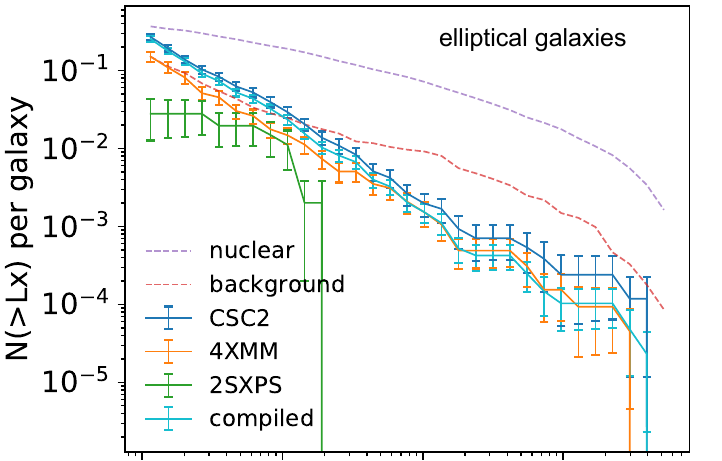}
    \includegraphics[width=8.5cm]{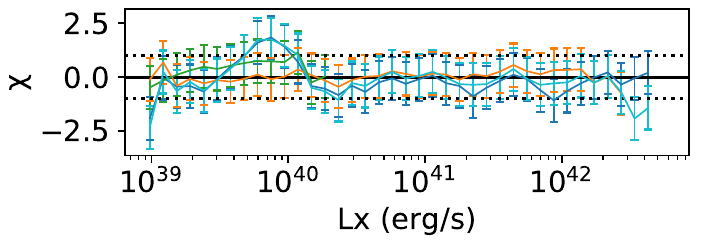}
    \caption{Cumulative X-ray luminosity functions of ULX candidates from 4XMM, CSC2 and 2SXPS. (Top panel) Deconvolved XLF of spiral galaxies, and residuals from the single (middle) and broken (bottom) power-law fits detailed in Table \ref{tab:bpl_params}. (Bottom panel) Deconvolved XLF of elliptical  galaxies, and residuals from the single power-law fits detailed in Table \ref{tab:pl_params}. For comparison, we overlay the cumulative distributions of sources matching their host nucleus, and of ULX candidates classified as background contaminants.}
    \label{fig:xlf}
\end{figure}

Figure \ref{fig:xlf} shows the deconvolved cumulative XLF (complete in each bin) for spiral  (top panel) and elliptical  (bottom panel) galaxies. It includes selected ULXs below $10^{41}$~erg~s$^{-1}$ from the complete sample, and complete robust HLXs for the bright end. Poisson errors are assumed in each bin on the number of X-ray sources and the number of complete galaxies. The three datasets are in good agreement (consistent at the 90\% level except at faint luminosities where the different resolution capabilities lead to substantially different numbers of detections, Section \ref{sec:instbias}) after making a correction on the CSC2 sample. Indeed, we noted a deficit of \textit{Chandra} sources at medium luminosities ($\gtrsim 10^{40}$~erg~s$^{-1}$  ), which after a thorough inspection of 4XMM-CSC2 intersecting fields was found to be caused by genuine ULXs that were flagged in CSC2 as extended or confused, or not detected by the \textit{Chandra} pipeline which is less sensitive close to the edges of the field of view. After visual inspection of 480 flagged sources, we retrieved 199 such sources including 51 at $L_X>10^{40}$~erg~s$^{-1}$  .  

We verified that both deconvolved and ULX-complete XLF had the same shape. Notably, we considered sources at low luminosities $10^{39}- 10^{40}$~erg~s$^{-1}$ in spiral galaxies, so that the sample becomes comparable to that of \cite{Wang2016} (who have only 20 candidates above this luminosity). Using a single power-law model, as they did in this luminosity range, we retrieve an XLF slope consistent with their result, $\alpha = 0.93 \pm 0.03$ instead of $\alpha = 0.96 \pm 0.05$ (1$\sigma$ errors). In elliptical  galaxies, we generally find flatter slopes than in the literature; this point is discussed in Section \ref{sec:xlfslop}.

Each XLF is then fitted with two models, a single power law 
\begin{align*}
~~~~~~~~~~~~~~~~~~~~~~~~n(>L)=n_{39} (L/10^{39})^{\alpha}    
\end{align*}

and a broken power law
\begin{align*}
~~~~~~~~~~~~~~~~~~~~~~~~&n(>L)=n_{39} (L/10^{39})^{\alpha_1}~\mathrm{if~}L<L_b\\
    &n(>L)=n_{39} (L/10^{39})^{\alpha_2}~\mathrm{~otherwise}
\end{align*}

where $n_{39}$ is the total ULX rate and $L_b$ is the break luminosity. Parameters of the XLF fits and their uncertainties are probabilistically estimated in a bayesian framework, by sampling the XLF with 40000 Monte-Carlo trials using observed values and errors. Flat priors are applied on each parameter, in the ranges $n_{39} \in [0,1]$, $\alpha \in [0,2]$ (single power law) and $n_{39} \in [0,1]$, $\alpha_1 \in [0.3,2]$, $\alpha_2 \in [0,2]$, $\log(L_{b}) \in [39,41]$ (broken power law). Results from the single power law and broken power law fits of the deconvolved cumulative XLF are detailed in Tables \ref{tab:pl_params} and \ref{tab:bpl_params}. Unlike the XLF of elliptical galaxies, in spiral galaxies, the broken power-law fit is always preferred over a single power-law fit. However, the different catalogues disagree on the exact location of the break, even at the three sigma level. This discrepancy is further discussed in Section \ref{sec:4.1.2}. However, it is to be noted that all three datasets are well-fitted by a power law breaking at $L_{X,break}=5\times 10^{39}$~erg~s$^{-1}$  (fixed parameter), with $\chi^2_r<1$.

Unlike the lower luminosity break seen in the XLF of elliptical  galaxies at a few $10^{38}$ erg~s$^{-1}$ which is consistent with the Eddington limit of neutron star binaries \citep{KimFabbiano2010}, the physical origin of this higher luminosity break has been poorly understood. Previous works consider it as the suggestion for a different class of objects above the break luminosity, in particular intermediate mass black holes, as it corresponds to the Eddington luminosity of a $\sim 80 \mathrm{M}_\odot$ black hole \citep{Swartz2011,Wang2016,Kaaret2017}. The reasons for this break are further examined in Section \ref{sec:xlfbk}.


\begin{table}
\caption{Parameters of the fits of the deconvolved XLF with a single power-law model.}
    \centering
    \begin{tabular}{cccc}
        \hline
                Sample & $n_{39}$ & $\alpha$ & $\chi^2_r$ (dof)\\ 
                \hline\\ [-1.5ex]
                CSC2, ETG & $0.31^{+0.01}_{-0.02}$ & $0.99^{+0.02}_{-0.02}$ & 0.54 (28)\\ 
                4XMM, ETG & $0.16^{+0.02}_{-0.03}$ & $0.99^{+0.05}_{-0.04}$ &0.09 (27)\\ 
                2SXPS, ETG & $0.03^{+0.02}_{-0.01}$ & $1.02^{+0.30}_{-0.24}$ &0.68 (9)\\ 
                compiled, ETG & $0.29^{+0.01}_{-0.01}$ & $1.06^{+0.02}_{-0.02}$ &0.67 (28)\\ 
                \\ [-1.5ex] \hline
    \end{tabular}
    \tablefoot{ `ETG' refer to the sample of complete ULXs in elliptical galaxies. $n_{39}$ and $\alpha$ are the amplitude and slope of the power-law model, respectively. See the text for more details.}
    \label{tab:pl_params}
\end{table}

\begin{table*}
\caption{Parameters of the fits of the deconvolved XLF of spiral galaxies with a broken power-law model.}
    \centering
    \begin{tabular}{cccccc}
        \hline
                Sample & $n_{39}$ & $\alpha_1$ & $\alpha_2$ & $\log(L_{break})$ & $\chi^2_r$ (dof)\\ 
                \hline \\ [-1.5ex]
                CSC2, LTG & $0.45^{+0.02}_{-0.02}$ & $0.67^{+0.09}_{-0.10}$ & $1.36^{+0.06}_{-0.05}$ & $39.45^{+0.07}_{-0.06}$ & 0.37 (27)\\ 
                4XMM, LTG & $0.35^{+0.02}_{-0.02} $ & $0.81^{+0.06}_{-0.07}$ & $1.24^{+0.05}_{-0.04}$ & $39.73^{+0.11}_{-0.11} $  & 0.19 (33)\\ 
                2SXPS, LTG & $0.29^{+0.02}_{-0.02} $  & $0.80^{+0.04}_{-0.05}$ & $1.73^{+0.12}_{-0.10}$ & $39.97^{+0.07}_{-0.08}$ & 0.33 (24)\\ 
                compiled, LTG & $0.41^{+0.01}_{-0.01} $& $0.86^{+0.04}_{-0.06}$ & $1.31^{+0.03}_{-0.03}$ & $39.66^{+0.08}_{-0.09}$ & 0.77 (33)\\ \\ [-1.5 ex]
                CSC2, LTG39 & $240^{+12}_{-12} $ & $0.80^{+0.08}_{-0.10}$ & $1.90^{+0.07}_{-0.14}$ & $39.67^{+0.08}_{-0.11}$ & 1.13 (12) \\ 
                4XMM, LTG39 & $168^{+10}_{-10} $ & $0.71^{+0.08}_{-0.09}$ & $1.80^{+0.14}_{-0.17}$ & $39.74^{+0.10}_{-0.10}$ & 0.17 (14)\\ 
                2SXPS, LTG39 & $150^{+10}_{-10} $ & $0.65^{+0.10}_{-0.12}$ & $1.39^{+0.25}_{-0.17}$ & $39.70^{+0.20}_{-0.19}$ & 0.13 (18)\\ 
        compiled, LTG39 & $424^{+15}_{-15} $ & $0.89^{+0.05}_{-0.06}$ & $1.97^{+0.03}_{-0.06}$ & $39.71^{+0.05}_{-0.06}$ & 0.60 (12)\\ 
                        \\ [-1.5ex] \hline
    \end{tabular}
    \tablefoot{`LTG' refer to the samples of complete ULXs in spiral galaxies. `LTG39' stands for the samples of spiral galaxies having $L_{lim}<10^{39}$~erg~s$^{-1}$; they correspond to the classical, non-renormalised XLF. $n_{39}$, $\alpha_1$, $\alpha_2$ and $\log(L_{break})$ are the amplitude, faint-end slope, bright end slope and break luminosity of the broken power-law model, respectively. See the text for more details.}
    \label{tab:bpl_params}
\end{table*}

\subsection{ULX rates in different environments}
\label{sec:ulxenv}

Generally speaking, precursory studies found that spiral galaxies were more prone to host ULXs, in particular in their star-forming regions (e.g. \citealt{King2001,Swartz2004,LiuBreg2005}). Elliptical galaxies also host a significant ULX population, especially those that have undergone a recent star-formation event \citep{KimFabbiano2004}. The luminosity function of ULXs and their location in their host (distance to the centre) seem to be consistent with the extrapolation of HMXB in spiral galaxies and LMXB in elliptical galaxies, at higher luminosities \citep{Gilfanov2004,Swartz2011,Mineo2012,Kovlakas2020}. Other demographic studies showed an excess of ULXs in dwarf galaxies and in low-metallicity galaxies (e.g. \citealt{Swartz2008}, \citealt{Kovlakas2020}). The scaling relations linking ULX rates with galaxy mass, SFR, and metallicity were even calibrated in recent works \citep{Anastasopoulo2019,Kovlakas2020,Lehmer2021}.

\begin{figure}
    \centering
    \includegraphics[width=8.5cm]{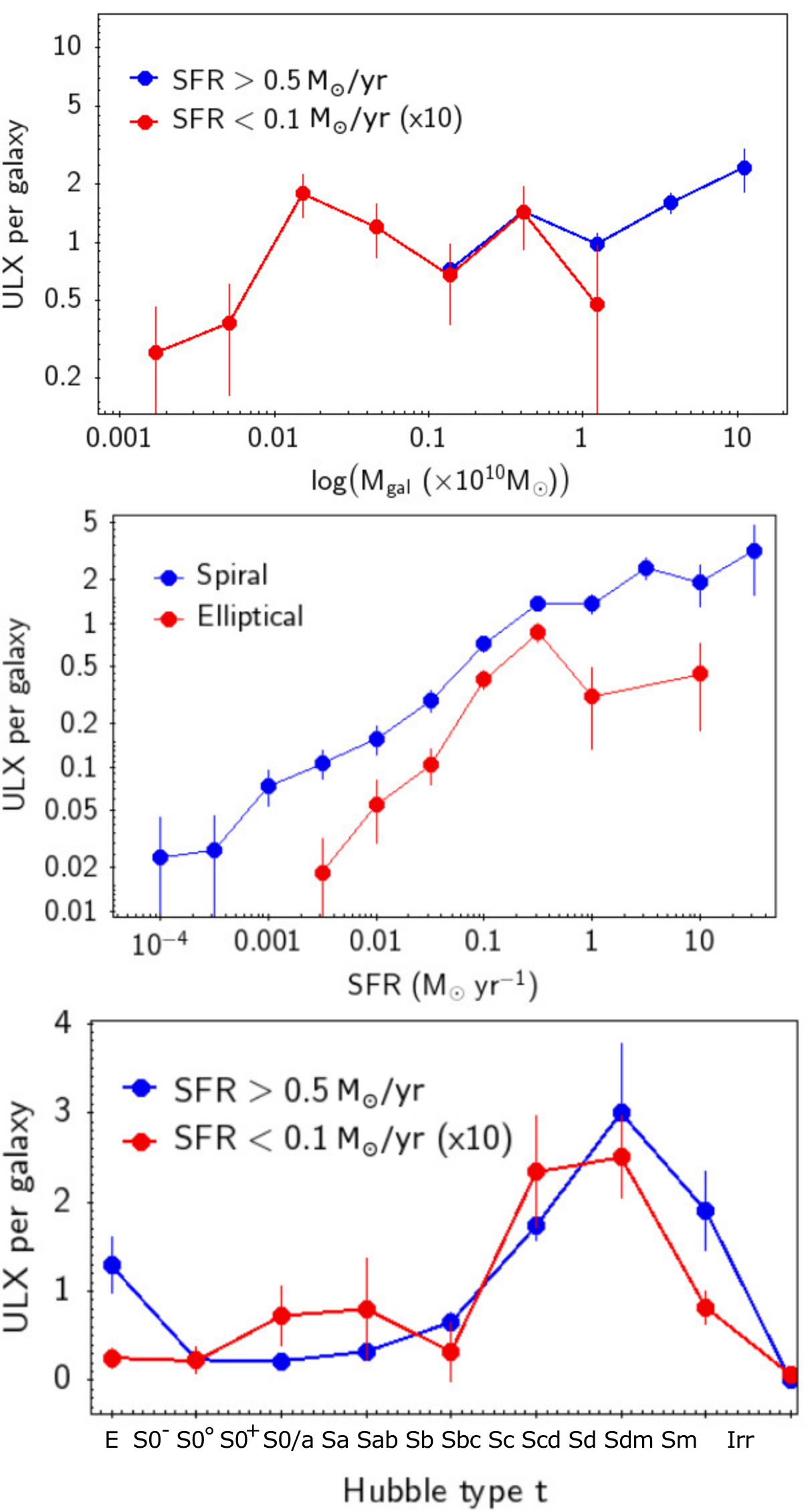}
    \caption{Rates of ULX as a function of different properties characterising galaxy environment.}
    \label{fig:env_ulx}
\end{figure}

Figure \ref{fig:env_ulx} shows the evolution of ULX rates $n_{ULX}$ with various galaxy parameters. The qualitative trends reported in the literature are also present in our cleaned ULX sample. We find that ULX rates globally increase with the galaxy stellar mass, regardless of the SFR. Besides, ULX rates increase with the SFR, at least in spiral galaxies. The rate of ULX is higher in spiral galaxies, regardless of the SFR; and a significant ULX population does exist in elliptical galaxies.

Modelling this ULX rate -- SFR relation with a power law, $\log(n_{ULX}) =  \alpha \log(\mathrm{SFR}) + \beta$ where $\alpha$ is the slope and $\beta$ the normalisation, we obtain $\alpha = 0.43 \pm 0.04$ for spiral galaxies shown in the middle panel of Figure \ref{fig:env_ulx}. This is in excellent agreement with the value obtained by \cite{Kovlakas2020} on average over all types of spiral galaxies: $\alpha=0.45^{+0.06}_{-0.09}$. Elliptical galaxies seem to present a sharper scaling relation, with a slope $\alpha=0.84 \pm 0.07$ at $\mathrm{SFR}<1~\mathrm{M}_\odot$~yr$^{-1}$ and a drop above this value. However, the SFR estimator we used is unreliable for elliptical  galaxies, being degenerate with the dust mass, so this result is inconclusive. In the top panel of Figure \ref{fig:env_ulx}, we can see an excess of ULXs in dwarf galaxies (hardly significant since it disappears with a different binning) and a tenfold increase of the ULX rate over five orders of magnitude in stellar masses covered by the sample: the specific ULX rate $n_{ULX}/M_*$ thus decreases with stellar mass, in agreement with \cite{Walton2011} and \cite{Kovlakas2020}.

\begin{figure}
    \centering
    \includegraphics[width=8.5cm]{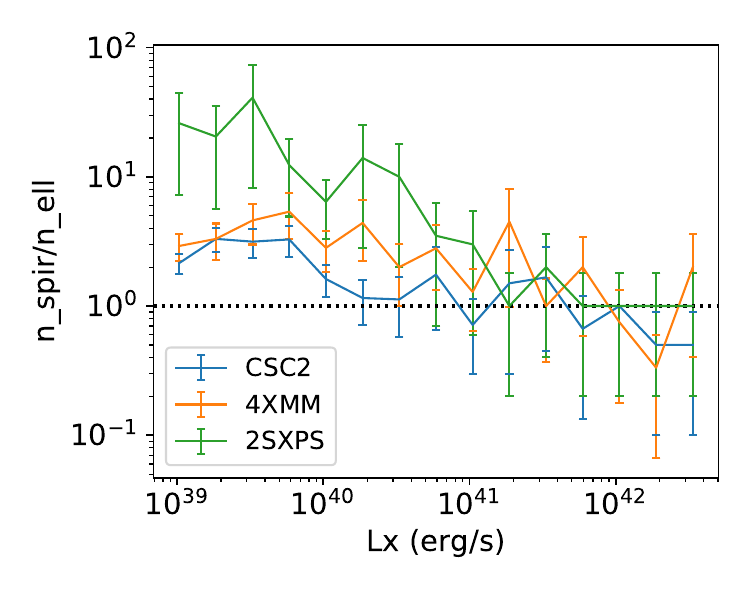}
    \caption{Evolution with X-ray luminosity of the ratio between ULX rates $N(L_X)$ in spiral  and elliptical  galaxies.}
    \label{fig:spiell_ratio}
\end{figure}

We find tentative evidence that ULXs and HLXs do not share exactly the same environment: the latter tend to be hosted in equal rates in spiral and elliptical galaxies (Figure \ref{fig:spiell_ratio}). We note that this trend was reported in \cite{Bernadich2021}, but not found in \cite{Kovlakas2020}. This trend can be interpreted in terms of the mean slope of the XLF of ULXs, which (in our sample) is shallower for elliptical  galaxies than for spiral galaxies. This is in tension with the trend reported in the literature (e.g. \citealt{Wang2016}). This point is discussed in Section \ref{sec:xlfslop}.

\begin{figure}
    \centering
    \includegraphics[width=8.5cm]{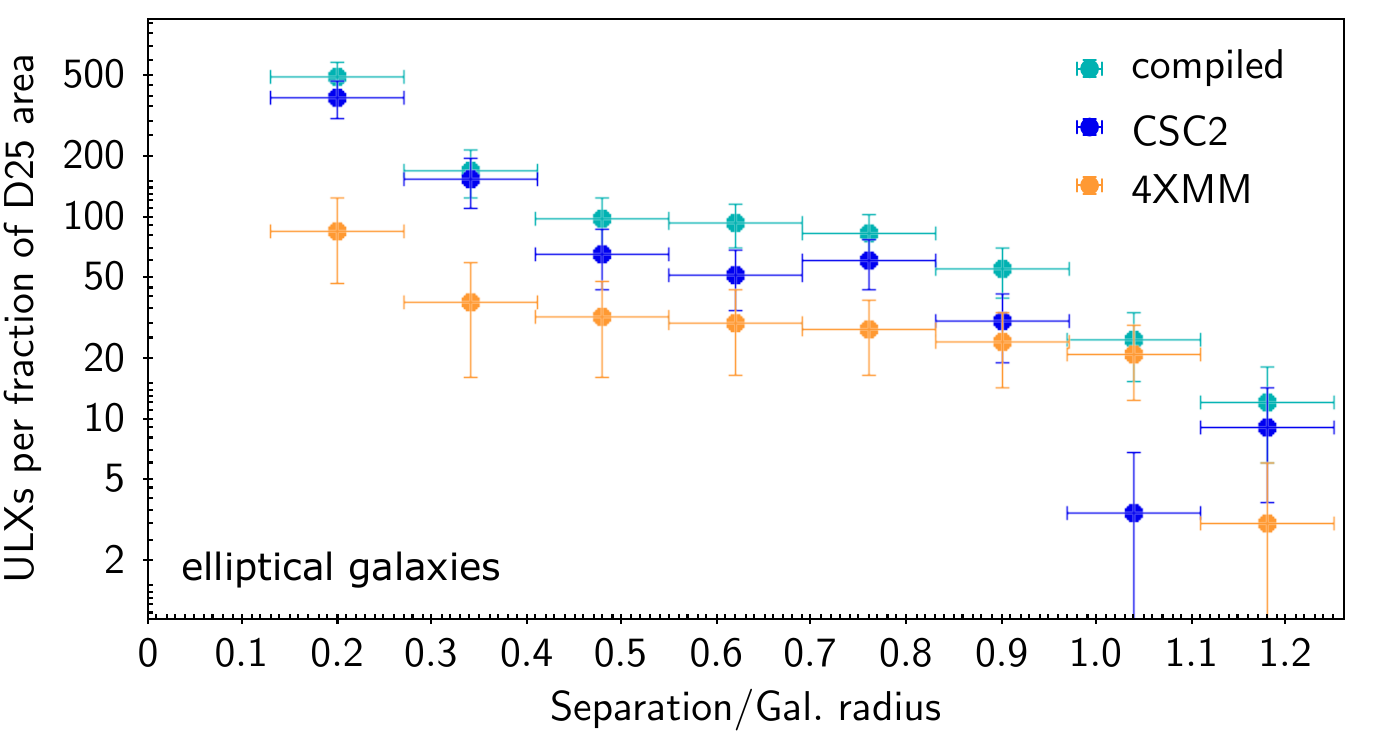}
    \includegraphics[width=8.5cm]{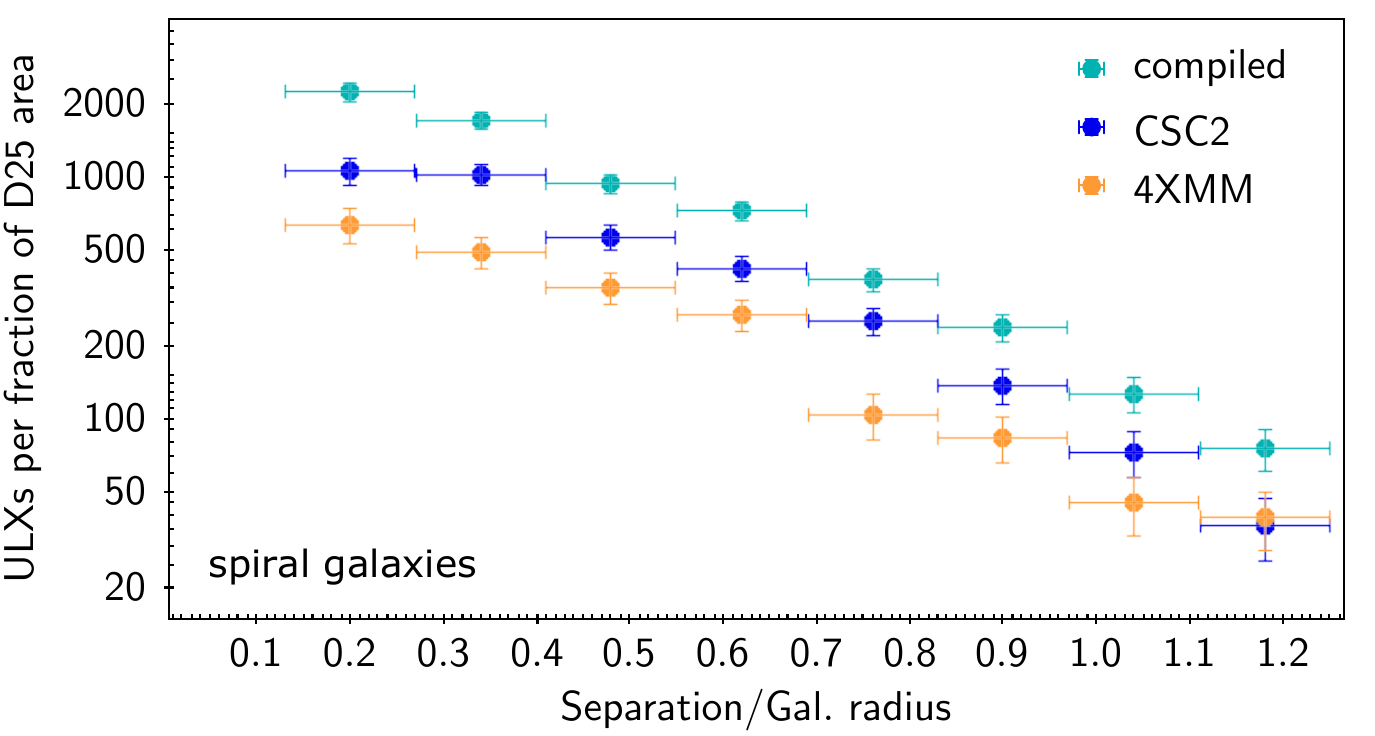}
    \includegraphics[width=8.5cm]{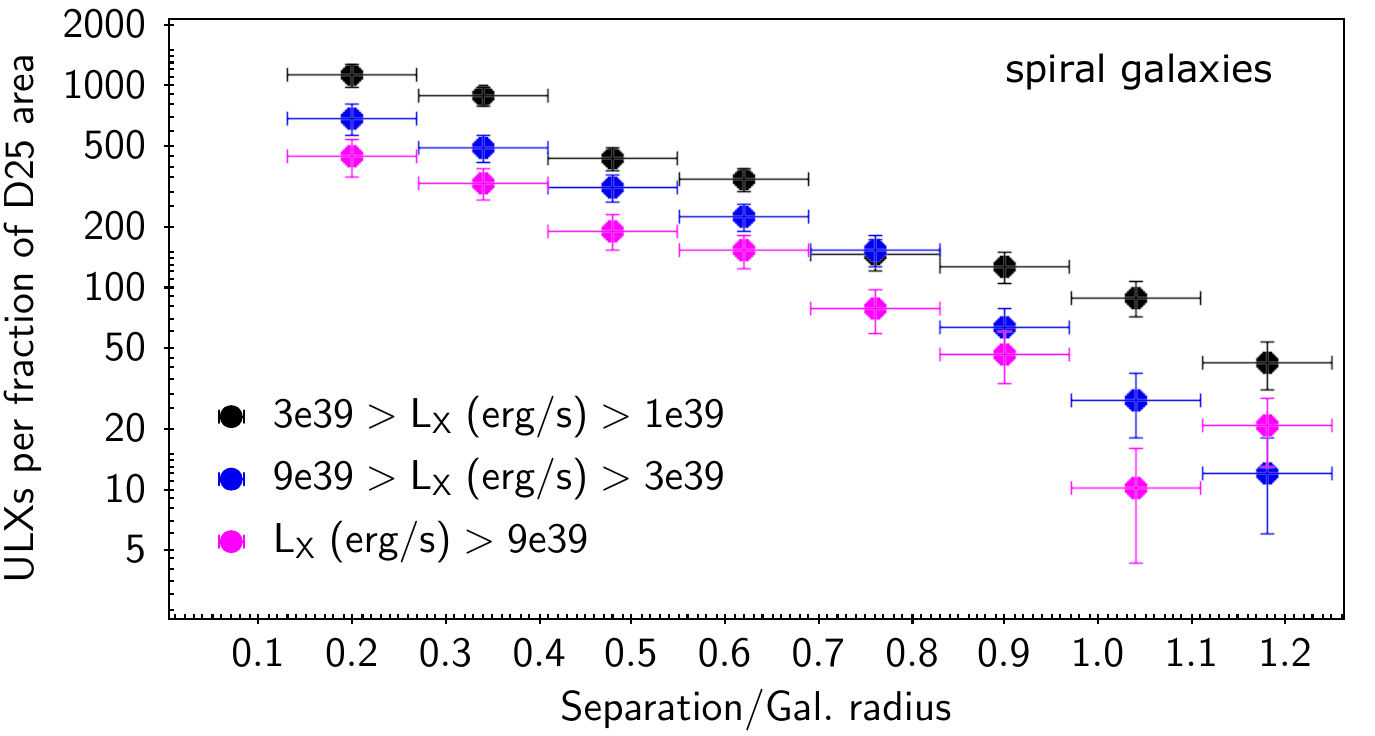}
    \caption{Radial distribution of ULXs in elliptical  (top) and spiral  (middle, bottom) galaxies, showing their surface density as a function of galactocentric distance. The last panel shows the surface density of different luminosity bins in the compilation of \textit{Chandra}, \textit{XMM-Newton} and \textit{Swift} selected ULXs.}
    \label{fig:surfdens}
\end{figure}

The radial distribution of ULXs in their host also gives constraints on their favourable environments. However, it cannot be studied in galaxies of small angular size or large distance, due to source confusion issues and the exclusion of ULXs close to the nucleus. We thus examine the number of ULXs in spiral and elliptical galaxies having a major axis $D_{25}>30$~arcsec, as a function of their galactocentric distance. The result is shown in Figure \ref{fig:surfdens}, where the first bin is not shown since the ULX census in this inner part of the galaxy ($\lesssim 0.15\times D_{25}$) is incomplete. Regardless of the sample in use, the surface density of ULXs in spiral galaxies smoothly decreases towards larger separations, while it becomes flatter in elliptical galaxies before being cut off beyond $D_{25}$. In the sub-sample of spiral galaxies, containing enough ULXs to probe several luminosity bins, the form of the distribution seems independent of the source luminosity. This is in agreement with the results of \cite{Kovlakas2020}, performing the same analysis in galaxies closer than 40~Mpc. However, our ULX selection extending beyond the $D_{25}$ ellipse as well as our filtering of contaminants allow us to observe for the first time the drop in ULX density in the outer part of elliptical galaxies.

\subsection{Comparison of XRB and ULX}
\label{sec:3.3}
It is now widely accepted that most ULXs below $\sim 10^{41}$~erg~s$^{-1}$  contain either neutron stars or stellar-mass black holes accreting above the Eddington limit. However, the geometry of their accretion flow, as well as the mechanisms explaining their spectral states are still poorly understood. 
From the few nearby individual ULXs that have been observed multiple times with large exposures, two decades of detailed spectral study led to the results cited in the introduction. 

For the three X-ray catalogues studied in this work, XRB are well-detected only in nearby galaxies ($D\lesssim 20$~Mpc) which have larger apparent sizes and thus overlap numerous background sources. For this reason, and because they correspond to a source type actually present in the training sample, we select them using a different criterion to avoid background and foreground contaminants: sources following $P_{XRB}>0.7$, $sep<0.8$ and $L_X<10^{39}$~erg~s$^{-1}$  make up our sample of XRB candidates. The contamination rate among these selected candidates is 10--12\% from visual inspection.

\subsubsection{Hardness}

\begin{figure}
    \centering
    \includegraphics[width=8.5cm]{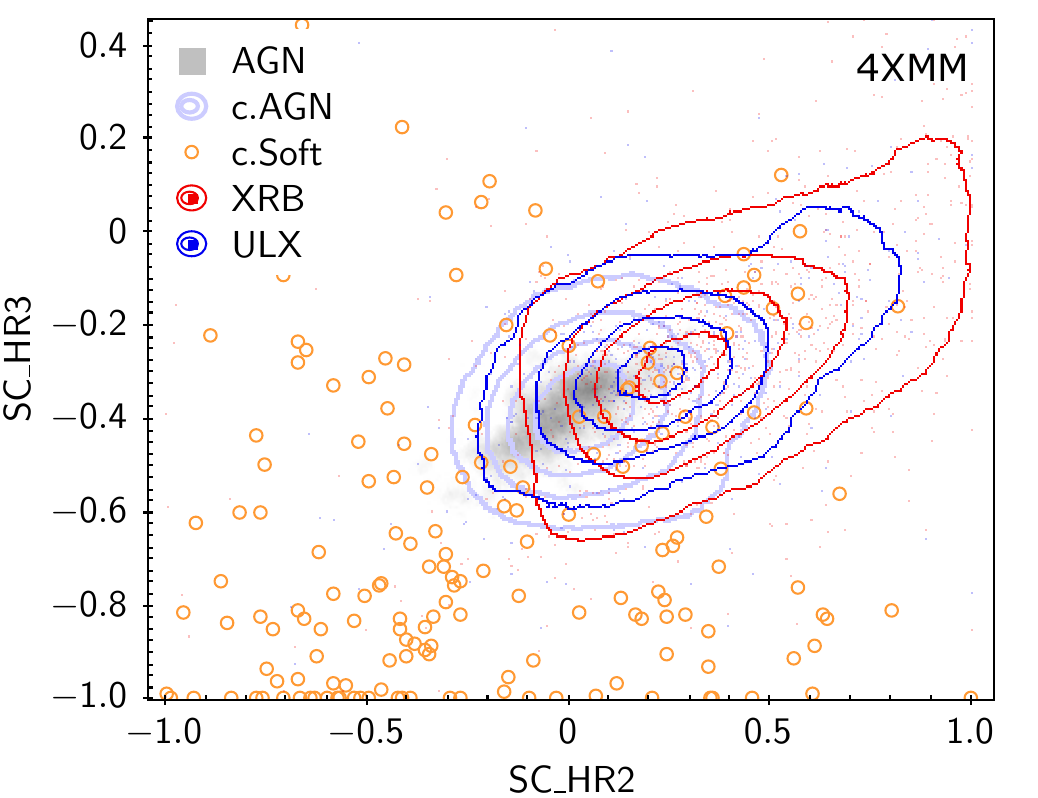}
    \includegraphics[width=8.5cm]{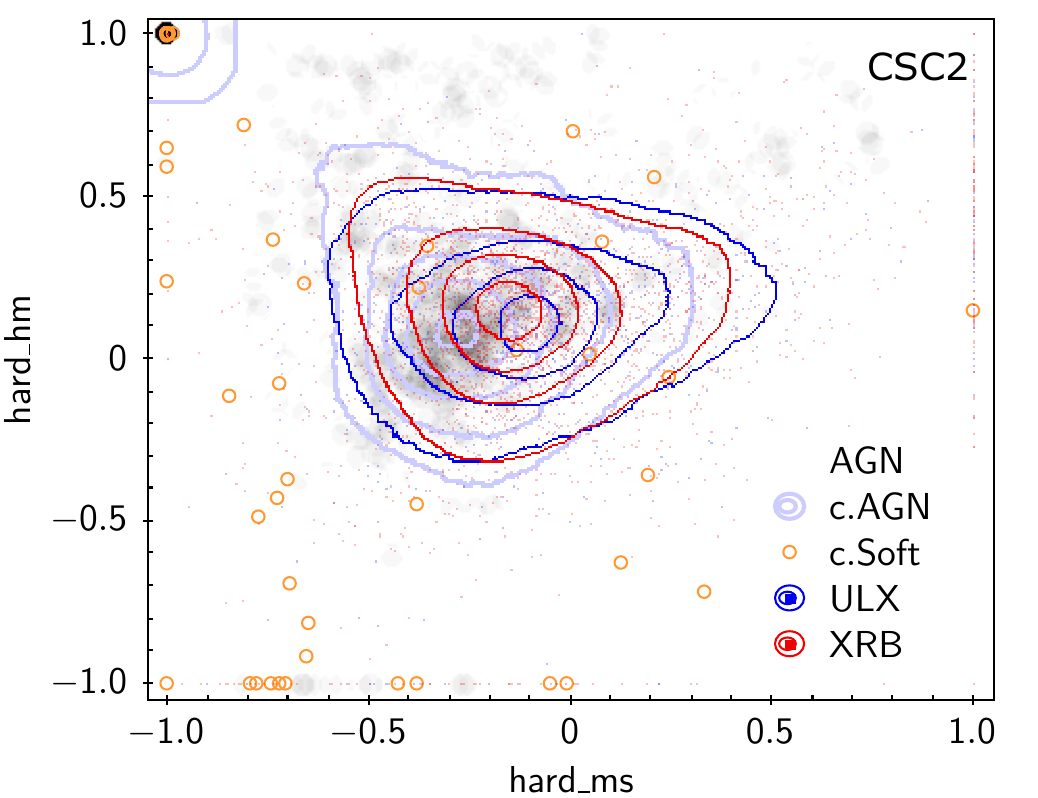}
    \includegraphics[width=8.5cm]{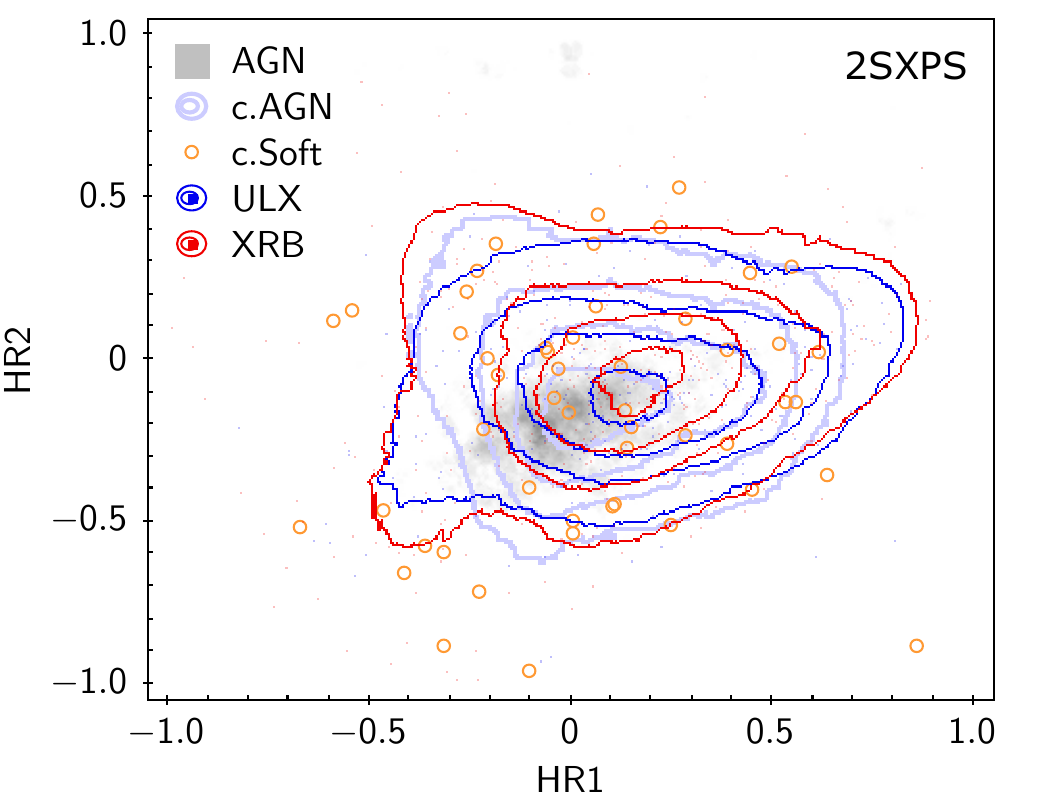}
    \caption{Hardness--hardness diagrams of XRB and ULX candidates from 4XMM (top panel), CSC2 (middle panel) and 2SXPS (bottom panel). Contours are shown to ease visualisation. The greyscale density in the background refers to known AGN. Candidates classified as foreground and background contaminants are shown as orange circles and pale blue contours, respectively. The hardness ratios plotted here correspond to the following energy ranges: between $0.5-1$ and $1-2$ keV (\texttt{SC\_HR2}), between $1-2$ and $2-4.5$ keV (\texttt{SC\_HR3}), between $0.5-1.2$ and $1.2-2$ keV (\texttt{hard\_ms}), between $1.2-2$ and $2-7$ keV (\texttt{hard\_hm}), between $0.3-1$ and $1-2$ keV (HR1), and between $1-2$ and $2-10$ keV (HR2).}
    \label{fig:hr_xrb_ulx}
\end{figure}

Figure \ref{fig:hr_xrb_ulx} shows three hardness -- hardness diagrams comparing ULX and XRB from each survey. In all three samples, their distributions are mostly overlapping, consistent with previous studies. Except for the minority of nearby ULX sources observed with long exposure times, this hardness information is not sufficient to probe differences in the spectral shape. 
We also plot ULX candidates classified as background ($P_{AGN}>P_{Soft}$) and soft ($P_{AGN}<P_{Soft}$) contaminants. The locus of sources classified as background contaminants is significantly offset from these populations, and as expected, it matches precisely the locus of AGN. This gives further credit to their classification as contaminants, and is in agreement with previous studies on AGN colours. Soft contaminants are fewer and dispersed at different loci of this parameter space.

\begin{figure*}[h!]
    \centering
     \includegraphics[width=16cm]{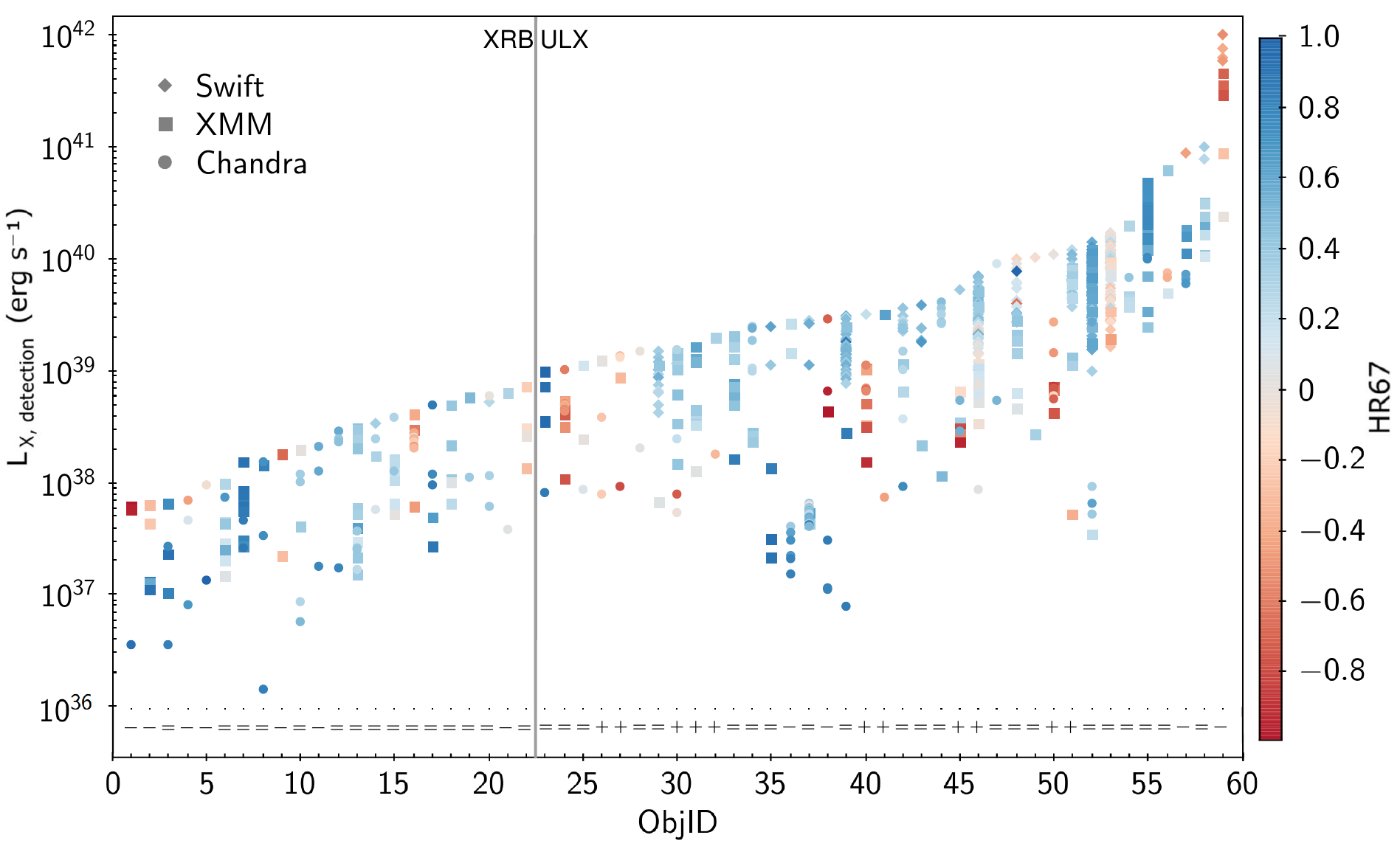}
   \caption{Luminosity-hardness evolution of the subset of 59 variable XRB and ULX. \texttt{HR67} is the hardness ratio between $0.2-2$ and $2-12$ keV bands. All detections having $\texttt{HR67\_err}<0.25$ are shown. To ease visualisation, for each source, the correlation, anti-correlation, or absence of correlation between $L_{X,detection}$ and \texttt{HR67} are highlighted at the bottom of the graph by the "+", "$-$" and "=" signs, respectively.}
    \label{fig:lx_hr_evol}
\end{figure*}

More conclusive results may be found by exploring the hardness evolution when the flux varies. Indeed, hardness-luminosity studies of black hole XRBs have shown a hysteresis cycle between two canonical states, high soft and low hard \citep{ShakuraSunyaev1973, Remillard2006} while neutron star and some black hole ULXs globally harden when they brighten \citep{Kaaret2017, Gurpide2021}. In ULXs, a softening occurring when the ULXs brighten may indicate the presence of a black hole (e.g. \citealt{Narayan2017}), as found in black hole XRBs; however this does not apply to at least some ULXs (e.g. NGC~5907 ULX1, \citealt{Gurpide2021}). We thus looked at the hardness evolution of a sample of XRB and ULXs having varied significantly between detections. While such a study would require a detailed spectral modelling of each source, we implement a simple approach to identify sources showing large hardness variations. To combine the detections coming from all three instruments, we built a flux hardness ratio \texttt{HR67} between the bands 0.2--2 and 2--12 keV: 
$$ \texttt{HR67} = \frac{F_{2-12}-F_{0.2-2}}{F_{2-12}+F_{0.2-2}}$$
This corresponds to existing \textit{XMM-Newton} energy bands, and an extrapolation of \textit{Swift} and \textit{Chandra} energy bands, for which we used fixed conversion factors. To this end, although this is a severe approximation, we assume an absorbed power-law model with ($\Gamma=1.7$, $n_H=3\times 10^{20}$~cm$^{-2}$) as already done in CSC2, 4XMM and part of \textit{Swift} catalogues to convert count rates into fluxes. From the ULX and XRB samples in all three surveys, we select 59 unique sources that varied by a factor of at least ten during their follow-up, and we consider their 799 detections that have $\texttt{HR67\_err}<0.25$. This selection targets high S/N sources having a good follow-up, thus most of them are already well-studied (32 have an entry in \textit{Simbad}). Sources having their peak flux below (resp. above) $10^{39}$~erg~s$^{-1}$  are considered as XRB (resp. ULX).
Figure \ref{fig:lx_hr_evol} shows the detected luminosities of these 59 sources, with hardness being colour-coded.

While a substantial fraction ($\sim 50\%$) of sources do not present a significant hardness evolution ($\Delta HR<0.2$), most of the other sources -- which have variable hardness -- follow the expected trend: from the lower luminosity state to the higher luminosity state, a significant fraction of XRBs become softer and most ULXs harder. A few outliers are to be noted: some ULXs become softer instead, such as M101 ULX-1 (ObjID=38) which is a well-known supersoft ULX thought to host a black hole \citep{Liu2013,Shen2015}, or ESO 243-49 HLX-1 (ObjID=59) having a well-studied high soft state (e.g. \citealt{Servillat2011,Godet2009,Godet2012}). NGC 4490 ULX-3 (ObjID=57) has a very low HR at peak (\textit{Swift}) detection, which is unreliable ($L_{X,det}>10^{41}$~erg~s$^{-1}$  ), probably due to its confusion with the ambient hot gas and the vicinity of ULX-1 which is softer -- indeed the \textit{Swift} coordinates are offset in this direction, by 5~arcsec from the \textit{Chandra} source. While 4XMM~J095524.8+690113 (ObjID=24) is always very soft, it matches SN~1993J. NGC 5907 ULX-1 (ObjID=55) and 4XMM~J022239.1+422328 (ObjID=25) are always very hard (this is confirmed by an inspection of their \textit{XMM-Newton} spectra), as expected from their locations right inside a gas-rich, edge-on host galaxy suggesting an important absorption. This sample of 59 sources represents a large sample of uniformly selected, highly variable extragalactic XRBs and ULXs.


\subsubsection{Variability}
\label{sec:3.3.2}

The variability within an observation is hard to characterise on a systematic basis, as it takes various forms depending on the source type and is strongly affected by the binning. Additionally, automated processing of the detailed light curve is time consuming and subject to biases. However, some variability indicators are made available as columns of the X-ray catalogues: for instance, the fractional variability $F_{var}$ represents the dispersion of the flux between snapshots, weighted by the flux error, and is given in 4XMM-DR11. It can be used to isolate the most significantly variable objects. The probability that the source is constant within the observation is available as well in all three catalogues, using various tests. 2SXPS records the maximum rate among snapshots in each band, which is valuable to spot short-term outbursts. For each 2SXPS source, we compute the ratio between the maximum and the median rate among all snapshots in the most variable band. Since \textit{Swift} snapshots are shorter (20-30 minutes) than the typical duration of our multi-mission observations ($\sim 10$ks), this indicator probes variability on shorter timescales.

\begin{figure}
    \centering
    \includegraphics[width=8.5cm]{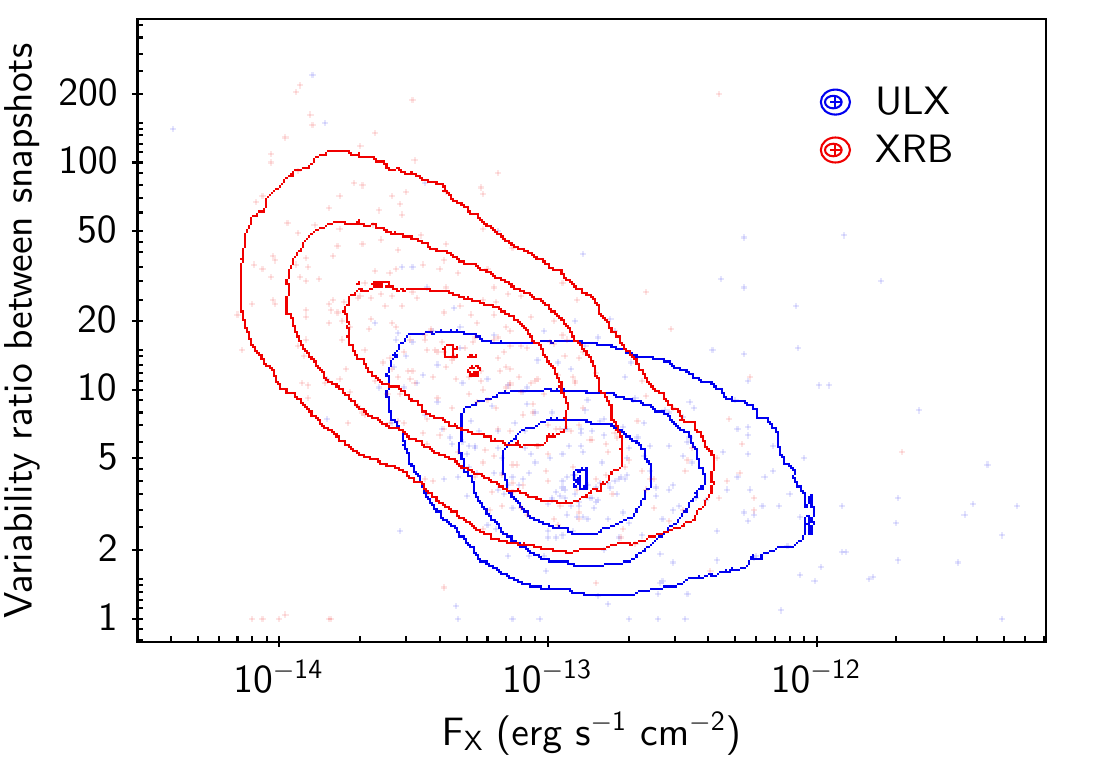}
    \includegraphics[width=8.5cm]{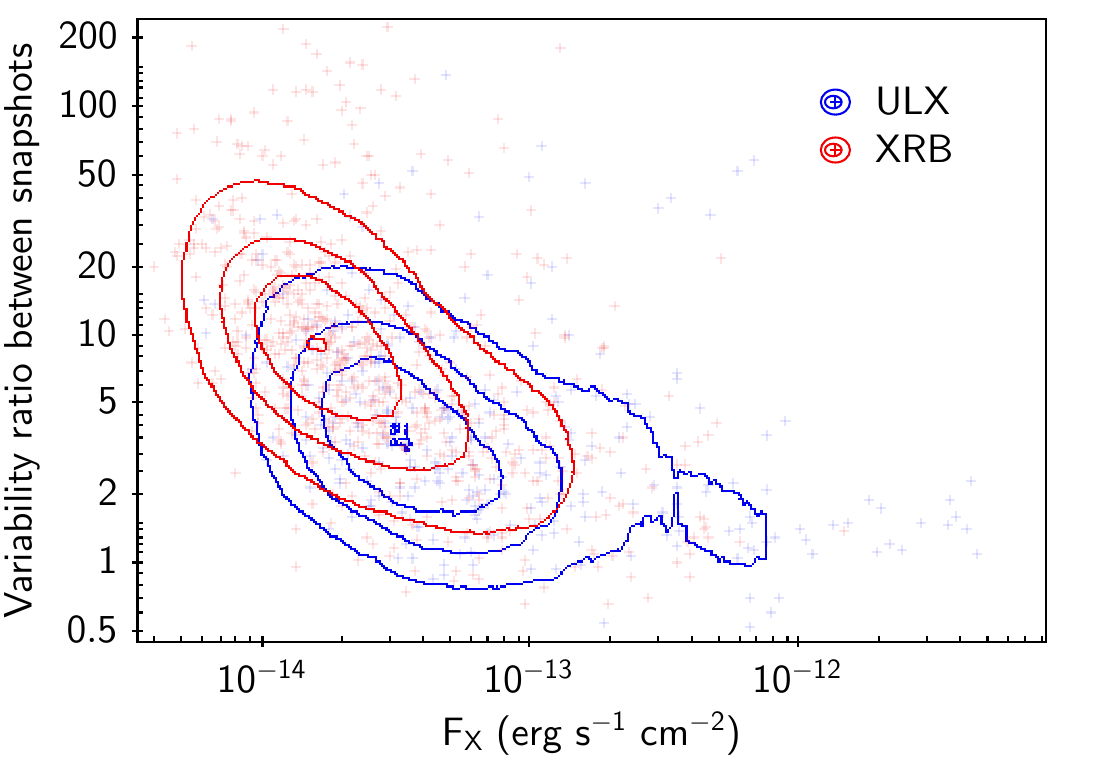}
    \caption{Variability between snapshots of XRB and ULX candidates from 2SXPS (top panel) and 4XMM (bottom panel), as a function of their flux. Contours are shown to ease visualisation.}
    \label{fig:var_xrb_ulx}
\end{figure}

We find no significant difference between the two populations for \textit{XMM-Newton} and \textit{Chandra} variability indicators. However, 2SXPS ULXs seem less variable than XRBs between observations and (especially) snapshots (Figure \ref{fig:var_xrb_ulx}, y-axis). On average, the former varied by 0.71~dex between snapshots, while the latter varied by 1.15~dex (the standard deviation are 0.4 and 0.5~dex, respectively). 
This difference may be intrinsic, due to the physics behind these populations; but it may also be an observer bias, because unlike other telescopes, the \textit{Swift} monitoring of an XRB is generally performed to track its particular variability. To remove this doubt, we also computed the variability ratio between snapshots for \textit{XMM-Newton} sources, by rebinning the 4XMM light curves to 20 minute long bins, and recording the peak rate of each band. The result is shown in the bottom panel of Figure \ref{fig:var_xrb_ulx}: the same offset is visible, with ULXs varying of 0.56 dex between snapshots on average and XRB varying of 0.93 dex.

However, this indicator is subject to at least two biases: first, the exposure time, because episodes of high flux variations are more likely to be detected in longer exposures. XRB have to be at lower distances to be detected, in particular in better-studied galaxies, targeted by longer or more numerous observations. This is why they are detected at lower fluxes in our samples. In the \textit{Swift} sample, 112 out of 293 XRB are located in just four galaxies observed for more than 200~ks: M51, M81, M101 and NGC~300. After removal of these galaxies, the ratio between the mean variability of XRBs and ULXs is reduced by 30\%. The second important parameter is the flux, because the noise contribution to the source flux is more significant at low flux and equal exposure. At equal fluxes and exposure time, both populations have similar variability distributions in both surveys. This is tentative evidence that both populations undergo a short-term flux variability of equal amplitude, however this result would need a detailed treatment of the light curves to be confirmed, which is beyond the scope of our study.



\subsection{The hyperluminous X-ray sample}
\label{sec:hlxsample}

\begin{figure*}[!htp]
    \centering
    \includegraphics[width=16.7cm]{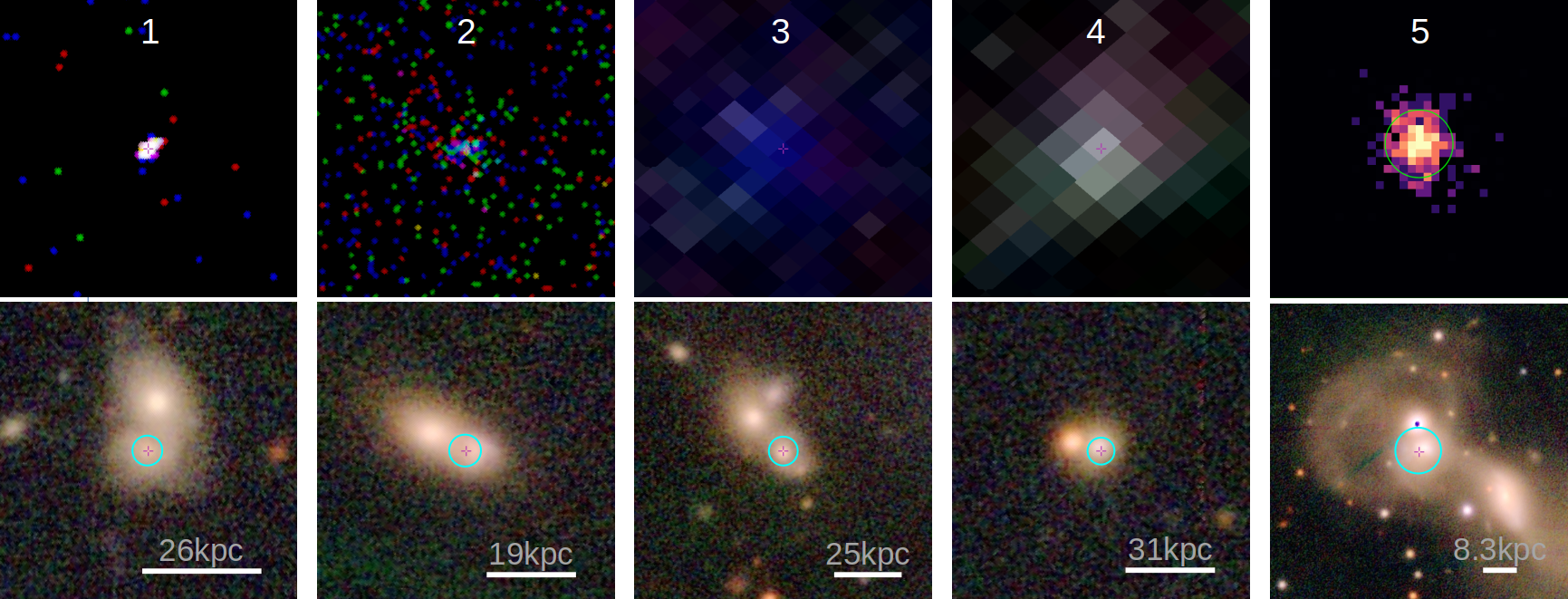}\vspace{0.05cm}
    \includegraphics[width=16.7cm]{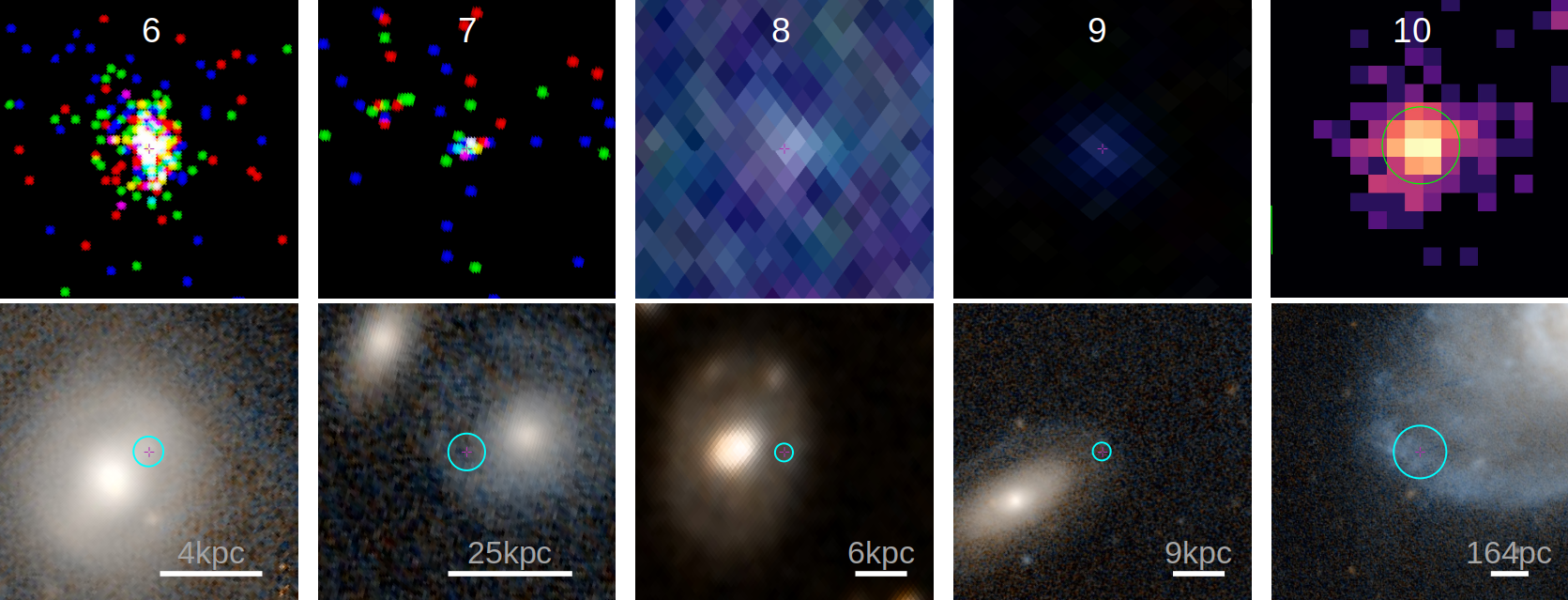}\vspace{0.05cm}
    \includegraphics[width=16.7cm]{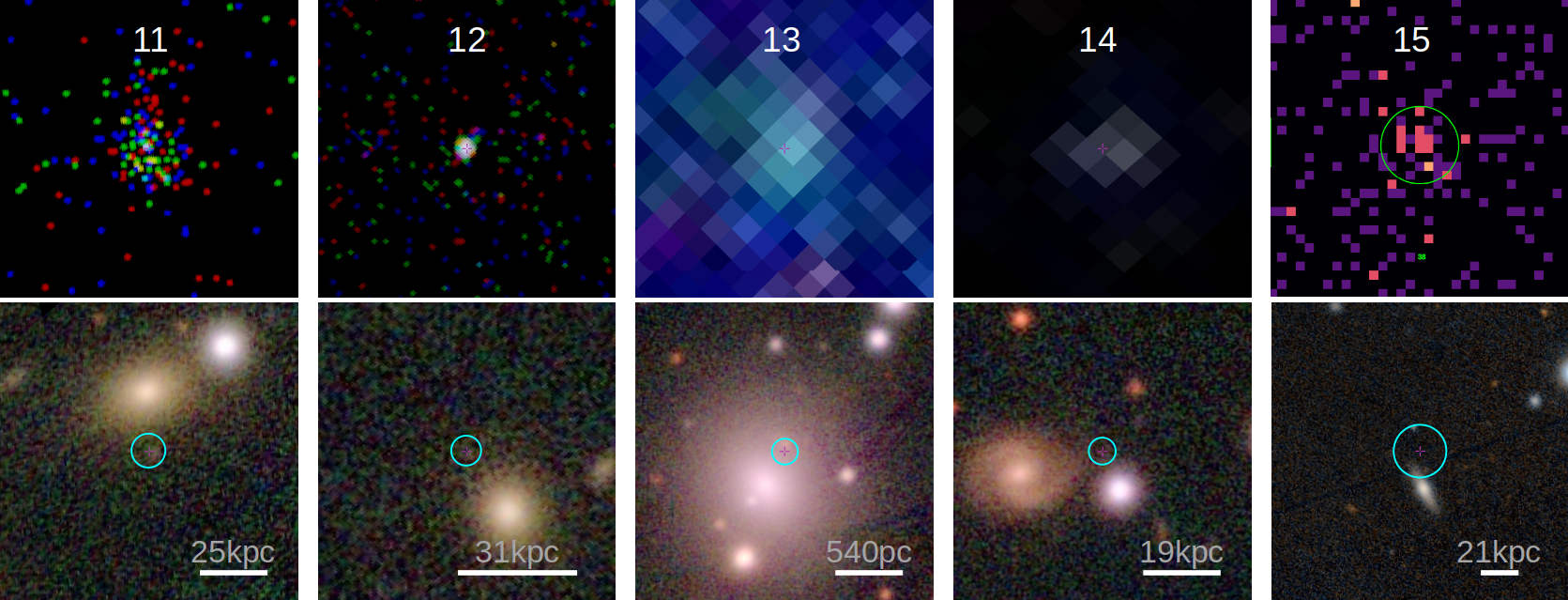}
    \caption{Images of a random sample of new HLX candidates. The images are structured as follows: column (1,2), (3,4) and (5) display CSC2, 4XMM-DR11 and 2SXPS candidates, respectively. (odd numbers) and (even numbers) rows show the X-ray and optical image at the same scale -- the white line corresponds to 10 arcsec. Rows (1,2), (3,4) and (5,6) correspond to galaxy pair, robust and weak candidates, respectively. These candidates have the following names: 1--2CXO J124208.4+331854, 2--2CXO J025921.5+132913, 3--4XMM J123441.0+020846, 4--4XMM J104444.7-012018, 5--2SXPS J164652.0+234011, 6--2CXO J082215.9+210535, 7--2CXO J105210.4+552243, 8--4XMM J062447.9-372122, 9--4XMM J125708.6-044144, 10--2SXPS J111416.1+481833, 11--2CXO J005151.7+474019, 12--2CXO J131133.3-011656, 13--4XMM J083235.2-225804, 14--4XMM J145753.6-113959 and 15--2SXPS J115109.2+570340. X-ray images correspond to the \textit{Chandra} and XMM-EPIC images displayed in ESASky, or the science exposures of \textit{Swift}-XRT where the green circle has a 10~arcsec radius. Optical images come from the PanSTARRS survey \citep{Chambers2016}, except candidate 8 (Digitized Sky Survey, \citealt{McLean2000}) as it lies in a region not covered by PanSTARRS. The 3$\sigma$ X-ray position error circle is shown in cyan.}
    \label{fig:line_hlx}
\end{figure*}

In our work, thanks to the high completeness of GLADE even at several hundred Mpc, we are able to retrieve 13, 115 and 75 HLX candidates from \textit{Swift}, \textit{XMM-Newton} and \textit{Chandra} samples, respectively. Of these 191 unique candidates, 76 sources have a bright optical counterpart and are split in two categories: `galaxy pair candidates' (22 objects), if the source matches the nucleus of a known galaxy in interaction with the host; and `weak HLX candidates' (54 objects), for other point-like optical sources or sources considerably offset from the galaxy area. The latter could indeed be background AGN, as suggested by the identification of four such sources as mid-infrared-detected AGN \citep{Secrest2015}. However, they could also be dwarf satellite galaxies of the host, which are a favourable environment to look for IMBH \citep{Webb2010,Greene2020,Barrows2019}; and actually seven of them have photometric redshifts consistent with the host distance. The 115 remaining HLXs with no optical counterpart, or a faint counterpart overlapping the area of the galaxy, are qualified as `robust HLX candidates'. Figure \ref{fig:line_hlx} shows a random selection of 15 candidates belonging to one of these three categories, taken from the three X-ray surveys. For clarity, we summarise the number counts of each sub-sample of initial and selected HLX candidates in Table \ref{tab:nHLX}.

\begin{figure}
    \centering
    \includegraphics[width=4.4cm]{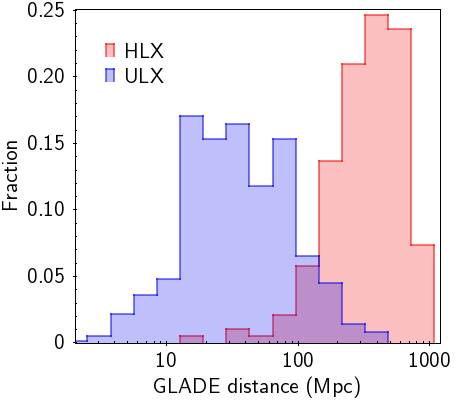}
    \includegraphics[width=4.4cm]{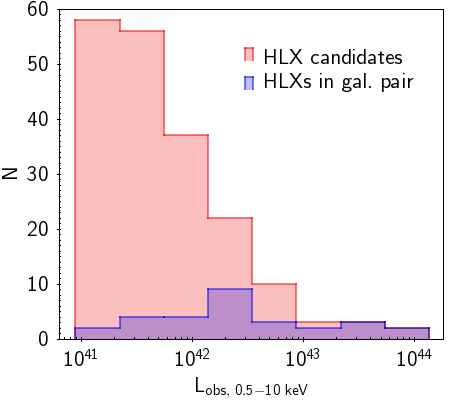}
    \caption{Distribution of some properties of selected samples of ULXs and HLXs. (Left) Distance of the host. ULXs are detected up to 500~Mpc and HLXs up to 1100~Mpc. (Right) 0.5--10~keV observed mean X-ray luminosity of the 191 HLXs candidates, including the 22 galaxy pair candidates.}
    \label{fig:dist_lx_hlx}
\end{figure}

The distributions of the distances and mean observed 0.5--10~keV X-ray luminosities of the 191 HLX candidates are shown in Figure \ref{fig:dist_lx_hlx}. As usually found, HLXs are detected much further away than ULXs. Most of them have luminosities in the range $10^{41}-10^{42}$~erg~s~$^{-1}$, with galaxy pair candidates being the most luminous. By construction, all candidates have $S/N>3$. However, we note that only 17\% of the sample of 169 (robust+weak) candidates have $S/N>10$. The median $S/N$ for this sample is 4.9.

Eight candidates that were removed, but are present in the literature are discussed here. Some sources are simply not present in our sample: NGC 2276 ULX-1 is located outside the GLADE $1.26\times D_{25}$ ellipse, which does not fit the actual galaxy area (this is the case for a small proportion of GLADE galaxies). Reported in \cite{Zolotukhin2016} as a reliable candidate, XMM0838+24 is here located outside the extent of its GLADE association, its separation being 1.5 times the galaxy radius at its position angle. The other reliable candidate cited by this study, XMM1226+12, was discarded during visual inspection as lying visibly outside the extent of the host. M82 X-1 is flagged as confused, or extended, in all three surveys. Thus these two sources are not present in our sample. The candidate in IC~4320 (an elliptical galaxy at $\sim$93~Mpc), proven to be a background AGN \citep{Sutton2015}, was discarded due to a PanSTARRS photometric redshift higher than the host ($z_{ph}=0.18\pm 0.06$, \citealt{Tarrio2020}). Because we use the mean luminosity in each survey for selection criterion, the candidates located in NGC~5907, NGC~4077, UGC~6697 and Cartwheel are not in our HLX sample, due to their $\langle L_X\rangle < 10^{41}$~erg~s$^{-1}$. We still retrieve some well-known candidates in our work: they include ESO~243-49 HLX-1, NGC~470 HLX-1 \citep{Walton2011,Sutton2012} and 3XMM~J161604.0-223726 in IC~4596 \citep{Earnshaw2019}.

\begin{table}[]
\caption{Sub-samples of HLXs under study.}
    \centering
    \resizebox{\columnwidth}{!}{\begin{tabular}{c|cccc}
        & CSC2 & 4XMM & 2SXPS & Total (unique) \\\hline
        Initial candidates & 195  & 360  & 110  &  665 (619) \\
        With redshift &  96 & 187  & 41  & 324 (298)  \\
        Background & 82  & 165  & 35  & 282 (259)  \\
        Foreground &  2 & 3  & 0  & 5 (5) \\
        Distance match &  12 & 19  & 6  &  37 (34) \\ \\ [-1 ex]
        Selected candidates &  75 & 115  & 13  &  203 (191) \\
        Robust &  51 & 66  & 5  & 122 (115)  \\
        Weak &  15 & 35  & 7  & 57 (54)  \\
        Galaxy pair &  9 & 14  & 1  & 24 (22)  \\
        With opt. counterpart & 29  & 63  & 8  & 100 (94)  \\
        Soft ($\texttt{HR67}<-0.6$) & 2  & 5  & 1  & 8 (6)  \\
        Hard ($\texttt{HR67}>-0.6$) &  67 & 110  & 12  &  189 (179) \\\hline
    \end{tabular}}
    \tablefoot{Number counts of initial and selected HLXs in different sub-samples (see Section \ref{sec:hlx_def} and Section \ref{sec:hlxsample}, respectively, for details on the sub-samples). The sample of `soft HLXs' is defined in Section \ref{sec:hlxhardness}, and is the complementary of `hard HLXs' (note that six \textit{Chandra} candidates are not included in these samples, because their \texttt{HR67} could not be calculated).}
    \label{tab:nHLX}
\end{table}

\subsection{Comparison of ULXs and HLXs}

\begin{figure}
    \centering
    \includegraphics[width=8.5cm]{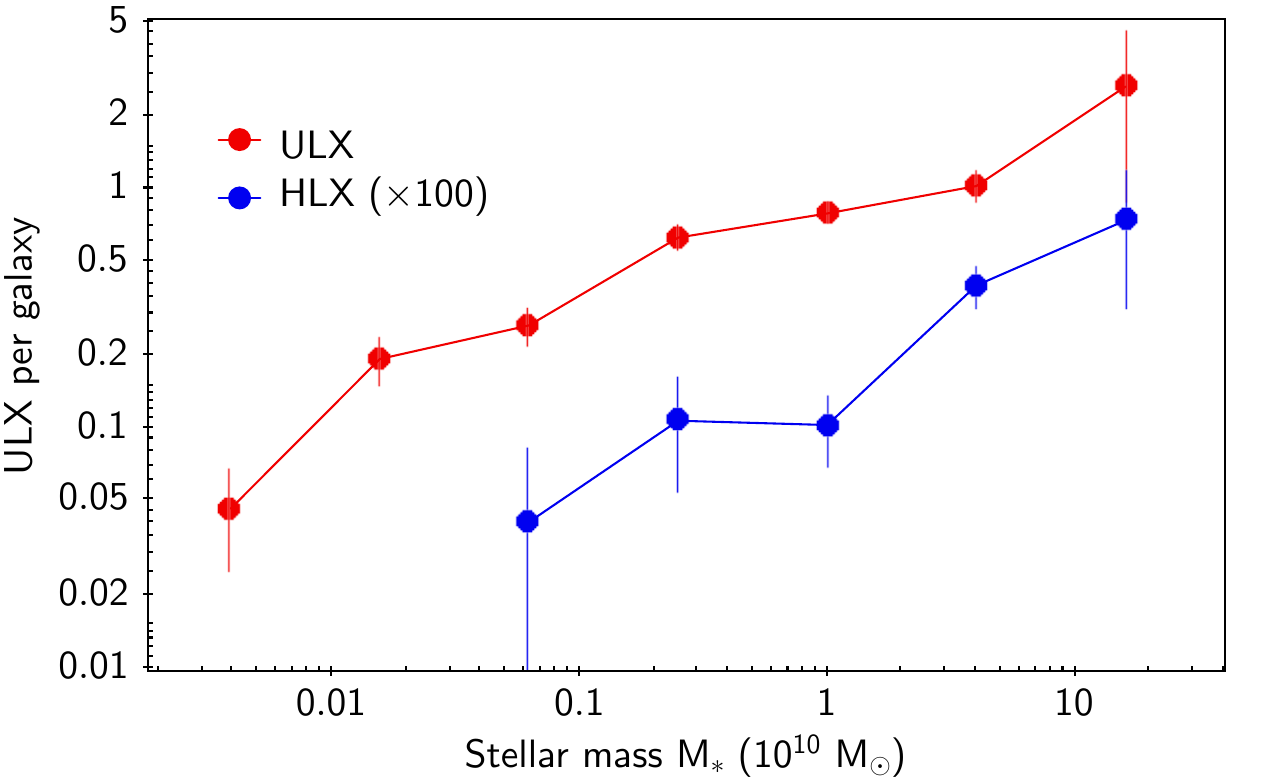}
    \includegraphics[width=8.5cm]{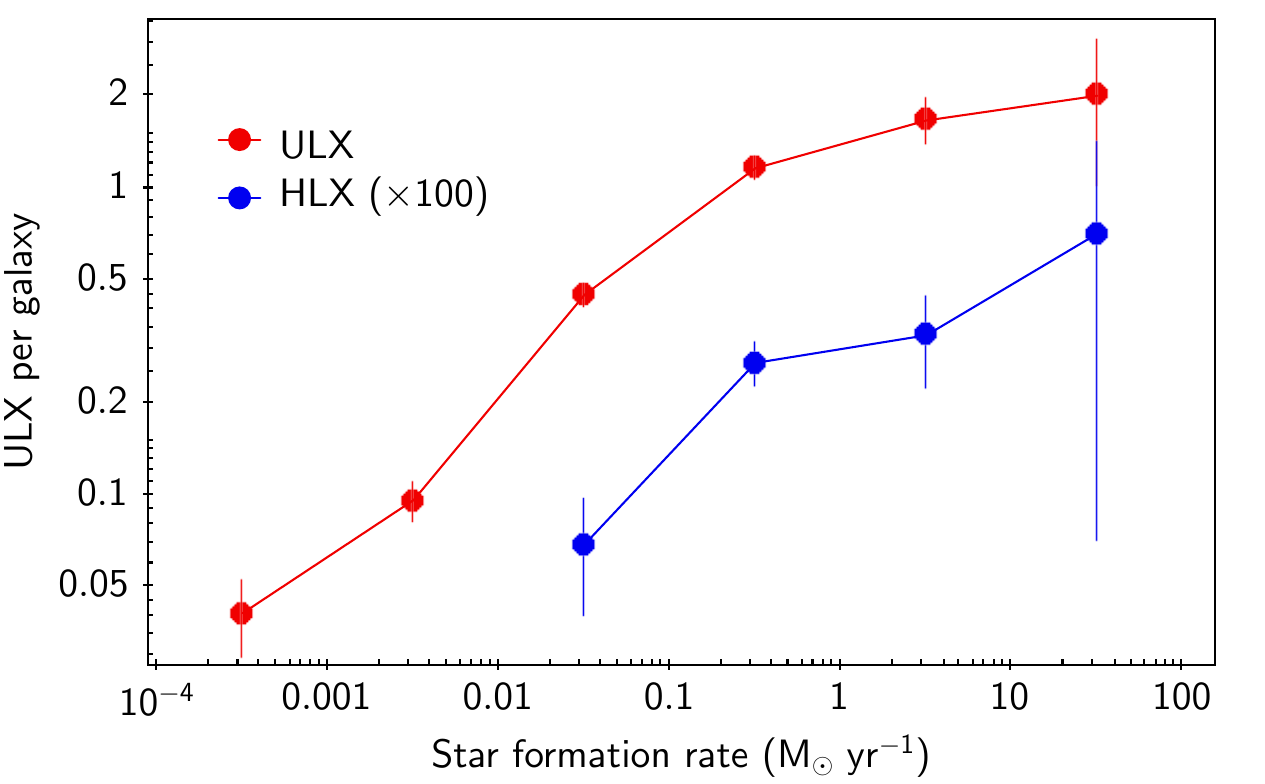}
    \includegraphics[width=8.5cm]{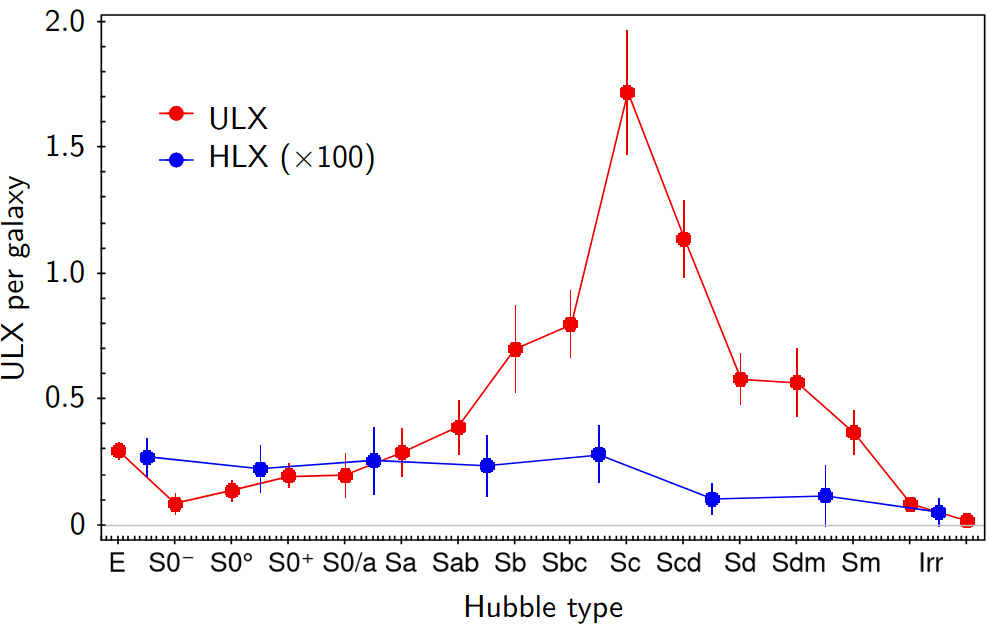}
    \caption{Rates of ULX and HLX as a function of different properties characterising galaxy environment.}
    \label{fig:env_ulx_hlx}
\end{figure}

Because of the low number of known HLXs,  few studies have attempted to characterise their typical environment. The location of ESO 243-49 HLX-1 outside the disc of its host may be due to it being embedded in a stellar cluster or the stripped core of a dwarf galaxy (e.g. \citealt{Webb2010}) but ESO 243-49 HLX-1 in itself seems to be an outlier among HLXs (e.g. \citealt{Sutton2015}). Other HLX discoveries led to the suggestion that HLXs and ULXs share the same type of environment, namely star-forming, spiral galaxies \citep{Sutton2015,Barrows2019}. This is at odds with the ratio of our HLX rates between spiral and elliptical (Figure \ref{fig:spiell_ratio}), suggesting an equal rate of robust HLXs in these two morphologies. We compare the behaviour followed by ULX and HLX rates as a function of three environmental parameters (Figure \ref{fig:env_ulx_hlx}): the galaxy stellar mass, its SFR and its Hubble type. Unlike ULXs, our HLXs tend indeed to be hosted equally in (regular) spiral and elliptical galaxies. However we note that only 38 HLXs have a Hubble type value provided in HyperLEDA, and that 40\% of them have errors on this type $e_t \geq 3$ (equivalent to 1.5 major graduations in Figure \ref{fig:env_ulx}). On the other hand, similarly to the trends obtained on ULXs, we find a positive correlation between the HLX rate and the stellar mass and SFR.


\subsubsection{Hardness}
\label{sec:hlxhardness}

\begin{figure}
    \centering
    \includegraphics[width=8.5cm]{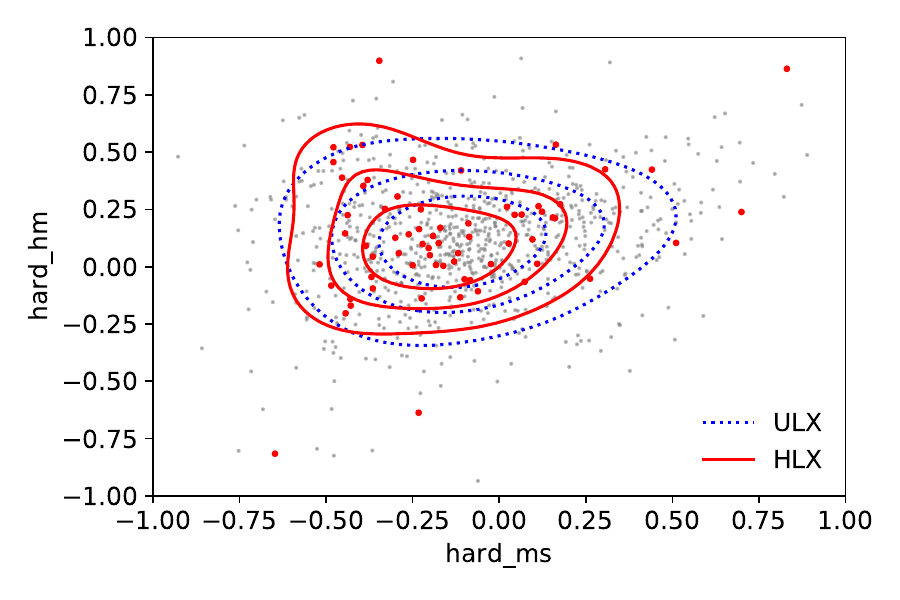}
    \includegraphics[width=8.5cm]{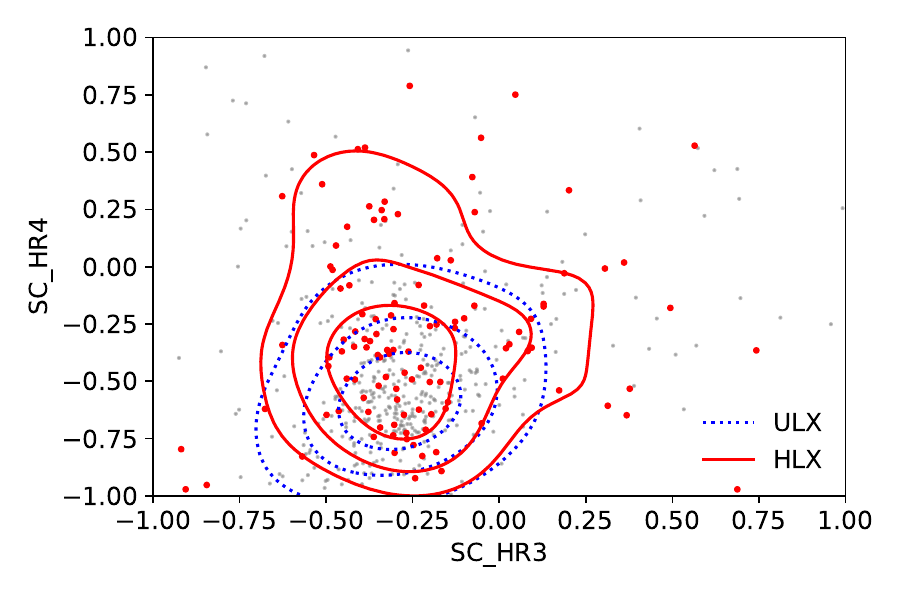}
    \includegraphics[width=8.5cm]{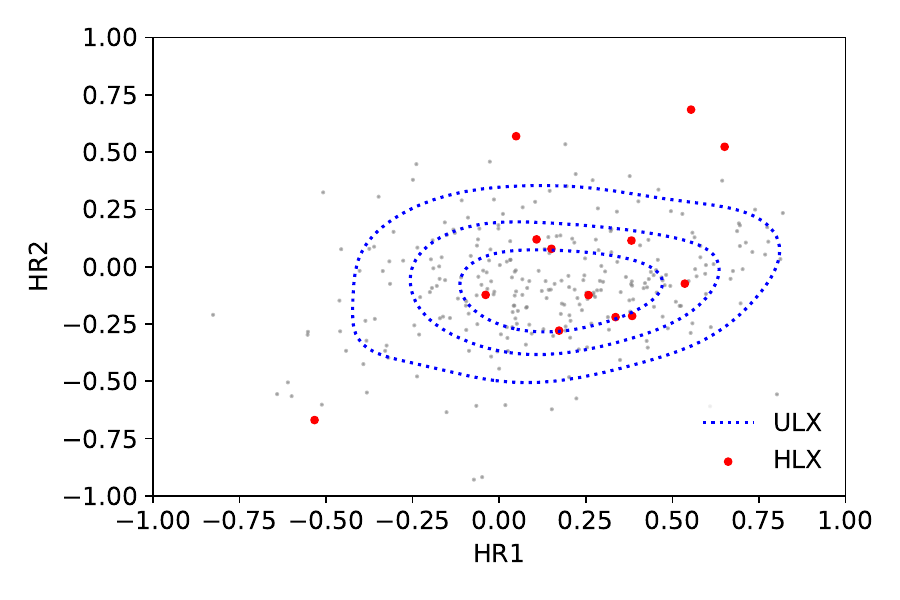}
    \caption{Hardness--hardness diagrams of ULX and HLX candidates from \textit{Chandra} (top), \textit{XMM-Newton} (middle) and \textit{Swift} (bottom). Grey and red dots correspond to ULXs and HLXs, respectively. Contours at 25th, 50th and 75th percentiles are shown to ease visualisation.}
    \label{fig:hrhr_ulxhlx}
\end{figure}

We investigate the locus of the 191 HLX candidates in a hardness--hardness diagram. Figure \ref{fig:hrhr_ulxhlx} shows the diagrams obtained from \textit{XMM-Newton}, \textit{Chandra} and \textit{Swift} data, after discarding sources with $HR=-1$~or~1 to limit unreliable measurements. The HLX and ULX distributions are mainly overlapping, with only a fraction of \textit{XMM-Newton} HLX candidates looking harder above 4.5 keV (\texttt{SC\_HR4}). They consist mainly of 15 robust HLX candidates having $\texttt{SC\_HR4}>0$ on average among detections; their low signal-to-noise ratio ($S/N<10$) does not allow a more in depth spectral study. However, the energy range covered by \texttt{SC\_HR4} essentially probes the photon index of the spectrum if it were an absorbed power-law spectrum, and in this case such values would only be produced by $\Gamma \lesssim 2$. 

\begin{figure}
    \centering
    \includegraphics[width=8.5cm]{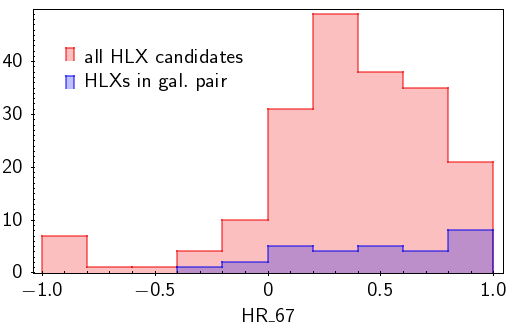}
    \caption{Distribution of the hardness ratio \texttt{HR67} between the 0.2-2 and 2-12 keV energy bands for selected HLX candidates.}
    \label{fig:hr67_hlx}
\end{figure}

Besides, in the \textit{Swift} sample, ESO 243-49 HLX-1 is an outlier located far in the lower left of the distribution. Three other outliers are found in 4XMM: 4XMM J215022.4-055109, a well-known IMBH candidate which underwent a tidal disruption event \citep{Lin2018}, 4XMM J161534.3+192707, a variable HLX possibly consistent with another offnuclear tidal disruption event (Soria et al., in prep), and 4XMM J085253.8+180110, a weak candidate located in a cluster of galaxies. Another soft outlier (not visible in Figure \ref{fig:hrhr_ulxhlx} because of having $\texttt{SC\_HR4}=-1$) is 4XMM~J231818.7-422237, a variable soft source already cited as a candidate IMBH in \cite{Lin2014}. It could actually be associated with two GLADE galaxies, and would be of ULX luminosity if belonging to the one closer to us. It is however softer than all other ULXs ($\texttt{SC\_HR2}=-0.73$, $\texttt{SC\_HR3}=-0.85$). The outlier nature of such soft HLXs is even more visible in the distribution of \texttt{HR67}, the hardness ratio between bands $0.2-2$ and $2-12$ keV, where they form the peak close to -1 ($\texttt{HR67}<-0.6$) in Figure \ref{fig:hr67_hlx}. All but two of them are also seen as outliers by the source classification, with an outlier measure $>12$ \citep{Tranin2022}, higher than 99\% of ULX candidates.  



\subsubsection{Variability}

\begin{figure}
    \centering
    \includegraphics[width=8.5cm]{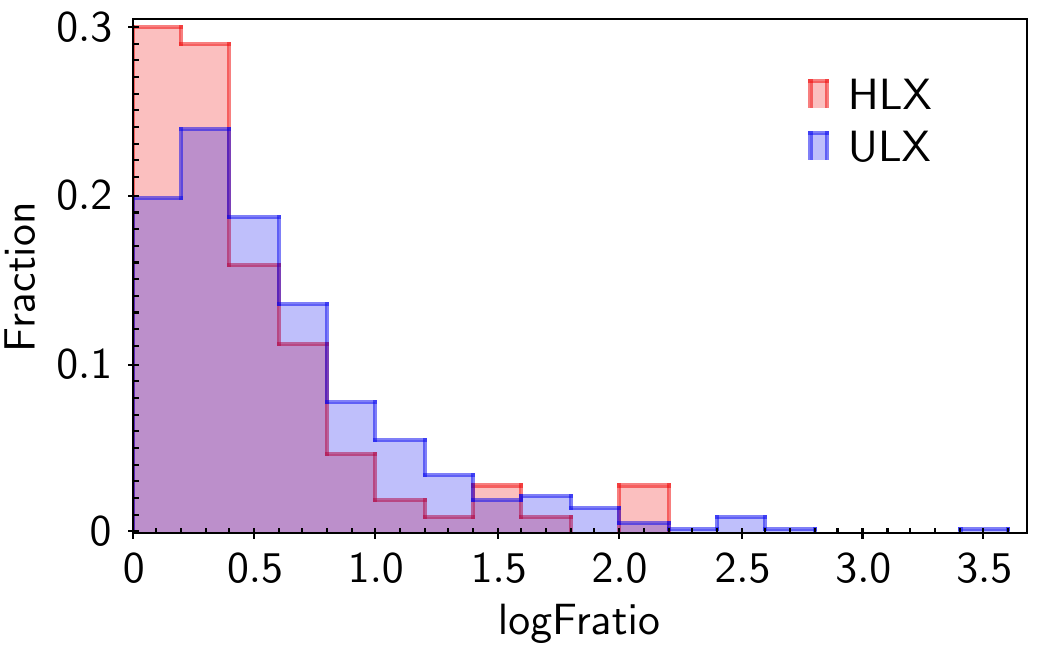}
    \caption{Distribution of the variability ratio between observations ($\texttt{logFratio}= \log\big(\frac{F_{max}}{F_{min}}\big)$), for ULXs and HLXs.}
    \label{fig:var_hlx}
\end{figure}

Some extreme ULXs were seen to approach or overcome the luminosity threshold of HLXs for transient periods. Super-Eddington radiation well above the Eddington limit seems possible for the timescale of an observation \citep{Israel2017}, and in this case the Eddington limit cannot give accurate mass estimations of the central object. HLX candidates detected only once could thus be powered by stellar-mass accretors. However, we find that out of the 191 HLX candidates, 96 were detected several times (up to 20 times, and even more for ESO 243-49 HLX-1, NGC 470 HLX-1 and the AGN in a galaxy pair XMMU~J134736.6+173403). Only 26 of them have a detection below $10^{41}$~erg~s$^{-1}$ -- including the known candidates in ESO 243-49, IC 4596 and NGC 470 -- suggesting that a majority of our HLX sample is actually persistent. The time between the first and the last observation is $>1$~yr for 75\% of persistent candidates. Figure \ref{fig:var_hlx} shows the distribution of their variability ratio between observations, along the one of ULXs. No significant variability excess is seen in either of these two categories.

\section{Discussion}
\label{sec:4}

\subsection{Comparison to other studies}

Recent releases of large X-ray surveys and complete galaxy catalogues enabled the discovery of large samples of ULXs and HLXs. We discuss our findings in the context of previous studies on this subject.

\cite{Kovlakas2020} found 629 ULX candidates in 309 galaxies in the local 40~Mpc volume, from \textit{Chandra} CSC2 catalogue matched with HECATE \citep{HECATE}. They estimated a contamination rate of 20\%, using the $\log N-\log S$ method. Because of the high \textit{Chandra} sensitivity, their sample is likely to be complete. In our work, despite the use of a different galaxy catalogue and slightly different selection criteria, we find a similar number of \textit{Chandra} ULX candidates (567) in galaxies closer than 40~Mpc. However, we remove 58 candidates of this sample from their classification as contaminants, 16 others from their identification as another type in the literature, and 41 are also flagged during visual inspection. This leaves a sample of 442 sources located within 40~Mpc, representing 78\% of the initial sample. Adding \textit{Swift} and \textit{XMM-Newton} candidates, the size of the sample is almost doubled (839). This outlines the need to cross-correlate several X-ray catalogues to find these sources, since focussing on one catalogue decreases the survey coverage and misses some candidates as they may have been fainter than $10^{39}$~erg~s$^{-1}$ at the time of the observation. Besides, we were able to confirm key results present in their work, namely the predominance of contaminants in uncleaned samples at luminosities $L_X>10^{40.5}$ erg~s$^{-1}$, the evolution of the ULX rate with SFR and the decrease of specific ULX rate with galaxy stellar mass.

\cite{Bernadich2021} used a recent release of the 4XMM catalogue, 4XMM-DR10, matched with HECATE galaxies, to build a clean ULX sample. Similarly to our work, their cleaning pipeline relies on removing the known AGN and foreground contaminants, as well as visual inspection of the sources having optical counterparts (from PanSTARRS and \textit{Gaia}). However, without an automatic classification, they were not able to remove contaminants having $F_X>F_{Opt}$ or no optical counterparts. Besides, we were able to improve this cleaning by visual inspection of faint counterparts in the southern hemisphere, using the DES survey. Overall, their sample of 779 candidates is somewhat larger than our cleaned sample (667 candidates). They however consider a source as ULX as soon as $L_X+L_{X,err}>10^{39}$ erg~s$^{-1}$ in at least one detection, and only 632 of their candidates have $\langle L_X\rangle >10^{39}$ erg~s$^{-1}$. 

Their work also focusses on the statistical study of bright ULXs, having $L_X+L_{X,err}>5\times 10^{40}$~erg~cm$^{-2}$~s$^{-1}$ in at least one detection. While this bright sample contains 94 such sources, it is significantly different from our sample of HLXs for several reasons: first, because we use the mean luminosity to select them, and because we use the more conventional threshold of $10^{41}$ erg~s$^{-1}$ (only 25 of their bright candidates have $\langle L_X \rangle > 10^{41}$ erg~s$^{-1}$). A large fraction of their sample is thus likely to be similar to luminous ULXs powered by a stellar-mass accretor. Second, because we use GLADE, which is more complete than HECATE as it is a compilation of catalogues (including HyperLEDA), and because HECATE is limited to galaxies at $D<200$~Mpc, our HLX sample contains significantly more candidates. As a result, we find significant differences in the preferred environments and counterparts of ULXs and HLXs (Section \ref{sec:4.4}). We do not find evidence in our sample for the softer average spectra for HLXs reported in \cite{Bernadich2021}.

\cite{Walton2022} is similar to our work in the sense that they included the same three X-ray surveys. However, their galaxy sample is different, compiling HyperLEDA and the Catalogue of Neighbouring Galaxies. While they used the same matching limit as we do, $1.26\times D_{25}$, the contamination rate of their sample must be high (Figure \ref{fig:cont_frac}) given that they only cleaned out known contaminants. 
Their final sample contains 1843 unique candidates, compiled from the 4XMM-DR10, 2SXPS and CSC2 samples containing 641, 501 and 1031 sources, respectively. This total is similar to the 1901 ULXs contained in our compilation, with however slightly different proportions from each survey (667, 304 and 1185). Our slightly larger sample, despite our strict cleaning, was possible thanks to the use of GLADE. Our sample thus constitutes both the largest and the cleanest ULX catalogue to date. 

We examine the classification of X-ray sources that we have in common with these three studies. We find 601 sources of \cite{Kovlakas2020} in our sample, of which only 423 pass the offnuclear and $\langle L_X\rangle >10^{39}$~erg~s$^{-1}$ criteria. The rest are mostly discarded because of insufficient luminosities, due to their use of CSC2 fluxes from the PSF 90\%-energy width (\texttt{flux\_aper90\_b}) instead of the source region aperture (\texttt{flux\_aper\_b}). 359 sources are selected as ULXs after classification and manual screening, hence $\sim$15\% of candidates are discarded as contaminants. Similarly, in the clean ULX catalogue of \cite{Earnshaw2019} claiming a contamination rate of 24\%, we find 17\% of candidates removed in our pipeline. From the ULX sample of \cite{Walton2022}, 1734 sources are included in this study, of which 1291 pass the first criteria (their luminosity criterion being less stringent). We obtain 27\% of these sources discarded as contaminants, which is higher than their estimated background contamination rate (18--23\%). Most of the ULXs of \cite{Bernadich2021} are present in our sample (746), and 574 have ULX-consistent positions and luminosities in our sample. However, 27\% of these sources are discarded by the classification and manual screening, considerably higher than expected from their contamination estimation ($\sim$2\%). Their estimation is however calculated for their entire X-ray sample (including non-ULXs), perhaps explaining this discrepancy. 

Last but not least, \cite{Barrows2019} produced a catalogue of 169 HLX candidates from the CSC2 catalogue and SDSS images. This unprecedented sample size is allowed by their detailed treatment of optical images to detect galaxies up to greater distances and fainter surface brightness; as a result, most of their candidates are located at distances $>1000$~Mpc, and present X-ray luminosities $>10^{42}$~erg~s$^{-1}$. Few of their galaxies are present in the GLADE catalogue: as a result, only ten of their candidates are present in our CSC2 sample. Two of them have $\langle L_X\rangle < 10^{41}$~erg~s$^{-1}$, and one is discarded when excluding the galaxy central region. Interestingly, 2CXO~J122148.2+491131 is flagged as background contaminant due to its photometric redshift $z_{ph}=0.46\pm 0.1$ (\citealt{Beck2016}, also consistent with PanSTARRS and DESI estimates), leaving only six candidates out of the possible ten in our sample. Whilst their study has the advantage of a low contamination rate (7--8\%) and probes the HLX population to greater distances in the Universe, our sample is more likely to contain candidates well-suited for a follow-up as well as HLXs of lower luminosity, probing lower IMBH masses and possibly earlier stages of black hole growth. These points are further discussed in Section \ref{sec:4.4}.

\subsection{Number of contaminants}
\label{sec:nbcontam}

 The cosmic X-ray background (CXB) has been extensively studied for several decades by several X-ray missions, revealing that it is essentially due to radiation from unresolved point-like sources, in particular AGN (e.g. \citealt{Moretti2003} and references therein). As such, its flux distribution has often been used in ULX and HLX studies to assess the number of background contaminants behind a given galaxy. For instance, the $\log N-\log S$ relation of \cite{Moretti2003}, gathering wide-field and pencil-beam surveys probing six orders of magnitude in flux, was used by \cite{Walton2011}, \cite{Zolotukhin2016} and \cite{Barrows2019} to compute their number of contaminants. To this aim, they converted the broad band flux of their sample to the hard energy band of \cite{Moretti2003} (2--10 keV) which yields the most contaminants, using an absorbed power-law spectrum with $\Gamma=1.69$ \& $n_H\sim 3\times 10^{20}$~cm$^{-2}$. However, a possible caveat of this method is that CXB sources present a variety of spectra \citep{Mushotzky2000} that cannot be accurately represented by this model. To alleviate this problem, we also compute the number of contaminants using $\log N-\log S$ distributions calibrated in an energy band similar to the broad bands of our sample, namely 0.2--12~keV (\textit{XMM-Newton}), 0.3--10~keV (\textit{Swift}) and 0.5--7~keV (\textit{Chandra}). This is the case of the relations of \cite{Georgakakis2008} and \cite{Kim2007}, respectively used in the ULX studies of \cite{Zolotukhin2016} and \cite{Kovlakas2020}. After converting our fluxes in the proper energy band, we find excellent agreement between the results obtained from each relation.

 The number of contaminants in each galaxy is computed by multiplying the cumulative flux density, $\log(N(>S))$ at a given flux $S$, by the extent of the galaxy used for finding ULXs (i.e. after the centre exclusion criterion has been applied). The flux to consider is directly given by the galaxy distance (such that the contaminant is found in the ULX luminosity range) and the X-ray sensitivity (so that the contaminant is detectable), as $\min(f_{39},f_{lim})$ where $f_{39} = \frac{10^{39} \text{erg~s}^{-1}}{4 \pi D^2}$.

 Because our centre exclusion radius $3(\texttt{POSERR}+0.5)$ depends on the source position error, we estimate the number of contaminants using the 10\%, 50\% and 90\% percentiles of the position error distribution in each sample of candidates. Results are given in Table \ref{tab:ncontam} for the ULX and HLX contaminants, where error bars correspond to the variations in position error. For ULXs, it is comparable to the number of candidates known or classified as AGN, or removed by visual inspection of the optical source. For HLXs, it is similar to the number of background sources (using photometric redshifts) added to the number of HLX candidates flagged as weak. This is a further indication that the selection of ULX and HLX candidates, and thus the automatic classification and the photometric redshifts estimates, are reliable and robust.

 Except in the 4XMM sample, our filtering process identifies more ULX candidates as background contaminants than expected from the CXB. Using the CXB curves of \cite{Moretti2003}, yielding the larger contamination rates, the 41 remaining background contaminants in our sample thus represent 2.2\% of the sample of selected ULXs.

 \begin{table}[]
 \caption{Number of contaminants estimated using different published $\log N-\log S$ relations, in the ULX and HLX samples.}
     \centering
     \resizebox{\columnwidth}{!}{\begin{tabular}{ccccc}
        Survey & Moretti+03 & Georgakakis+08 & Kim+07 & Identified\\\hline \\ [-1.5 ex]
        \vspace{1mm}
         2SXPS ULXs & $140^{+7}_{-15}$ & $141^{+7}_{-14}$ & $139^{+7}_{-15}$ & 171 \\\vspace{1mm}
         2SXPS HLXs & $13^{+2}_{-6}$ & $13^{+3}_{-5}$ & $13^{+2}_{-6}$ & 25\\\vspace{1mm}
         4XMM ULXs & $582^{+55}_{-73}$ & $558^{+53}_{-70}$ & $554^{+53}_{-69}$ & 541\\\vspace{1mm}
         4XMM HLXs & $136^{+34}_{-44}$ & $137^{+33}_{-44}$ & $135^{+33}_{-43}$ & 121\\\vspace{1mm}
         CSC2 ULXs & $483^{+38}_{-67}$ & $451^{+35}_{-62}$ & $452^{+35}_{-62}$ & 482\\\vspace{1mm}
         CSC2 HLXs & $75^{+17}_{-31}$ & $73^{+17}_{-30}$ & $73^{+17}_{-30}$ & 64\\
     \end{tabular}}
     \tablefoot{Positive and negative errors correspond to the estimates at 10\% and 90\% of the exclusion radius ($3(\texttt{POSERR}+0.5)$) distribution. The last column gives the number of candidates filtered out by our pipeline as background contaminants, or identified as HLX weak candidates.}
     \label{tab:ncontam}
 \end{table}


\subsection{Heterogeneity of the sample and instrumental biases}
 \label{sec:instbias}

Because of the different capabilities of \textit{Swift}-XRT, \textit{XMM-Newton} and \textit{Chandra}, we do not expect to probe exactly the same ULX populations in each survey. \textit{Chandra} on-axis observations are more sensitive and have a higher angular resolving power, so it can probe larger distances where the galaxies look smaller. \textit{XMM-Newton} and \textit{Swift} have covered larger sky areas, and  the superior follow-up capabilities of \textit{Swift} can also reveal populations of variable ULXs.  The design of \textit{Swift} as a gamma-ray burst monitor has led to a more uniform sky coverage, less biased towards galaxies of special interest, so we can expect a lower observer bias of its ULX population. However, the centre-excluding criterion (Equation (\ref{eq:0}))  leads to exclude a significant fraction of actual ULXs at low galaxy angular separation. This is particularly the case of \textit{Swift} HLX candidates, which are all offset from their host centre by $>10$~arcsec, while half of \textit{XMM-Newton} and \textit{Chandra }HLX candidates have lower separations. This probably causes the steeper slope of the 2SXPS XLF above $L_X>10^{40}$~erg~s$^{-1}$.

Another bias in our selection of ULXs and HLXs is the significant difference between \textit{Swift}, \textit{XMM-Newton} and \textit{Chandra} energy bands: in particular, the narrower energy band of the latter probably removes valid candidates just below the ULX luminosity threshold. Assuming a phenomenological absorbed power-law model for ULX spectra, with typical parameters $\Gamma=2.4$, $n_H=2\times 10^{21}$ cm$^{-2}$ \citep{Gladstone2009}, the conversion factor from the \textit{XMM-Newton} 0.2--12~keV band to the \textit{Chandra} 0.5--7 keV band is 0.85, and becomes as low as 0.7 for a photon index $\Gamma\lesssim 2$. If this conversion is valid on average, the \textit{Chandra} XLF break in spiral galaxies would be shifted to up to $\sim 4\times 10^{39}$~erg~s$^{-1}$, better matching the other XLFs.

Nevertheless, results obtained from the XLF and the comparison with XRBs, ULXs and HLXs are comparable between instruments, showing the robustness of both our cleaning pipeline and our statistical study with respect to instrumental biases.

The XLF inferred from 2SXPS, 4XMM and CSC2 are compatible at the $2-\sigma$ level except at the faint end $L_X<5\times 10^{39}$~erg~s$^{-1}$ . At these luminosities, most of the sample is made up of sources in galaxies in the distance range of 5--60~Mpc: those sources are most often resolved by \textit{Chandra}, whereas a substantial fraction of \textit{XMM-Newton} and \textit{Swift}-XRT sources are confused (Figure \ref{fig:frac_ambiguous}). Moreover, because of the higher noise level and the poorer spatial resolution, sources hidden in diffuse X-ray emission are harder to detect with those facilities. Consequently, the intrinsic slope of the \textit{XMM-Newton} and \textit{Swift} XLF are substantially underestimated at the faint end.

A similar problem occurs at fainter luminosities for \textit{Chandra} sources: as explained in \cite{Wang2016}, sources fainter than $10^{38}$~erg~s$^{-1}$  are more likely to be below the detection threshold, in particular when located in diffuse emission. This leads to the flattening of the XLF at these luminosities, so that diagnostics on the XLF faint end of XRB must rely on deep observations of close individual galaxies, as done in \cite{Mineo2012} or include a completeness correction as in \cite{Wang2016}.

Another moderately significant difference is the steeper slope in the bright end of 2SXPS XLFs, above $\sim 2\times 10^{40}$~erg~s$^{-1}$, while 4XMM and CSC2 are consistent with each other. A closer look at each sample above this luminosity reveals that half of 4XMM and CSC2 candidates are detected in galaxies of apparent diameter $D_{25}<20$~arcsec, whereas 2SXPS candidates in such galaxies are systematically discarded by the centre exclusion criterion due to their larger position error. There is thus a lack of HLXs in the 2SXPS sample. As a result, the XLF based on \textit{Chandra} sources is probably the most representative of the intrinsic, `universal' XLF of spiral galaxies.

\subsection{X-ray luminosity function}
\label{sec:4.1.2}

\subsubsection{XLF break}
\label{sec:xlfbk}

Our work shows significant evidence of the existence of a break between $3\times 10^{39}$ and $10^{40}$~erg~s$^{-1}$  . This somewhat agrees with the findings of \cite{Swartz2011} and \cite{Mineo2012}, suggesting a cutoff luminosity of $1-2\times 10^{40}$~erg~s$^{-1}$  . In contrast, \cite{Wang2016} reported that this break is spuriously caused by the lack of statistics in these studies, and stated that no break is needed at this point of their XLF. We see two reasons for this inconsistency: first, their sample is made up by searching X-ray sources within twice the $D_{25}$ of the galaxy, and they do not implement any procedure to remove the abundant contaminants this implies. These contaminants have an important influence at high luminosities (Figure \ref{fig:cont_frac}) and thus can flatten the XLF around $10^{40}$~erg~s$^{-1}$  . Second, they only considered the differential luminosity function in their XLF fits, showing higher dispersion, so that a break can be buried in this "noise". In their Figure \ref{fig:spiell_ratio} showing the cumulative XLF, there might be an hint for a break around $10^{40}$~erg~s$^{-1}$  .

The precise location of the break varies between samples, with 2SXPS sources yielding the largest value. We argue that the lack of low-luminosity ULXs due to source confusion issues, given the lower spatial resolution of \textit{Swift}-XRT, and of HLXs due to the large \textit{Swift} exclusion radius, are likely to distort the XLF and shift the break towards higher luminosities. This effect was already pointed out by \cite{Wang2016} when considering more distant elliptical galaxies, increasing the population of unresolved sources. Similarly, if ULXs and HLXs have different nature (see Section \ref{sec:4.4}), adding HLXs in the ULX cumulative XLFs will distort it so that the break will shift towards fainter luminosities and there will be a flattening at the high-luminosity end.

We test how the location of the break is changed when sources $>10^{41}$~erg~s$^{-1}$ are removed before fitting the XLF: while it is unchanged for 2SXPS ULXs, the best-fit break luminosity is found increased to $(5.7\pm 1.2)\times 10^{39}$~erg~s$^{-1}$ ($\log(L_{break}\mathrm{[erg~s^{-1}]}) = 39.75$)  and $(9.1\pm 2.6)\times 10^{39}$~erg~s$^{-1}$ ($\log(L_{break}\mathrm{[erg~s^{-1}]}) = 39.96$)  for the CSC2 and 4XMM samples, respectively. The single power-law fit is still excluded. Once converted to the \textit{XMM-Newton} broad energy band of 0.2--12~keV (assuming an absorbed power-law spectrum with $\Gamma=1.7$, $n_H=3\times 10^{20}$~cm$^{-2}$), the CSC2 break is further moved to $\sim 8\times 10^{39}$~erg~s$^{-1}$ ($\log(L_{break}\mathrm{[erg~s^{-1}]}) = 39.9$), consistent with the break fitted to the 4XMM sample.

More generally, although the use of cumulative XLFs provides higher statistics and may reveal more features, it is also subject to more biases. In particular, the faint end of the cumulative XLF is impacted by all the selection biases, leading to its flattening, while having the smallest poissonian error bars. This is likely to force the faint end of the XLF fit to flatten, potentially shifting the fitted break luminosity towards fainter luminosities to alleviate the tension between the model and the data. When the differential XLF is fitted (in the form $\mathrm{d}N/\mathrm{d}\ln L_X$ so that the slopes of the cumulative version are conserved), the break is shifted to $(6.4\pm2.4)\times 10^{39}$, $(6.5\pm4.2)\times 10^{39}$, $(1.21\pm0.27)\times 10^{40}$ and $(7.9 \pm 2.5)\times 10^{39}$~erg~s$^{-1}$ for the samples of CSC2, 4XMM, 2SXPS and all ULXs, respectively. As a result, the break is probably the cutoff feature reported by \cite{Mineo2012} and \cite{Swartz2011} for spiral galaxies. Most recently, this feature was also found in the XLF from deep \textit{Chandra} observations of the Virgo cluster \citep{Soria2022}, with a normalisation jump occurring in the range $4-5\times 10^{39}$~erg~s$^{-1}$  .


We investigate the physical reasons for this break. The first possibility is that a different galaxy population is probed in each luminosity bin of the XLF. Indeed, the XLF shape was found to significantly depend on the galaxy properties, with, for example, a flatter slope for more star-forming galaxies \citep{Wang2016}. We find a moderate increase of the mean SFR of ULX hosts in bins of increasing luminosity, with typical values $<0.5$~M$_\odot$~yr$^{-1}$ below $5\times 10^{39}$~erg~s$^{-1}$ and $>3$~M$_\odot$~yr$^{-1}$ above $10^{41}$~erg~s$^{-1}$. Extrapolating the result of \cite{Wang2016} to higher luminosities, this should produce a flattening of our XLF with luminosity, whereas we find the opposite behaviour. The stellar mass follows the same trend as the SFR, so that each bin probed similar specific SFRs. The XLF break is thus unlikely to be a selection effect of different environments.

Another possibility is that the luminosity break corresponds to the Eddington luminosity of a given class of objects, such as neutron stars or a certain population of black holes. For example, the Eddington luminosity of $\sim 2 \mathrm{M}_\odot$ neutron stars was argued to be consistent with the $5\times 10^{38}$~erg~s$^{-1}$ break seen in the luminosity function of LMXB, when accounting for beaming effects and the accretion of helium-rich material \citep{KimFabbiano2004,Wang2016}. With the same arguments, we could expect the break found in this study to be due to the Eddington luminosity of $\sim 25$~M$_\odot$ black holes. Recent modelling works show that radiative processes allow a luminosity as high as $10^{40}$~erg~s$^{-1}$  from supercritically accreting $20 \mathrm{M}_\odot$ stellar-mass black holes \citep{Krticka2022}. However, the $>50$~M$_\odot$ black holes detected by gravitational wave facilities rule out this break to match the maximum mass of stellar-mass black holes, unless the mass distribution of stellar-mass black holes is bimodal (which could be the case if black holes in XRBs and black holes in mergers do not share the same formation channels, \citealt{Fishbach2022}, but see \citealt{Belczynski2021}). Moreover, the same luminosity could be reached by the most magnetised neutron stars \citep{Mushtukov2015}, but the discovery of pulsars reaching even greater luminosities (e.g. \citealt{Israel2017}) may have challenged this interpretation. 

\begin{figure}
    \centering
    \includegraphics[width=8.5cm]{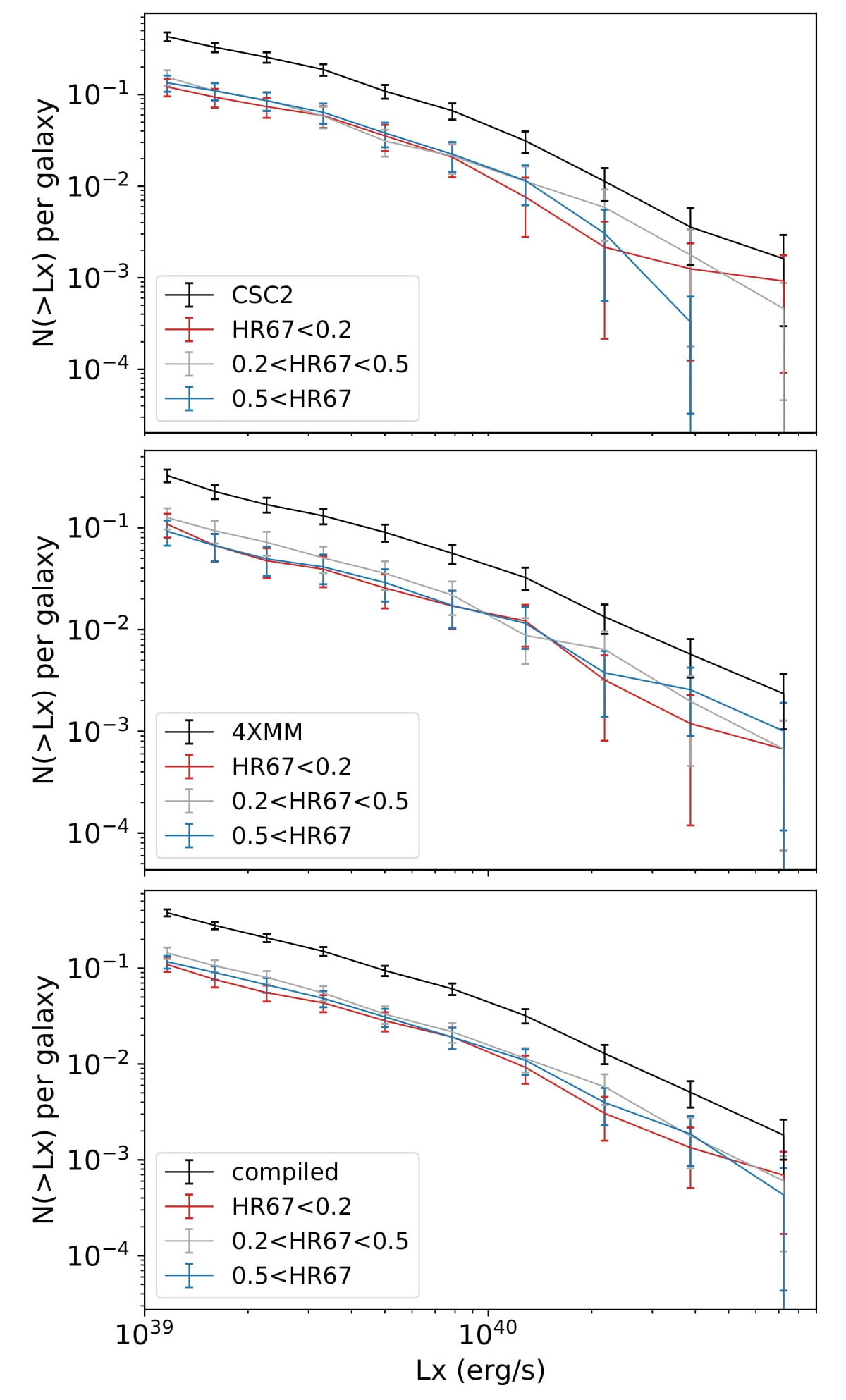}
    \caption{Deconvolved XLFs of \textit{Chandra}, \textit{XMM-Newton} and all selected ULXs in spiral galaxies, after removal of HLX candidates. Three bins of hardness ratio \texttt{HR67} (see the text for details) are also shown, representing soft ($\texttt{HR67}<0.2$), medium ($0.2<\texttt{HR67}<0.5$) and hard ($\texttt{HR67}>0.5$) sources.}
    \label{fig:xlf_hr}
\end{figure}

On the other hand, the source spectrum, and hardness ratios, may allow better diagnostics of the source nature. Using a sample of 17 ULXs including six PULXs, \cite{Gurpide2021} find that the secure neutron stars of their sample are among the hardest sources. To probe which type of accretor dominate at different luminosities, and if they all present the same XLF features, we divide our samples in three bins of \texttt{HR67}: soft ($\texttt{HR67}<0.2$), medium ($0.2<\texttt{HR67}<0.5$) and hard ($\texttt{HR67}>0.5$) ULXs. The result is plotted in Figure \ref{fig:xlf_hr}, showing no significant difference between each XLF. The possible reason is that \texttt{HR67} is inefficient to distinguish neutron star and black hole ULXs, or perhaps more interestingly, that a similar break should exist in both populations even if for different physical reasons. For instance, since the contribution of LMXB in spiral galaxies is known to be non-negligible even at ULX luminosities (e.g. \citealt{Mineo2012,Lehmer2019}), the break could be the manifestation of a normalisation jump in the XLF due to a luminosity cutoff at a few $10^{39}$~erg~s$^{-1}$  for LMXBs, as mentioned in \cite{Soria2022}. 

Alternatively, the HMXB XLF was found to present a large dispersion in the range $10^{38}-10^{40}$~erg~s$^{-1}$  related to galaxy metallicity \citep{Lehmer2021}, with a high luminosity cutoff located between $3\times 10^{39}$~erg~s$^{-1}$  (low metallicity galaxies, 12+log(O/H)$\simeq$7.6) and $10^{40}$~erg~s$^{-1}$  (galaxies of solar metallicity, 12+log(O/H)$\simeq$8.7). The metallicity information is absent from GLADE, but we find 106 galaxies in our ULX sample that have a match in HECATE with a metallicity value: 85\% of them are in the range 8.4$<$12+log(O/H)$<$8.9, close to solar metallicity, supporting this explanation for the break. In this scenario, the XLF break we obtained may be due to the combination of an excess of luminous HMXB in low metallicity environments followed by a cutoff, and an increasingly large contribution of another type of object at the bright end, possibly intermediate-mass black holes, corresponding to the bulk of HLXs.

\subsubsection{XLF slopes}
\label{sec:xlfslop}
Throughout this Section, we compare the best-fit slopes obtained on cumulative XLFs to slope values reported in the literature. Since the latter are often obtained on differential XLF, e.g. $\mathrm{d}N/\mathrm{d}L_X\propto L_X^{-\alpha}$ for a simple power-law model, we convert these values to the cumulative case whenever needed ($N(>L_X)\propto L_X^{-\alpha'}$ where $\alpha'=\alpha-1$).

As discussed in the previous Section, the fits to the cumulative XLFs are subject to a number of biases which affect the fitted parameters, including the XLF slopes. For example, the inclusion of HLXs flattens the bright end of the distribution. Analysing the cumulative XLF with HLXs removed, we obtain best-fit high-luminosity slopes ($\alpha_2$) of $1.82\pm 0.1$, $1.66\pm 0.13$, $1.79\pm 0.14$ and $1.80\pm 0.10$ for CSC2, 4XMM, 2SXPS and all ULXs, respectively, in better agreement with the values inferred on the ULX-complete galaxies and close to the value $1.73^{+1.58}_{-0.54}$ obtained by \cite{Mineo2012} for HMXBs in the same luminosity range. 

In addition, resolution effects are also likely to distort the XLF, flattening it at the faint end when two low-luminosity ULXs are mistaken for a unique, brighter source. This also explains why the number of ULXs per galaxy decreases between CSC2, 4XMM and 2SXPS samples, with decreasing spatial resolution. Similarly, given the centre exclusion criterion, the larger position errors of \textit{Swift} lead to the removal of some `bona fide' HLXs, steepening the high-luminosity slope of the XLF.

Some of these biases can be alleviated by fitting differential XLFs, rather than their cumulative version. The slopes we obtain are indeed different from the cumulative XLF fits, with $(\alpha_1,\alpha_2)$ being $(0.64\pm 0.16,1.69\pm 0.17)$, $(0.88\pm 0.27, 1.16\pm 0.12)$, $(0.58\pm 0.11, 1.70\pm 0.18)$ and $(0.82\pm 0.09, 1.38\pm 0.09)$ for CSC2, 4XMM, 2SXPS and all ULXs, respectively. These low luminosity slope values better agree with the slope of the HMXB XLF in \cite{Mineo2012} and \cite{Lehmer2019}, however the high luminosity slopes are not well-constrained.

Unlike spiral galaxies, the XLF we obtained for elliptical  galaxies are in disagreement with the steeper ones reported in the literature. For instance, \cite{Walton2011} find a best-fit slope $\alpha=1.5\pm 0.4$ for their elliptical  galaxies, while \cite{Wang2016} obtain steeper slopes $1.80\pm 0.03$, $2.74\pm 0.10$ and $2.29\pm 0.18$ for elliptical, lenticular and all early-type galaxies, respectively. Note that the definition of elliptical galaxies in the present paper includes lenticular galaxies. In nearby galaxies, the XLF of LMXBs is found to present a break at a few $10^{38}$~erg~s$^{-1}$  followed by a power-law decrease of slope $\sim 1.8$ (\citealt{Humphrey2008,KimFabbiano2010}).

\begin{figure}
    \centering
    \includegraphics[width=8.5cm]{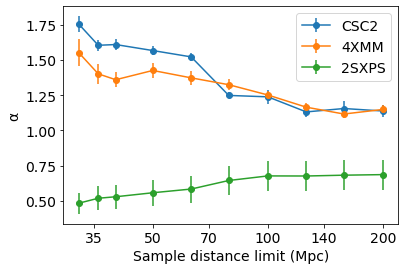}
    \caption{Best-fit slope of the deconvolved cumulative XLF of elliptical  galaxies, as a function of the distance limit of the sample.}
    \label{fig:alpha_ell}
\end{figure}

The flatter slope of the XLF for elliptical  galaxies in our study is examined. One possibility is that the hot gas content, prominent in these galaxies, is drowning the signal of ULXs well above the limiting luminosity we compute, which does not depend on the background level. In that case the census of ULXs is largely incomplete, up to a higher luminosity. For instance, M86 has a CSC2 limiting luminosity of $2\times 10^{38}$~erg~s$^{-1}$ from Equation (\ref{eq:2})  but the only sources detected within the $D_{25}/2$ ellipse have $L_X>2\times 10^{39}$~erg~s$^{-1}$  and $S/N<3$. At fixed hot gas density, the hot gas flux in a typical aperture increases with galaxy distance, affecting the detection of increasingly bright sources. To test this effect, we fit the deconvolved XLFs for samples with a range of distance limits. Figure \ref{fig:alpha_ell} shows a clear flattening of the XLF when extending the sample to further distances. Below 70~Mpc, the XLF of 4XMM ULXs becomes flatter than that of CSC2, probably because of the lower spatial resolution of \textit{XMM-Newton}. Even the closest elliptical  galaxies have few low-luminosity ULXs detected by \textit{Swift}-XRT because of its lower spatial resolution, hence the flat faint end of the XLF and the inconsistent slope. For a distance limit of $\sim 40$~Mpc, as adopted in previous studies of this XLF (e.g. \citealt{Wang2016}), we find a slope in the range $1.6-1.8$, in good agreement with values reported in the literature.

Actually, \cite{Mineo2012} noted a large diversity of XLFs from different spiral galaxies. This diversity of XLF is arguably due to different star formation histories, metallicities and initial mass functions. Ongoing efforts to constrain these quantities for a large fraction of galaxies is thus valuable in this regard. The assumption of a universal XLF thus cannot hold regardless of the galaxy environment, and ideally the XLF should be considered in specific bins of (morphology, specific SFR, $M_*$, metallicity), which is beyond the scope of this study. 

\subsection{Nature of HLXs}
\label{sec:4.4}
\defcitealias{Barrows2019}{B19}

Taking advantage of the large sky coverage and sensitivities of the catalogues we used, we were able to constitute a sample of 191 HLX candidates. Unlike ULXs, their classification by the classifier (mostly as AGN) cannot be used to identify their nature. Their high luminosities and galactocentric distances tend to bias the classifier towards the AGN class, particularly given that no HLX is present in the training sample of the classifier. However, \cite{Barrows2019} (hereafter \citetalias{Barrows2019}) found that their HLX population is at least partly consistent with accreting black holes at the centre of dwarf galaxies, satellites of the matched host.  
On the other hand, some HLXs are seen to have soft spectra, with most events detected below 1~keV, some of which being explained by the total or partial tidal disruption of a star by an intermediate-mass black hole \citep{godet2014,Lin2018}. Their classification as soft sources is therefore expected. The filtering of HLXs thus relies on visual inspection and the removal of known background and foreground sources.

A few contaminants may be left in the sample of selected HLXs. From the $\log N - \log S$ estimations reported in Table \ref{tab:ncontam}, the fraction of contaminants in the 2SXPS, 4XMM and CSC2 samples are $\sim 20$\%, $\sim 70$\% and $\sim 60$\%, after removal of spurious sources and before removal of known background and foreground sources. The shallower sensitivity of 2SXPS results in HLX candidates in nearby galaxies only, hence the lower contamination rate. The high contamination rates obtained for other surveys agree with the 70\% estimate of \cite{Zolotukhin2016}, but are in contrast with the 7\% of contaminants reported in \citetalias{Barrows2019}. As they explain, this difference is mainly driven by the lower angular sizes of their galaxies, most of them being located beyond 1000 Mpc, a distance range which is not probed by our sample. By using large and deep photometric redshift surveys, we were able to drastically reduce our high initial contamination fraction (background sources represent 85-95\% of sources reported in the last column of Table \ref{tab:ncontam}). Following the reasoning of Section \ref{sec:nbcontam} on ULX candidates, we conservatively expect the contamination rate among robust candidates (resp. robust and weak candidates) to be $\sim 20\%$ (resp. $\sim 30$\%). Thus, our slightly larger HLX sample is the result of the larger sky coverage of GLADE, \textit{Swift}-XRT and \textit{XMM-Newton}, to provide local analogues to their HLX population, better suited for future individual follow-ups.

\begin{figure}
    \centering
    \includegraphics[width=8.5cm]{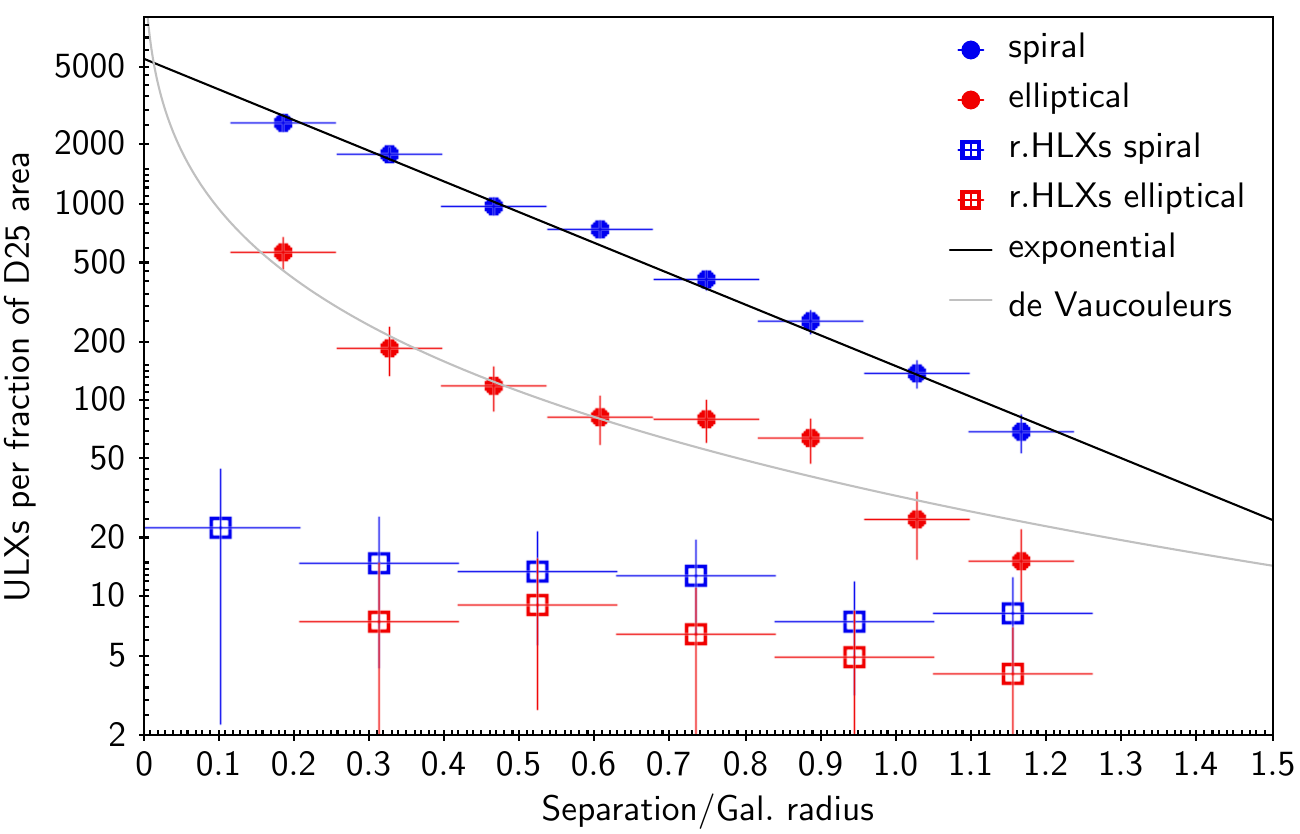}
    \caption{Radial distribution of all ULXs (filled circles) and robust HLXs (empty squares), showing their surface density as a function of galactocentric distance. Examples of typical galaxy light profiles are shown in solid lines.}
    \label{fig:surfdens_hlx}
\end{figure}

We investigate the possible nature of our HLX candidates. First of all, the bimodality of \texttt{HR67} as shown in Figure \ref{fig:hr67_hlx} suggests a difference of nature between soft and hard HLXs, the former being rarer and sometimes transient, with at least some of them consistent with partial or full tidal disruption events (e.g. \citealt{godet2014,Lin2018}). No such candidates were present in the sample of \citetalias{Barrows2019}, rather showing intrinsically hard spectra or important absorption. By construction, some AGN in galaxy pairs are also included in our HLX sample, and flagged as ``galaxy pair'' candidates. Since such objects are discarded when their true host is in the GLADE sample, the remaining sources represent a minority of our HLX candidates, but their luminosity distribution is consistent with the one of `nuclear' sources (Figure \ref{fig:xlf}). In contrast, robust HLXs  candidates (having no clear optical counterpart and visibly overlapping the area of the galaxy) and weak HLX candidates (having an optical counterpart in \textit{Gaia}, PanSTARRS, or DES, or visibly offset from the area of the galaxy), taken together or separately, present a steeper luminosity function, consistent with the bright end slope fitted to ULXs. This maybe supports a common mechanism behind the X-ray emission of these objects, such as accreting black holes of stellar to intermediate mass. However, following the method described in Section \ref{sec:ulxenv} for ULXs, we find a moderately significant difference between the radial distribution of ULXs and HLXs. Figure \ref{fig:surfdens_hlx} illustrates this discrepancy: unlike ULXs, HLXs may not follow typical light profiles and may be equally distributed in the extent of their host. This behaviour perhaps supports the scenario in which they are the central accretor of a galaxy satellite or globular cluster.

\begin{figure}
    \centering
    \includegraphics[width=8.5cm]{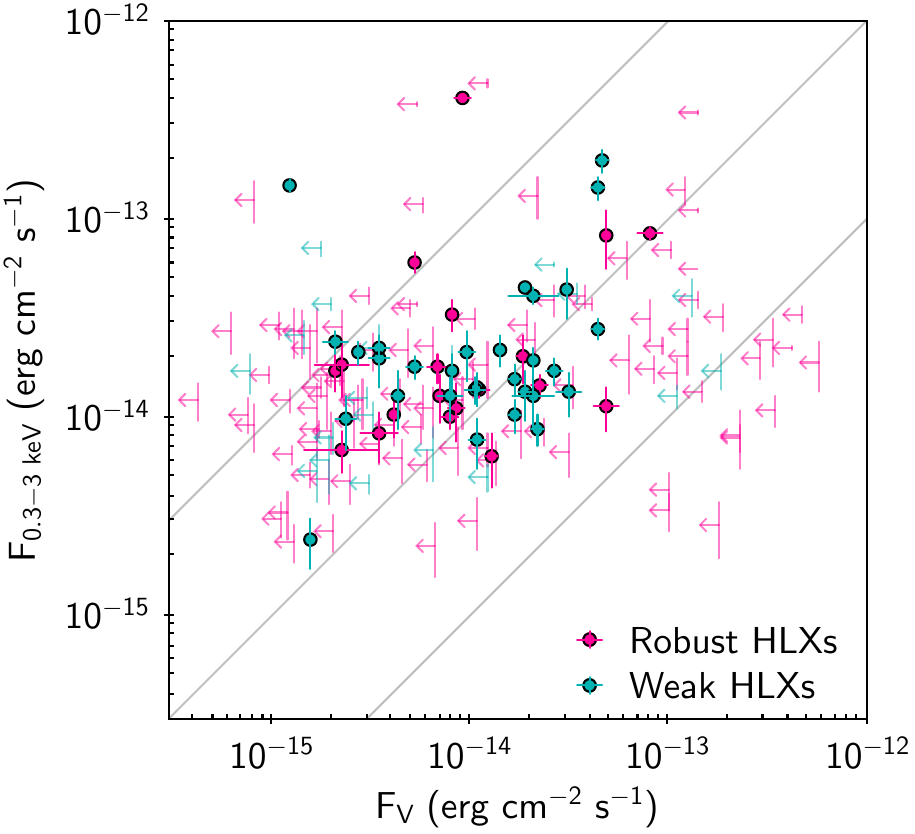}
    \caption{$F_X-F_V$ plane of robust and weak HLX candidates, showing their observed 0.3--3.5~keV X-ray flux as a function of the V-band flux of their optical counterpart (filled circles), or the upper limit on $F_V$ at this location (arrows). The grey solid lines indicate ratios of $F_X/F_V=0.1$, 1 and 10.}
    \label{fig:fxfv}
\end{figure}

Unlike ULXs, a significant fraction ($\sim 35$\%) of HLXs is found to have an optical counterpart. The X-ray to optical flux ratio can often give clues on the source nature. For instance, in AGN, the optical-UV emission of the disc scales with the soft X-ray emission, so that their ratio is generally in the range 0.1--10 (e.g. \citealt{Maccacaro1988,Barrows2019, Tranin2022}). Following the methodology of \citetalias{Barrows2019}, we first convert the B- and R-band fluxes given different matched optical surveys (PanSTARRS, DES, \textit{Gaia}, SDSS and the Beijing Arizona Sky Survey) to the V-band, using the conversion of \cite{Jester2005}. X-ray broad band fluxes are converted to the 0.3--3.5~keV band using the same fixed spectrum as above. While this is surely a limitation of our study, the heterogeneity of our sample and the low signal-to-noise levels do not allow for a more detailed spectral study. For HLXs present in both \citetalias{Barrows2019} and this work, we find a mean bias of our CSC2 fluxes of -0.2~dex with respect to their rest-frame unabsorbed fitted fluxes ($\pm$0.3~dex deviation), so this estimation is still reasonable for a population analysis. Finally, for X-ray sources having no counterpart, we attempt to compute an upper limit on their optical flux. \citetalias{Barrows2019} estimated the local background level directly from optical images, using an annulus of width being twice the radius of the host. However, we argue that such a large area is likely to contain other optical sources and thus bias the sensitivity at the HLX location towards shallower values, as suggested by the bright locus of flux upper limits in their $F_X-F_V$ plane. We adopt a simpler approach in which this sensitivity is assimilated to the flux of the faintest optical source having $S/N>5$ within a radius of 15~arcsec around the HLX candidate. Flux limits obtained with this method are found to scale well with the limits computed in \citetalias{Barrows2019} ($\pm 0.5$~dex deviation, bias-subtracted), but are on average ten times fainter. Figure \ref{fig:fxfv} shows the $F_X-F_{V}$ plane resulting from this analysis. In agreement with the finding of \citetalias{Barrows2019}, most of $F_X/F_V$ values in our sample are consistent with expectations for AGN. The use of more optical surveys allows us to obtain more stringent constraints on the optical flux upper limit.

Considering the scenario that HLX candidates are indeed accreting black holes associated with their host, we estimate their mass and the mass of their stellar counterpart. Black hole masses are estimated using bolometric corrections of \cite{Marconi2004} and a fixed Eddington ratio. Unlike \citetalias{Barrows2019} who compute bolometric luminosities from their unabsorbed hard (2--10~keV) X-ray flux obtained after spectral fitting, we rather use the softer (0.5--2~keV) part of the spectrum which better matches the union of several energy bands in CSC2, 4XMM and 2SXPS, and is less affected by background emission. Since we do not correct for absorption, this approach is likely to yield some underestimated masses, but we estimate this bias to be $\lesssim 0.6$~dex from a comparison of CSC2 fluxes with \citetalias{Barrows2019} fluxes. The derived bolometric corrections are in the range 10--30 depending on the source luminosity (for comparison, the band 2--10~keV would yield corrections in the range 7--20, and \citetalias{Barrows2019} use the fixed value 10). Assuming spectral properties similar to their sample, we adopt a value $f_{Edd}=0.24$ to derive black hole masses, and the standard formula $L_{Edd}\mathrm{[erg~s^{-1}]}\simeq 1.26\times 10^{38}M_{BH} [\mathrm{M}_\odot]$. The resulting masses are in the range $2\times 10^3-2\times 10^6 \mathrm{M}_\odot$ for robust and weak HLXs, with a median value of $4\times 10^4 \mathrm{M}_\odot$. In particular, 122 of these 169 (robust+weak) candidates fall in the IMBH mass range, below $10^5\mathrm{M}_\odot$. Being more luminous, HLXs in galaxy pairs have an estimated black hole mass $\gtrsim 10^4 \mathrm{M}_\odot$ with a median at $2\times 10^5 \mathrm{M}_\odot$.

Concerning galaxy stellar masses, as done in \citetalias{Barrows2019}, we use the mass to light ratio calibrated on optical colours \citep{Bell2003}. Since some optical surveys in our sample do not provide $i$-band magnitudes, the relation giving $M_*/L_r$ as a function of the $(g-r)$ colour is used, allowing us to compute $M_*$. Point-like optical counterparts of HLXs are assimilated to their stellar associations (we note as a caveat that this neglects the optical contribution from the putative AGN, while on the contrary we previously interpreted $F_V$ as the AGN flux).  Another perhaps important bias is that the point source flux $F_r$ is used in this step, instead of the flux integrated over the S\'ersic component. Except for galaxy pair candidates, for which $M_*$ may well be largely underestimated, we do not expect this effect to have a large impact on the computed stellar masses. We obtain that most of stellar counterparts of robust and weak HLXs are in the dwarf regime, $M_*<3\times 10^9\mathrm{M}_\odot$.

\begin{figure}
    \centering
    \includegraphics[width=8.7cm]{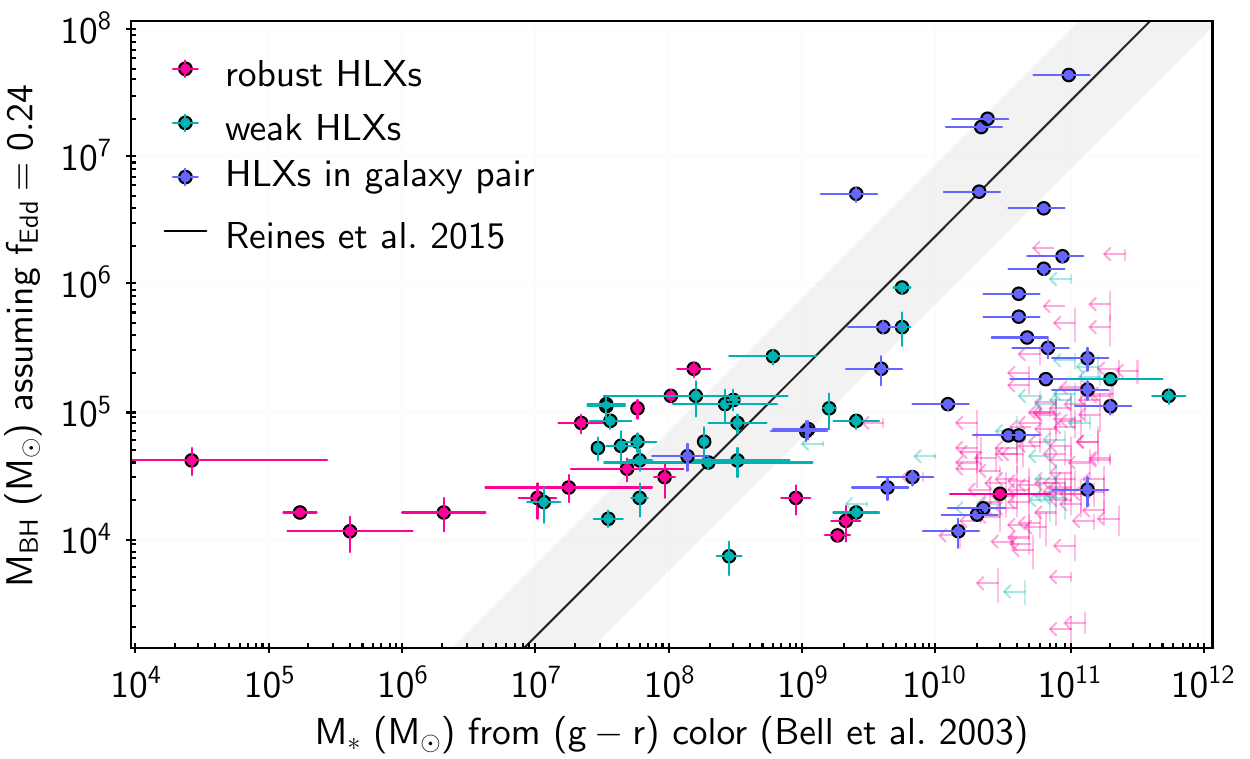}
    \caption{Black hole mass -- stellar mass plane, estimated from the 0.5-2keV X-ray flux and the $(g-r)$ colour as detailed in the text. Different types of HLX candidates are shown. Upper limits correspond to the stellar mass of the primary host.}
    \label{fig:mbhmg}
\end{figure}

The resulting $M_{BH}-M_*$ plane is shown in Figure \ref{fig:mbhmg}, also showing the scaling relation calibrated by \cite{Reines2015} on local AGN in low-mass galaxies. Upper limits corresponding to the stellar mass of the host are shown for HLXs having no optical counterpart detected in both $g$ and $r$ bands. While some robust and weak HLXs appear under-massive, most of them are hard sources, with $F_{0.5-2~\mathrm{keV}}$ accounting for less than 30\% of their broad-band flux: their black hole mass is likely to be underestimated. A second estimation using the extrapolated band 2--10~keV gives black hole masses in the range $2\times 10^3 - 2\times 10^7 \mathrm{M}_\odot$ for robust and weak HLXs (median $9\times 10^4 \mathrm{M}_\odot$), with these sources all above $3\times 10^4 \mathrm{M}_\odot$. Some galaxy pair candidates are largely under-massive, probably because of an underestimated $M_{BH}$ (occurring for instance if $f_{Edd}<0.24$, as is generally observed in such massive galaxies). Interestingly, we find a significant proportion ($\sim 25$\%) of overmassive black holes among robust and weak HLXs, supporting a scenario in which these black holes have undergone a merger-free growth and/or that their stellar counterpart have been tidally stripped by the primary host, or is instead a globular cluster.

This contrasts with \citetalias{Barrows2019} results who find a broad alignment of their HLXs to the relation of \cite{Reines2015}, maybe because our work is focussing on the more local Universe and thus probes lower black hole masses and an earlier growth regime. The proportion of overmassive black hole is further increased if a lower Eddington ratio is applied to compute black hole masses. For example, \cite{Baldassare2017} find a median Eddington ratio of $f_{Edd}=0.05$ in their sample of active dwarf galaxies, that they obtain through the unabsorbed X-ray luminosity and the black hole mass inferred from the broad H$\alpha$ line width (e.g. \citealt{Greene2005, Reines2013}). In our study, nevertheless, a more detailed modelling of X-ray spectrum and optical emission, in particular to securely estimate galaxy masses, would be needed to confirm this overmassive locus.

\section{Summary}
\label{sec:5}

Using the three largest X-ray surveys to date, and a galaxy catalogue of unprecedented completeness for the field, we were able to build the largest and cleanest catalogue of ULXs. It contains 1901 unique sources, built from 304, 667, and 1185 reliable ULXs from 2SXPS, 4XMM, and CSC2, respectively. They result from a strict filtering pipeline involving off-centre selection, the removal of all known contaminants, the use of an automatic, probabilistic classification, and visual inspection. The contamination rate is estimated to be no greater than 2\%, which is a considerable improvement on recent ULX catalogues.

We conducted a statistical study of the ULX luminosity function and preferred environments, and compared them with a clean XRB sample in terms of hardness and variability. We also isolated a sample of 191 HLX candidates and investigated their properties. We report the following results:
\begin{enumerate}
    \item Thanks to the use of Malmquist-corrected XLFs, considerably enlarging the number of bright sources usable in the XLF, the presence of a significant break at $\sim10^{40}$ erg~s$^{-1}$ is confirmed in all three surveys, in contradiction with the recent claim of \cite{Wang2016}.
    \item We confirm the trends followed by the ULX rate with the galaxy environment, found in previous studies. The ULX rate is positively correlated with the SFR and stellar mass. While low-luminosity ULXs tend to be hosted more often in spiral galaxies, the occupation fraction of spiral and elliptical galaxies becomes similar at HLX luminosities.
    \item No clear hardness or variability difference is seen among the whole populations of ULXs and XRBs. However, from a sample of significantly variable sources, we confirm the opposite evolution of hardness with luminosity between ULXs and XRBs. Half of the sample shows no hardness evolution with luminosity.
    \item The 191 HLX candidates include 115 robust HLXs and 54 less certain HLXs. Their optical counterparts and radial distribution, together with the XLF break, suggest a diverse nature as to the X-ray sources at such luminosities. We note that 120 of them are consistent with an accreting IMBH in a dwarf galaxy satellite of the identified host. 
\end{enumerate}

In the future, we will study in greater detail the hardness-luminosity evolution of ULXs and HLXs to better assess their nature (Amato et al. in prep), and conduct individual studies of some HLX outliers. A statistical study of X-ray sources in dwarf galaxies is also in preparation, to seek IMBHs and wandering black holes in isolated galaxies and investigate the ULX excess previously reported in these galaxies. 

\begin{acknowledgements}
We thank the anonymous referee for his helpful advice and insightful comments that improved the quality of this article. This work benefited from support from the CNES. This project has received funding from the European Union's Horizon 2020 research and innovation programme under grant agreement n°101004168, the XMM2ATHENA project. Hugo Tranin acknowledges Xan Astiasarain and Simon Dupourqué for fruitful discussions during the writing of this paper.
\end{acknowledgements}

\bibliographystyle{aa}
\bibliography{44952corr}

\begin{thebibliography}{155}
\expandafter\ifx\csname natexlab\endcsname\relax\def\natexlab#1{#1}\fi

\bibitem[{{Abbott} {et~al.}(2020){Abbott}, {Abbott}, {Abraham}, {Acernese},
  {Ackley}, {Adams}, {Adhikari}, {Adya}, {Affeldt}, {Agathos}, {Agatsuma},
  {Aggarwal}, {Aguiar}, {Aich}, {Aiello}, {Ain}, {Ajith}, {Akcay}, {Allen},
  {Allocca}, {Altin}, {Amato}, {Anand}, {Ananyeva}, {Anderson}, {Anderson},
  {Angelova}, {Ansoldi}, {Antier}, {Appert}, {Arai}, {Araya}, {Areeda},
  {Ar{\`e}ne}, {Arnaud}, {Aronson}, {Arun}, {Asali}, {Ascenzi}, {Ashton},
  {Aston}, {Astone}, {Aubin}, {Aufmuth}, {AultONeal}, {Austin}, {Avendano},
  {Babak}, {Bacon}, {Badaracco}, {Bader}, {Bae}, {Baer}, {Baird}, {Baldaccini},
  {Ballardin}, {Ballmer}, {Bals}, {Balsamo}, {Baltus}, {Banagiri}, {Bankar},
  {Bankar}, {Barayoga}, {Barbieri}, {Barish}, {Barker}, {Barkett}, {Barneo},
  {Barone}, {Barr}, {Barsotti}, {Barsuglia}, {Barta}, {Bartlett}, {Bartos},
  {Bassiri}, {Basti}, {Bawaj}, {Bayley}, {Bazzan}, {B{\'e}csy}, {Bejger},
  {Belahcene}, {Bell}, {Beniwal}, {Benjamin}, {Bentley}, {Bergamin}, {Berger},
  {Bergmann}, {Bernuzzi}, {Berry}, {Bersanetti}, {Bertolini}, {Betzwieser},
  {Bhandare}, {Bhandari}, {Bidler}, {Biggs}, {Bilenko}, {Billingsley},
  {Birney}, {Birnholtz}, {Biscans}, {Bischi}, {Biscoveanu}, {Bisht},
  {Bissenbayeva}, {Bitossi}, {Bizouard}, {Blackburn}, {Blackman}, {Blair},
  {Blair}, {Blair}, {Bobba}, {Bode}, {Boer}, {Boetzel}, {Bogaert}, {Bondu},
  {Bonilla}, {Bonnand}, {Booker}, {Boom}, {Bork}, {Boschi}, {Bose},
  {Bossilkov}, {Bosveld}, {Bouffanais}, {Bozzi}, {Bradaschia}, {Brady},
  {Bramley}, {Branchesi}, {Brau}, {Breschi}, {Briant}, {Briggs}, {Brighenti},
  {Brillet}, {Brinkmann}, {Brockill}, {Brooks}, {Brooks}, {Brown}, {Brunett},
  {Bruno}, {Bruntz}, {Buikema}, {Bulik}, {Bulten}, {Buonanno}, {Buscicchio},
  {Buskulic}, {Byer}, {Cabero}, {Cadonati}, {Cagnoli}, {Cahillane},
  {Calder{\'o}n Bustillo}, {Callaghan}, {Callister}, {Calloni}, {Camp},
  {Canepa}, {Cannon}, {Cao}, {Cao}, {Carapella}, {Carbognani}, {Caride},
  {Carney}, {Carullo}, {Casanueva Diaz}, {Casentini}, {Casta{\~n}eda},
  {Caudill}, {Cavagli{\`a}}, {Cavalier}, {Cavalieri}, {Cella},
  {Cerd{\'a}-Dur{\'a}n}, {Cesarini}, {Chaibi}, {Chakravarti}, {Chan}, {Chan},
  {Chandra}, {Chao}, {Charlton}, {Chase}, {Chassande-Mottin}, {Chatterjee},
  {Chaturvedi}, {Chatziioannou}, {Chen}, {Chen}, {Chen}, {Cheng}, {Cheong},
  {Chia}, {Chiadini}, {Chierici}, {Chincarini}, {Chiummo}, {Cho}, {Cho}, {Cho},
  {Christensen}, {Chu}, {Chua}, {Chung}, {Chung}, {Ciani}, {Ciecielag},
  {Cie{\'s}lar}, {Ciobanu}, {Ciolfi}, {Cipriano}, {Cirone}, {Clara}, {Clark},
  {Clearwater}, {Clesse}, {Cleva}, {Coccia}, {Cohadon}, {Cohen}, {Colleoni},
  {Collette}, {Collins}, {Colpi}, {Constancio}, {Conti}, {Cooper}, {Corban},
  {Corbitt}, {Cordero-Carri{\'o}n}, {Corezzi}, {Corley}, {Cornish}, {Corre},
  {Corsi}, {Cortese}, {Costa}, {Cotesta}, {Coughlin}, {Coughlin}, {Coulon},
  {Countryman}, {Couvares}, {Covas}, {Coward}, {Cowart}, {Coyne}, {Coyne},
  {Creighton}, {Creighton}, {Cripe}, {Croquette}, {Crowder}, {Cudell},
  {Cullen}, {Cumming}, {Cummings}, {Cunningham}, {Cuoco}, {Curylo}, {Canton},
  {D{\'a}lya}, {Dana}, {Daneshgaran-Bajastani}, {D'Angelo}, {Danilishin},
  {D'Antonio}, {Danzmann}, {Darsow-Fromm}, {Dasgupta}, {Datrier}, {Dattilo},
  {Dave}, {Davier}, {Davies}, {Davis}, {Daw}, {DeBra}, {Deenadayalan},
  {Degallaix}, {De Laurentis}, {Del{\'e}glise}, {Delfavero}, {De Lillo}, {Del
  Pozzo}, {DeMarchi}, {D'Emilio}, {Demos}, {Dent}, {De Pietri}, {De Rosa}, {De
  Rossi}, {DeSalvo}, {de Varona}, {Dhurandhar}, {D{\'\i}az}, {Diaz-Ortiz},
  {Dietrich}, {Di Fiore}, {Di Fronzo}, {Di Giorgio}, {Di Giovanni}, {Di
  Giovanni}, {Di Girolamo}, {Di Lieto}, {Ding}, {Di Pace}, {Di Palma}, {Di
  Renzo}, {Divakarla}, {Dmitriev}, {Doctor}, {Donovan}, {Dooley}, {Doravari},
  {Dorrington}, {Downes}, {Drago}, {Driggers}, {Du}, {Ducoin}, {Dupej},
  {Durante}, {D'Urso}, {Dwyer}, {Easter}, {Eddolls}, {Edelman}, {Edo}, {Edy},
  {Effler}, {Ehrens}, {Eichholz}, {Eikenberry}, {Eisenmann}, {Eisenstein},
  {Ejlli}, {Errico}, {Essick}, {Estelles}, {Estevez}, {Etienne}, {Etzel},
  {Evans}, {Evans}, {Ewing}, {Fafone}, {Fairhurst}, {Fan}, {Farinon}, {Farr},
  {Farr}, {Fauchon-Jones}, {Favata}, {Fays}, {Fazio}, {Feicht}, {Fejer},
  {Feng}, {Fenyvesi}, {Ferguson}, {Fernandez-Galiana}, {Ferrante}, {Ferreira},
  {Ferreira}, {Fidecaro}, {Fiori}, {Fiorucci}, {Fishbach}, {Fisher},
  {Fittipaldi}, {Fitz-Axen}, {Fiumara}, {Flaminio}, {Floden}, {Flynn}, {Fong},
  {Font}, {Forsyth}, {Fournier}, {Frasca}, {Frasconi}, {Frei}, {Freise},
  {Frey}, {Frey}, {Fritschel}, {Frolov}, {Fronz{\`e}}, {Fulda}, {Fyffe},
  {Gabbard}, {Gadre}, {Gaebel}, {Gair}, {Galaudage}, {Ganapathy}, {Ganguly},
  {Gaonkar}, {Garc{\'\i}a-Quir{\'o}s}, {Garufi}, {Gateley}, {Gaudio},
  {Gayathri}, {Gemme}, {Genin}, {Gennai}, {George}, {George}, {Gergely},
  {Ghonge}, {Ghosh}, {Ghosh}, {Ghosh}, {Giacomazzo}, {Giaime}, {Giardina},
  {Gibson}, {Gier}, {Gill}, {Glanzer}, {Gniesmer}, {Godwin}, {Goetz}, {Goetz},
  {Gohlke}, {Goncharov}, {Gonz{\'a}lez}, {Gopakumar}, {Gossan}, {Gosselin},
  {Gouaty}, {Grace}, {Grado}, {Granata}, {Grant}, {Gras}, {Grassia}, {Gray},
  {Gray}, {Greco}, {Green}, {Green}, {Gretarsson}, {Griggs}, {Grignani},
  {Grimaldi}, {Grimm}, {Grote}, {Grunewald}, {Gruning}, {Guidi}, {Guimaraes},
  {Guix{\'e}}, {Gulati}, {Guo}, {Gupta}, {Gupta}, {Gupta}, {Gustafson},
  {Gustafson}, {Haegel}, {Halim}, {Hall}, {Hamilton}, {Hammond}, {Haney},
  {Hanke}, {Hanks}, {Hanna}, {Hannam}, {Hannuksela}, {Hansen}, {Hanson},
  {Harder}, {Hardwick}, {Haris}, {Harms}, {Harry}, {Harry}, {Hasskew},
  {Haster}, {Haughian}, {Hayes}, {Healy}, {Heidmann}, {Heintze}, {Heinze},
  {Heitmann}, {Hellman}, {Hello}, {Hemming}, {Hendry}, {Heng}, {Hennes},
  {Hennig}, {Heurs}, {Hild}, {Hinderer}, {Hoback}, {Hochheim}, {Hofgard},
  {Hofman}, {Holgado}, {Holland}, {Holt}, {Holz}, {Hopkins}, {Horst}, {Hough},
  {Howell}, {Hoy}, {Huang}, {H{\"u}bner}, {Huerta}, {Huet}, {Hughey}, {Hui},
  {Husa}, {Huttner}, {Huxford}, {Huynh-Dinh}, {Idzkowski}, {Iess}, {Inchauspe},
  {Ingram}, {Intini}, {Isac}, {Isi}, {Iyer}, {Jacqmin}, {Jadhav}, {Jadhav},
  {James}, {Jani}, {Janthalur}, {Jaranowski}, {Jariwala}, {Jaume}, {Jenkins},
  {Jiang}, {Johns}, {Johnson-McDaniel}, {Jones}, {Jones}, {Jones}, {Jones},
  {Jones}, {Jonker}, {Ju}, {Junker}, {Kalaghatgi}, {Kalogera}, {Kamai},
  {Kandhasamy}, {Kang}, {Kanner}, {Kapadia}, {Karki}, {Kashyap}, {Kasprzack},
  {Kastaun}, {Katsanevas}, {Katsavounidis}, {Katzman}, {Kaufer}, {Kawabe},
  {K{\'e}f{\'e}lian}, {Keitel}, {Keivani}, {Kennedy}, {Key}, {Khadka},
  {Khalili}, {Khan}, {Khan}, {Khan}, {Khazanov}, {Khetan}, {Khursheed},
  {Kijbunchoo}, {Kim}, {Kim}, {Kim}, {Kim}, {Kim}, {Kim}, {Kim}, {Kimball},
  {King}, {Kinley-Hanlon}, {Kirchhoff}, {Kissel}, {Kleybolte}, {Klimenko},
  {Knowles}, {Knyazev}, {Koch}, {Koehlenbeck}, {Koekoek}, {Koley},
  {Kondrashov}, {Kontos}, {Koper}, {Korobko}, {Korth}, {Kovalam}, {Kozak},
  {Kringel}, {Krishnendu}, {Kr{\'o}lak}, {Krupinski}, {Kuehn}, {Kumar},
  {Kumar}, {Kumar}, {Kumar}, {Kumar}, {Kuo}, {Kutynia}, {Lackey}, {Laghi},
  {Lalande}, {Lam}, {Lamberts}, {Landry}, {Lane}, {Lang}, {Lange}, {Lantz},
  {Lanza}, {La Rosa}, {Lartaux-Vollard}, {Lasky}, {Laxen}, {Lazzarini},
  {Lazzaro}, {Leaci}, {Leavey}, {Lecoeuche}, {Lee}, {Lee}, {Lee}, {Lee}, {Lee},
  {Lehmann}, {Leroy}, {Letendre}, {Levin}, {Li}, {Li}, {li}, {Li}, {Li},
  {Linde}, {Linker}, {Linley}, {Littenberg}, {Liu}, {Liu},
  {Llorens-Monteagudo}, {Lo}, {Lockwood}, {London}, {Longo}, {Lorenzini},
  {Loriette}, {Lormand}, {Losurdo}, {Lough}, {Lousto}, {Lovelace}, {L{\"u}ck},
  {Lumaca}, {Lundgren}, {Ma}, {Macas}, {Macfoy}, {MacInnis}, {Macleod},
  {MacMillan}, {Macquet}, {Maga{\~n}a Hernandez}, {Maga{\~n}a-Sandoval},
  {Magee}, {Majorana}, {Maksimovic}, {Malik}, {Man}, {Mandic}, {Mangano},
  {Mansell}, {Manske}, {Mantovani}, {Mapelli}, {Marchesoni}, {Marion},
  {M{\'a}rka}, {M{\'a}rka}, {Markakis}, {Markosyan}, {Markowitz}, {Maros},
  {Marquina}, {Marsat}, {Martelli}, {Martin}, {Martin}, {Martinez}, {Martynov},
  {Masalehdan}, {Mason}, {Massera}, {Masserot}, {Massinger}, {Masso-Reid},
  {Mastrogiovanni}, {Matas}, {Matichard}, {Mavalvala}, {Maynard}, {McCann},
  {McCarthy}, {McClelland}, {McCormick}, {McCuller}, {McGuire}, {McIsaac},
  {McIver}, {McManus}, {McRae}, {McWilliams}, {Meacher}, {Meadors}, {Mehmet},
  {Mehta}, {Mejuto Villa}, {Melatos}, {Mendell}, {Mercer}, {Mereni}, {Merfeld},
  {Merilh}, {Merritt}, {Merzougui}, {Meshkov}, {Messenger}, {Messick},
  {Metzdorff}, {Meyers}, {Meylahn}, {Mhaske}, {Miani}, {Miao}, {Michaloliakos},
  {Michel}, {Middleton}, {Milano}, {Miller}, {Millhouse}, {Mills}, {Milotti},
  {Milovich-Goff}, {Minazzoli}, {Minenkov}, {Mishkin}, {Mishra}, {Mistry},
  {Mitra}, {Mitrofanov}, {Mitselmakher}, {Mittleman}, {Mo}, {Mogushi},
  {Mohapatra}, {Mohite}, {Molina-Ruiz}, {Mondin}, {Montani}, {Moore}, {Moraru},
  {Morawski}, {Moreno}, {Morisaki}, {Mours}, {Mow-Lowry}, {Mozzon},
  {Muciaccia}, {Mukherjee}, {Mukherjee}, {Mukherjee}, {Mukherjee}, {Mukund},
  {Mullavey}, {Munch}, {Mu{\~n}iz}, {Murray}, {Nagar}, {Nardecchia},
  {Naticchioni}, {Nayak}, {Neil}, {Neilson}, {Nelemans}, {Nelson}, {Nery},
  {Neunzert}, {Ng}, {Ng}, {Nguyen}, {Nguyen}, {Nichols}, {Nichols}, {Nissanke},
  {Nitz}, {Nocera}, {Noh}, {North}, {Nothard}, {Nuttall}, {Oberling},
  {O'Brien}, {Oganesyan}, {Ogin}, {Oh}, {Oh}, {Ohme}, {Ohta}, {Okada},
  {Oliver}, {Olivetto}, {Oppermann}, {Oram}, {O'Reilly}, {Ormiston}, {Ortega},
  {O'Shaughnessy}, {Ossokine}, {Osthelder}, {Ottaway}, {Overmier}, {Owen},
  {Pace}, {Pagano}, {Page}, {Pagliaroli}, {Pai}, {Pai}, {Palamos}, {Palashov},
  {Palomba}, {Pan}, {Panda}, {Pang}, {Pankow}, {Pannarale}, {Pant}, {Paoletti},
  {Paoli}, {Parida}, {Parker}, {Pascucci}, {Pasqualetti}, {Passaquieti},
  {Passuello}, {Patricelli}, {Payne}, {Pearlstone}, {Pechsiri}, {Pedersen},
  {Pedraza}, {Pele}, {Penn}, {Perego}, {Perez}, {P{\'e}rigois}, {Perreca},
  {Perri{\`e}s}, {Petermann}, {Pfeiffer}, {Phelps}, {Phukon}, {Piccinni},
  {Pichot}, {Piendibene}, {Piergiovanni}, {Pierro}, {Pillant}, {Pinard},
  {Pinto}, {Piotrzkowski}, {Pirello}, {Pitkin}, {Plastino}, {Poggiani}, {Pong},
  {Ponrathnam}, {Popolizio}, {Porter}, {Powell}, {Prajapati}, {Prasai},
  {Prasanna}, {Pratten}, {Prestegard}, {Principe}, {Prodi}, {Prokhorov},
  {Punturo}, {Puppo}, {P{\"u}rrer}, {Qi}, {Quetschke}, {Quinonez}, {Raab},
  {Raaijmakers}, {Radkins}, {Radulesco}, {Raffai}, {Rafferty}, {Raja}, {Rajan},
  {Rajbhandari}, {Rakhmanov}, {Ramirez}, {Ramos-Buades}, {Rana}, {Rao},
  {Rapagnani}, {Raymond}, {Razzano}, {Read}, {Regimbau}, {Rei}, {Reid},
  {Reitze}, {Rettegno}, {Ricci}, {Richardson}, {Richardson}, {Ricker},
  {Riemenschneider}, {Riles}, {Rizzo}, {Robertson}, {Robinet}, {Rocchi},
  {Rodriguez-Soto}, {Rolland}, {Rollins}, {Roma}, {Romanelli}, {Romano},
  {Romel}, {Romero-Shaw}, {Romie}, {Rose}, {Rose}, {Rose}, {Rosi{\'n}ska},
  {Rosofsky}, {Ross}, {Rowan}, {Rowlinson}, {Roy}, {Roy}, {Roy}, {Ruggi},
  {Rutins}, {Ryan}, {Sachdev}, {Sadecki}, {Sakellariadou}, {Salafia},
  {Salconi}, {Saleem}, {Salemi}, {Samajdar}, {Sanchez}, {Sanchez},
  {Sanchis-Gual}, {Sanders}, {Santiago}, {Santos}, {Sarin}, {Sassolas},
  {Sathyaprakash}, {Sauter}, {Savage}, {Savant}, {Sawant}, {Sayah}, {Schaetzl},
  {Schale}, {Scheel}, {Scheuer}, {Schmidt}, {Schnabel}, {Schofield},
  {Sch{\"o}nbeck}, {Schreiber}, {Schulte}, {Schutz}, {Schwarm}, {Schwartz},
  {Scott}, {Scott}, {Seidel}, {Sellers}, {Sengupta}, {Sennett}, {Sentenac},
  {Sequino}, {Sergeev}, {Setyawati}, {Shaddock}, {Shaffer}, {Sharifi},
  {Shahriar}, {Sharma}, {Sharma}, {Shawhan}, {Shen}, {Shikauchi}, {Shink},
  {Shoemaker}, {Shoemaker}, {Shukla}, {ShyamSundar}, {Siellez}, {Sieniawska},
  {Sigg}, {Singer}, {Singh}, {Singh}, {Singha}, {Singhal}, {Sintes}, {Sipala},
  {Skliris}, {Slagmolen}, {Slaven-Blair}, {Smetana}, {Smith}, {Smith},
  {Somala}, {Son}, {Soni}, {Sorazu}, {Sordini}, {Sorrentino}, {Souradeep},
  {Sowell}, {Spencer}, {Spera}, {Srivastava}, {Srivastava}, {Staats},
  {Stachie}, {Standke}, {Steer}, {Steinke}, {Steinlechner}, {Steinlechner},
  {Steinmeyer}, {Stevenson}, {Stocks}, {Stops}, {Stover}, {Strain}, {Stratta},
  {Strunk}, {Sturani}, {Stuver}, {Sudhagar}, {Sudhir}, {Summerscales}, {Sun},
  {Sunil}, {Sur}, {Suresh}, {Sutton}, {Swinkels}, {Szczepa{\'n}czyk}, {Tacca},
  {Tait}, {Talbot}, {Tanasijczuk}, {Tanner}, {Tao}, {T{\'a}pai}, {Tapia},
  {Tapia San Martin}, {Tasson}, {Taylor}, {Tenorio}, {Terkowski},
  {Thirugnanasambandam}, {Thomas}, {Thomas}, {Thompson}, {Thondapu}, {Thorne},
  {Thrane}, {Tinsman}, {Saravanan}, {Tiwari}, {Tiwari}, {Tiwari}, {Toland},
  {Tonelli}, {Tornasi}, {Torres-Forn{\'e}}, {Torrie}, {Tosta e Melo},
  {T{\"o}yr{\"a}}, {Travasso}, {Traylor}, {Tringali}, {Tripathee}, {Trovato},
  {Trudeau}, {Tsang}, {Tse}, {Tso}, {Tsukada}, {Tsuna}, {Tsutsui}, {Turconi},
  {Ubhi}, {Udall}, {Ueno}, {Ugolini}, {Unnikrishnan}, {Urban}, {Usman},
  {Utina}, {Vahlbruch}, {Vajente}, {Valdes}, {Valentini}, {van Bakel}, {van
  Beuzekom}, {van den Brand}, {Van Den Broeck}, {Vander-Hyde}, {van der
  Schaaf}, {Van Heijningen}, {van Veggel}, {Vardaro}, {Varma}, {Vass},
  {Vas{\'u}th}, {Vecchio}, {Vedovato}, {Veitch}, {Veitch}, {Venkateswara},
  {Venugopalan}, {Verkindt}, {Veske}, {Vetrano}, {Vicer{\'e}}, {Viets},
  {Vinciguerra}, {Vine}, {Vinet}, {Vitale}, {Vivanco}, {Vo}, {Vocca},
  {Vorvick}, {Vyatchanin}, {Wade}, {Wade}, {Wade}, {Walet}, {Walker},
  {Wallace}, {Wallace}, {Walsh}, {Wang}, {Wang}, {Wang}, {Ward}, {Warden},
  {Warner}, {Was}, {Watchi}, {Weaver}, {Wei}, {Weinert}, {Weinstein}, {Weiss},
  {Wellmann}, {Wen}, {We{\ss}els}, {Westhouse}, {Wette}, {Whelan}, {Whiting},
  {Whittle}, {Wilken}, {Williams}, {Willis}, {Willke}, {Winkler}, {Wipf},
  {Wittel}, {Woan}, {Woehler}, {Wofford}, {Wong}, {Wright}, {Wu}, {Wysocki},
  {Xiao}, {Yamamoto}, {Yang}, {Yang}, {Yang}, {Yap}, {Yazback}, {Yeeles}, {Yu},
  {Yu}, {Yuen}, {Zadro{\.Z}ny}, {Zadro{\.Z}ny}, {Zanolin}, {Zelenova},
  {Zendri}, {Zevin}, {Zhang}, {Zhang}, {Zhang}, {Zhao}, {Zhao}, {Zhou}, {Zhou},
  {Zhu}, {Zimmerman}, {Zucker}, {Zweizig}, {LIGO Scientific Collaboration}, \&
  {Virgo Collaboration}}]{GW190521}
{Abbott}, R., {Abbott}, T.~D., {Abraham}, S., {et~al.} 2020, \prl, 125, 101102

\bibitem[{{Abbott} {et~al.}(2018){Abbott}, {Abdalla}, {Allam}, {Amara},
  {Annis}, {Asorey}, {Avila}, {Ballester}, {Banerji}, {Barkhouse}, {Baruah},
  {Baumer}, {Bechtol}, {Becker}, {Benoit-L{\'e}vy}, {Bernstein}, {Bertin},
  {Blazek}, {Bocquet}, {Brooks}, {Brout}, {Buckley-Geer}, {Burke}, {Busti},
  {Campisano}, {Cardiel-Sas}, {Carnero Rosell}, {Carrasco Kind}, {Carretero},
  {Castander}, {Cawthon}, {Chang}, {Chen}, {Conselice}, {Costa}, {Crocce},
  {Cunha}, {D'Andrea}, {da Costa}, {Das}, {Daues}, {Davis}, {Davis}, {De
  Vicente}, {DePoy}, {DeRose}, {Desai}, {Diehl}, {Dietrich}, {Dodelson},
  {Doel}, {Drlica-Wagner}, {Eifler}, {Elliott}, {Evrard}, {Farahi}, {Fausti
  Neto}, {Fernandez}, {Finley}, {Flaugher}, {Foley}, {Fosalba}, {Friedel},
  {Frieman}, {Garc{\'\i}a-Bellido}, {Gaztanaga}, {Gerdes}, {Giannantonio},
  {Gill}, {Glazebrook}, {Goldstein}, {Gower}, {Gruen}, {Gruendl}, {Gschwend},
  {Gupta}, {Gutierrez}, {Hamilton}, {Hartley}, {Hinton}, {Hislop}, {Hollowood},
  {Honscheid}, {Hoyle}, {Huterer}, {Jain}, {James}, {Jeltema}, {Johnson},
  {Johnson}, {Kacprzak}, {Kent}, {Khullar}, {Klein}, {Kovacs}, {Koziol},
  {Krause}, {Kremin}, {Kron}, {Kuehn}, {Kuhlmann}, {Kuropatkin}, {Lahav},
  {Lasker}, {Li}, {Li}, {Liddle}, {Lima}, {Lin}, {L{\'o}pez-Reyes}, {MacCrann},
  {Maia}, {Maloney}, {Manera}, {March}, {Marriner}, {Marshall}, {Martini},
  {McClintock}, {McKay}, {McMahon}, {Melchior}, {Menanteau}, {Miller},
  {Miquel}, {Mohr}, {Morganson}, {Mould}, {Neilsen}, {Nichol}, {Nogueira},
  {Nord}, {Nugent}, {Nunes}, {Ogando}, {Old}, {Pace}, {Palmese},
  {Paz-Chinch{\'o}n}, {Peiris}, {Percival}, {Petravick}, {Plazas}, {Poh},
  {Pond}, {Porredon}, {Pujol}, {Refregier}, {Reil}, {Ricker}, {Rollins},
  {Romer}, {Roodman}, {Rooney}, {Ross}, {Rykoff}, {Sako}, {Sanchez}, {Sanchez},
  {Santiago}, {Saro}, {Scarpine}, {Scolnic}, {Serrano}, {Sevilla-Noarbe},
  {Sheldon}, {Shipp}, {Silveira}, {Smith}, {Smith}, {Smith}, {Soares-Santos},
  {Sobreira}, {Song}, {Stebbins}, {Suchyta}, {Sullivan}, {Swanson}, {Tarle},
  {Thaler}, {Thomas}, {Thomas}, {Troxel}, {Tucker}, {Vikram}, {Vivas},
  {Walker}, {Wechsler}, {Weller}, {Wester}, {Wolf}, {Wu}, {Yanny}, {Zenteno},
  {Zhang}, {Zuntz}, {DES Collaboration}, {Juneau}, {Fitzpatrick}, {Nikutta},
  {Nidever}, {Olsen}, {Scott}, \& {NOAO Data Lab}}]{DESDR1}
{Abbott}, T.~M.~C., {Abdalla}, F.~B., {Allam}, S., {et~al.} 2018, \apjs, 239,
  18

\bibitem[{{Abolmasov} {et~al.}(2008){Abolmasov}, {Fabrika}, {Sholukhova}, \&
  {Kotani}}]{Abolmasov2008}
{Abolmasov}, P., {Fabrika}, S., {Sholukhova}, O., \& {Kotani}, T. 2008, arXiv
  e-prints, arXiv:0809.0409

\bibitem[{{Aird} {et~al.}(2010){Aird}, {Nandra}, {Laird}, {Georgakakis},
  {Ashby}, {Barmby}, {Coil}, {Huang}, {Koekemoer}, {Steidel}, \&
  {Willmer}}]{Aird2010}
{Aird}, J., {Nandra}, K., {Laird}, E.~S., {et~al.} 2010, \mnras, 401, 2531

\bibitem[{{Anastasopoulou} {et~al.}(2019){Anastasopoulou}, {Zezas}, {Gkiokas},
  \& {Kovlakas}}]{Anastasopoulo2019}
{Anastasopoulou}, K., {Zezas}, A., {Gkiokas}, V., \& {Kovlakas}, K. 2019,
  \mnras, 483, 711

\bibitem[{{Bachetti} {et~al.}(2014){Bachetti}, {Harrison}, {Walton},
  {Grefenstette}, {Chakrabarty}, {F{\"u}rst}, {Barret}, {Beloborodov}, {Boggs},
  {Christensen}, {Craig}, {Fabian}, {Hailey}, {Hornschemeier}, {Kaspi},
  {Kulkarni}, {Maccarone}, {Miller}, {Rana}, {Stern}, {Tendulkar}, {Tomsick},
  {Webb}, \& {Zhang}}]{Bachetti2014}
{Bachetti}, M., {Harrison}, F.~A., {Walton}, D.~J., {et~al.} 2014, \nat, 514,
  202

\bibitem[{{Bachetti} {et~al.}(2013){Bachetti}, {Rana}, {Walton}, {Barret},
  {Harrison}, {Boggs}, {Christensen}, {Craig}, {Fabian}, {F{\"u}rst},
  {Grefenstette}, {Hailey}, {Hornschemeier}, {Madsen}, {Miller}, {Ptak},
  {Stern}, {Webb}, \& {Zhang}}]{Bachetti2013}
{Bachetti}, M., {Rana}, V., {Walton}, D.~J., {et~al.} 2013, \apj, 778, 163

\bibitem[{{Baldassare} {et~al.}(2017){Baldassare}, {Reines}, {Gallo}, \&
  {Greene}}]{Baldassare2017}
{Baldassare}, V.~F., {Reines}, A.~E., {Gallo}, E., \& {Greene}, J.~E. 2017,
  \apj, 836, 20

\bibitem[{{Barrows} {et~al.}(2019){Barrows}, {Mezcua}, \&
  {Comerford}}]{Barrows2019}
{Barrows}, R.~S., {Mezcua}, M., \& {Comerford}, J.~M. 2019, \apj, 882, 181

\bibitem[{{Beck} {et~al.}(2016){Beck}, {Dobos}, {Budav{\'a}ri}, {Szalay}, \&
  {Csabai}}]{Beck2016}
{Beck}, R., {Dobos}, L., {Budav{\'a}ri}, T., {Szalay}, A.~S., \& {Csabai}, I.
  2016, \mnras, 460, 1371

\bibitem[{{Belczynski} {et~al.}(2021){Belczynski}, {Done}, \&
  {Lasota}}]{Belczynski2021}
{Belczynski}, K., {Done}, C., \& {Lasota}, J.~P. 2021, arXiv e-prints,
  arXiv:2111.09401

\bibitem[{{Belfiore} {et~al.}(2020){Belfiore}, {Esposito}, {Pintore}, {Novara},
  {Salvaterra}, {De Luca}, {Tiengo}, {Caraveo}, {F{\"u}rst}, {Israel},
  {Magistrali}, {Marelli}, {Mereghetti}, {Papitto}, {Rodr{\'\i}guez Castillo},
  {Salvaggio}, {Stella}, {Walton}, {Wolter}, \& {Zampieri}}]{Belfiore2020}
{Belfiore}, A., {Esposito}, P., {Pintore}, F., {et~al.} 2020, Nature Astronomy,
  4, 147

\bibitem[{{Bell} {et~al.}(2003){Bell}, {McIntosh}, {Katz}, \&
  {Weinberg}}]{Bell2003}
{Bell}, E.~F., {McIntosh}, D.~H., {Katz}, N., \& {Weinberg}, M.~D. 2003, \apjs,
  149, 289

\bibitem[{{Berghea} {et~al.}(2010){Berghea}, {Dudik}, {Weaver}, \&
  {Kallman}}]{Berghea2010}
{Berghea}, C.~T., {Dudik}, R.~P., {Weaver}, K.~A., \& {Kallman}, T.~R. 2010,
  \apj, 708, 364

\bibitem[{{Berghea} {et~al.}(2020){Berghea}, {Johnson}, {Secrest}, {Dudik},
  {Hennessy}, \& {El-khatib}}]{Berghea2020}
{Berghea}, C.~T., {Johnson}, M.~C., {Secrest}, N.~J., {et~al.} 2020, \apj, 896,
  117

\bibitem[{{Bernadich} {et~al.}(2022){Bernadich}, {Schwope}, {Kovlakas},
  {Zezas}, \& {Traulsen}}]{Bernadich2021}
{Bernadich}, M.~C.~i., {Schwope}, A.~D., {Kovlakas}, K., {Zezas}, A., \&
  {Traulsen}, I. 2022, \aap, 659, A188

\bibitem[{{Binggeli} {et~al.}(1988){Binggeli}, {Sandage}, \&
  {Tammann}}]{Binggeli1988}
{Binggeli}, B., {Sandage}, A., \& {Tammann}, G.~A. 1988, \araa, 26, 509

\bibitem[{{Boch} \& {Fernique}(2014)}]{Boch2014}
{Boch}, T. \& {Fernique}, P. 2014, in Astronomical Society of the Pacific
  Conference Series, Vol. 485, Astronomical Data Analysis Software and Systems
  XXIII, ed. N.~{Manset} \& P.~{Forshay}, 277

\bibitem[{{Bolton} {et~al.}(2012){Bolton}, {Schlegel}, {Aubourg}, {Bailey},
  {Bhardwaj}, {Brownstein}, {Burles}, {Chen}, {Dawson}, {Eisenstein}, {Gunn},
  {Knapp}, {Loomis}, {Lupton}, {Maraston}, {Muna}, {Myers}, {Olmstead},
  {Padmanabhan}, {P{\^a}ris}, {Percival}, {Petitjean}, {Rockosi}, {Ross},
  {Schneider}, {Shu}, {Strauss}, {Thomas}, {Tremonti}, {Wake}, {Weaver}, \&
  {Wood-Vasey}}]{Bolton2012}
{Bolton}, A.~S., {Schlegel}, D.~J., {Aubourg}, {\'E}., {et~al.} 2012, \aj, 144,
  144

\bibitem[{{Bondi}(1952)}]{Bondi1952}
{Bondi}, H. 1952, \mnras, 112, 195

\bibitem[{{Bondi} \& {Hoyle}(1944)}]{BondiHoyle1944}
{Bondi}, H. \& {Hoyle}, F. 1944, \mnras, 104, 273

\bibitem[{{Bonnarel} {et~al.}(2000){Bonnarel}, {Fernique}, {Bienaym{\'e}},
  {Egret}, {Genova}, {Louys}, {Ochsenbein}, {Wenger}, \&
  {Bartlett}}]{Bonnarel2000}
{Bonnarel}, F., {Fernique}, P., {Bienaym{\'e}}, O., {et~al.} 2000, \aaps, 143,
  33

\bibitem[{{Bouwens} {et~al.}(2011){Bouwens}, {Illingworth}, {Oesch},
  {Labb{\'e}}, {Trenti}, {van Dokkum}, {Franx}, {Stiavelli}, {Carollo},
  {Magee}, \& {Gonzalez}}]{Bouwens2011}
{Bouwens}, R.~J., {Illingworth}, G.~D., {Oesch}, P.~A., {et~al.} 2011, \apj,
  737, 90

\bibitem[{{Carpano} {et~al.}(2018){Carpano}, {Haberl}, {Maitra}, \&
  {Vasilopoulos}}]{Carpano2018}
{Carpano}, S., {Haberl}, F., {Maitra}, C., \& {Vasilopoulos}, G. 2018, \mnras,
  476, L45

\bibitem[{{Chambers} {et~al.}(2016){Chambers}, {Magnier}, {Metcalfe},
  {Flewelling}, {Huber}, {Waters}, {Denneau}, {Draper}, {Farrow}, {Finkbeiner},
  {Holmberg}, {Koppenhoefer}, {Price}, {Rest}, {Saglia}, {Schlafly}, {Smartt},
  {Sweeney}, {Wainscoat}, {Burgett}, {Chastel}, {Grav}, {Heasley}, {Hodapp},
  {Jedicke}, {Kaiser}, {Kudritzki}, {Luppino}, {Lupton}, {Monet}, {Morgan},
  {Onaka}, {Shiao}, {Stubbs}, {Tonry}, {White}, {Ba{\~n}ados}, {Bell},
  {Bender}, {Bernard}, {Boegner}, {Boffi}, {Botticella}, {Calamida},
  {Casertano}, {Chen}, {Chen}, {Cole}, {Deacon}, {Frenk}, {Fitzsimmons},
  {Gezari}, {Gibbs}, {Goessl}, {Goggia}, {Gourgue}, {Goldman}, {Grant},
  {Grebel}, {Hambly}, {Hasinger}, {Heavens}, {Heckman}, {Henderson}, {Henning},
  {Holman}, {Hopp}, {Ip}, {Isani}, {Jackson}, {Keyes}, {Koekemoer}, {Kotak},
  {Le}, {Liska}, {Long}, {Lucey}, {Liu}, {Martin}, {Masci}, {McLean}, {Mindel},
  {Misra}, {Morganson}, {Murphy}, {Obaika}, {Narayan}, {Nieto-Santisteban},
  {Norberg}, {Peacock}, {Pier}, {Postman}, {Primak}, {Rae}, {Rai}, {Riess},
  {Riffeser}, {Rix}, {R{\"o}ser}, {Russel}, {Rutz}, {Schilbach}, {Schultz},
  {Scolnic}, {Strolger}, {Szalay}, {Seitz}, {Small}, {Smith}, {Soderblom},
  {Taylor}, {Thomson}, {Taylor}, {Thakar}, {Thiel}, {Thilker}, {Unger},
  {Urata}, {Valenti}, {Wagner}, {Walder}, {Walter}, {Watters}, {Werner},
  {Wood-Vasey}, \& {Wyse}}]{Chambers2016}
{Chambers}, K.~C., {Magnier}, E.~A., {Metcalfe}, N., {et~al.} 2016, arXiv
  e-prints, arXiv:1612.05560

\bibitem[{{Cluver} {et~al.}(2017){Cluver}, {Jarrett}, {Dale}, {Smith},
  {August}, \& {Brown}}]{Cluver2017}
{Cluver}, M.~E., {Jarrett}, T.~H., {Dale}, D.~A., {et~al.} 2017, \apj, 850, 68

\bibitem[{{Colbert} \& {Mushotzky}(1999)}]{Colbert1999}
{Colbert}, E. J.~M. \& {Mushotzky}, R.~F. 1999, \apj, 519, 89

\bibitem[{{Cutri} {et~al.}(2012)}]{Cutri2012}
{Cutri}, R.~M. {et~al.} 2012, VizieR Online Data Catalog, II/311

\bibitem[{{D{\'a}lya} {et~al.}(2022){D{\'a}lya}, {D{\'\i}az}, {Bouchet},
  {Frei}, {Jasche}, {Lavaux}, {Macas}, {Mukherjee}, {P{\'a}lfi}, {de Souza},
  {Wandelt}, {Bilicki}, \& {Raffai}}]{Glade2021}
{D{\'a}lya}, G., {D{\'\i}az}, R., {Bouchet}, F.~R., {et~al.} 2022, \mnras, 514,
  1403

\bibitem[{{D{\'a}lya} {et~al.}(2018){D{\'a}lya}, {Galg{\'o}czi}, {Dobos},
  {Frei}, {Heng}, {Macas}, {Messenger}, {Raffai}, \& {de Souza}}]{Glade2018}
{D{\'a}lya}, G., {Galg{\'o}czi}, G., {Dobos}, L., {et~al.} 2018, \mnras, 479,
  2374

\bibitem[{{de Vaucouleurs} {et~al.}(1991){de Vaucouleurs}, {de Vaucouleurs},
  {Corwin}, {Buta}, {Paturel}, \& {Fouque}}]{RC31991}
{de Vaucouleurs}, G., {de Vaucouleurs}, A., {Corwin}, Herold~G., J., {et~al.}
  1991, {Third Reference Catalogue of Bright Galaxies}

\bibitem[{{Earnshaw} {et~al.}(2019){Earnshaw}, {Roberts}, {Middleton},
  {Walton}, \& {Mateos}}]{Earnshaw2019}
{Earnshaw}, H.~P., {Roberts}, T.~P., {Middleton}, M.~J., {Walton}, D.~J., \&
  {Mateos}, S. 2019, \mnras, 483, 5554

\bibitem[{{Erkut} {et~al.}(2019){Erkut}, {Ek{\c{s}}i}, \& {Alpar}}]{Erkut2019}
{Erkut}, M.~H., {Ek{\c{s}}i}, K.~Y., \& {Alpar}, M.~A. 2019, \apj, 873, 105

\bibitem[{{Evans} {et~al.}(2019){Evans}, {Allen}, {Anderson}, {Budynkiewicz},
  {Burke}, {Chen}, {Civano}, {D'Abrusco}, {Doe}, {Evans}, {Fabbiano}, {Gibbs},
  {Glotfelty}, {Graessle}, {Grier}, {Hain}, {Hall}, {Harbo}, {Houck}, {Lauer},
  {Laurino}, {Lee}, {Martinez-Galarza}, {McCollough}, {McDowell}, {Miller},
  {McLaughlin}, {Morgan}, {Mossman}, {Nguyen}, {Nichols}, {Nowak}, {Paxson},
  {Plummer}, {Primini}, {Rots}, {Siemiginowska}, {Sundheim}, {Tibbetts}, {Van
  Stone}, \& {Zografou}}]{EvansCSC2019}
{Evans}, I.~N., {Allen}, C., {Anderson}, C.~S., {et~al.} 2019, in AAS/High
  Energy Astrophysics Division, Vol.~17, AAS/High Energy Astrophysics Division,
  114.01

\bibitem[{{Evans} {et~al.}(2010){Evans}, {Primini}, {Glotfelty}, {Anderson},
  {Bonaventura}, {Chen}, {Davis}, {Doe}, {Evans}, {Fabbiano}, {Galle}, {Gibbs},
  {Grier}, {Hain}, {Hall}, {Harbo}, {He}, {Houck}, {Karovska}, {Kashyap},
  {Lauer}, {McCollough}, {McDowell}, {Miller}, {Mitschang}, {Morgan},
  {Mossman}, {Nichols}, {Nowak}, {Plummer}, {Refsdal}, {Rots}, {Siemiginowska},
  {Sundheim}, {Tibbetts}, {Van Stone}, {Winkelman}, \&
  {Zografou}}]{EvansCSC2010}
{Evans}, I.~N., {Primini}, F.~A., {Glotfelty}, K.~J., {et~al.} 2010, \apjs,
  189, 37

\bibitem[{{Evans} {et~al.}(2020){Evans}, {Page}, {Osborne}, {Beardmore},
  {Willingale}, {Burrows}, {Kennea}, {Perri}, {Capalbi}, {Tagliaferri}, \&
  {Cenko}}]{Evans2020}
{Evans}, P.~A., {Page}, K.~L., {Osborne}, J.~P., {et~al.} 2020, \apjs, 247, 54

\bibitem[{{Fabbiano}(2006)}]{Fabbiano2006}
{Fabbiano}, G. 2006, \araa, 44, 323

\bibitem[{{Fabian}(2012)}]{Fabian2012}
{Fabian}, A.~C. 2012, \araa, 50, 455

\bibitem[{{Farrell} {et~al.}(2012){Farrell}, {Servillat}, {Pforr}, {Maccarone},
  {Knigge}, {Godet}, {Maraston}, {Webb}, {Barret}, {Gosling}, {Belmont}, \&
  {Wiersema}}]{Farrell2012}
{Farrell}, S.~A., {Servillat}, M., {Pforr}, J., {et~al.} 2012, \apjl, 747, L13

\bibitem[{{Farrell} {et~al.}(2009){Farrell}, {Webb}, \& {Barret et
  al.}}]{Farrell2009}
{Farrell}, S.~A., {Webb}, N.~A., \& {Barret et al.}, D. 2009, Nature, 460, 73

\bibitem[{{Fishbach} \& {Kalogera}(2022)}]{Fishbach2022}
{Fishbach}, M. \& {Kalogera}, V. 2022, \apjl, 929, L26

\bibitem[{{F{\"u}rst} {et~al.}(2016){F{\"u}rst}, {Walton}, {Harrison}, {Stern},
  {Barret}, {Brightman}, {Fabian}, {Grefenstette}, {Madsen}, {Middleton},
  {Miller}, {Pottschmidt}, {Ptak}, {Rana}, \& {Webb}}]{Fuerst2016}
{F{\"u}rst}, F., {Walton}, D.~J., {Harrison}, F.~A., {et~al.} 2016, \apjl, 831,
  L14

\bibitem[{{Gaia Collaboration} {et~al.}(2022){Gaia Collaboration},
  {Bailer-Jones}, {Teyssier}, {Delchambre}, {Ducourant}, {Garabato},
  {Hatzidimitriou}, {Klioner}, {Rimoldini}, {Bellas-Velidis}, {Carballo},
  {Carnerero}, {Diener}, {Fouesneau}, {Galluccio}, {Gavras}, {Krone-Martins},
  {Raiteri}, {Teixeira}, {Brown}, {Vallenari}, {Prusti}, {de Bruijne},
  {Arenou}, {Babusiaux}, {Biermann}, {Creevey}, {Evans}, {Eyer}, {Guerra},
  {Hutton}, {Jordi}, {Lammers}, {Lindegren}, {Luri}, {Mignard}, {Panem},
  {Pourbaix}, {Randich}, {Sartoretti}, {Soubiran}, {Tanga}, {Walton},
  {Bastian}, {Drimmel}, {Jansen}, {Katz}, {Lattanzi}, {van Leeuwen}, {Bakker},
  {Cacciari}, {Casta{\~n}eda}, {De Angeli}, {Fabricius}, {Fr{\'e}mat},
  {Guerrier}, {Heiter}, {Masana}, {Messineo}, {Mowlavi}, {Nicolas},
  {Nienartowicz}, {Pailler}, {Panuzzo}, {Riclet}, {Roux}, {Seabroke}, {Sordo},
  {Th{\'e}venin}, {Gracia-Abril}, {Portell}, {Altmann}, {Andrae}, {Audard},
  {Benson}, {Berthier}, {Blomme}, {Burgess}, {Busonero}, {Busso},
  {C{\'a}novas}, {Carry}, {Cellino}, {Cheek}, {Clementini}, {Damerdji},
  {Davidson}, {de Teodoro}, {Nu{\~n}ez Campos}, {Dell'Oro}, {Esquej},
  {Fern{\'a}ndez-Hern{\'a}ndez}, {Fraile}, {Garc{\'\i}a-Lario}, {Gosset},
  {Haigron}, {Halbwachs}, {Hambly}, {Harrison}, {Hern{\'a}ndez}, {Hestroffer},
  {Hodgkin}, {Holl}, {Jan{\ss}en}, {Jevardat de Fombelle}, {Jordan},
  {Lanzafame}, {L{\"o}ffler}, {Marchal}, {Marrese}, {Moitinho}, {Muinonen},
  {Osborne}, {Pancino}, {Pauwels}, {Recio-Blanco}, {Reyl{\'e}}, {Riello},
  {Roegiers}, {Rybizki}, {Sarro}, {Siopis}, {Smith}, {Sozzetti}, {Utrilla},
  {van Leeuwen}, {Abbas}, {{\'A}brah{\'a}m}, {Abreu Aramburu}, {Aerts},
  {Aguado}, {Ajaj}, {Aldea-Montero}, {Altavilla}, {{\'A}lvarez}, {Alves},
  {Anderson}, {Anglada Varela}, {Antoja}, {Baines}, {Baker},
  {Balaguer-N{\'u}{\~n}ez}, {Balbinot}, {Balog}, {Barache}, {Barbato},
  {Barros}, {Barstow}, {Bartolom{\'e}}, {Bassilana}, {Bauchet}, {Becciani},
  {Bellazzini}, {Berihuete}, {Bernet}, {Bertone}, {Bianchi}, {Binnenfeld},
  {Blanco-Cuaresma}, {Boch}, {Bombrun}, {Bossini}, {Bouquillon}, {Bragaglia},
  {Bramante}, {Breedt}, {Bressan}, {Brouillet}, {Brugaletta}, {Bucciarelli},
  {Burlacu}, {Butkevich}, {Buzzi}, {Caffau}, {Cancelliere}, {Cantat-Gaudin},
  {Carlucci}, {Carrasco}, {Casamiquela}, {Castellani}, {Castro-Ginard},
  {Chaoul}, {Charlot}, {Chemin}, {Chiaramida}, {Chiavassa}, {Chornay},
  {Comoretto}, {Contursi}, {Cooper}, {Cornez}, {Cowell}, {Crifo}, {Cropper},
  {Crosta}, {Crowley}, {Dafonte}, {Dapergolas}, {David}, {de Laverny}, {De
  Luise}, {De March}, {De Ridder}, {de Souza}, {de Torres}, {del Peloso}, {del
  Pozo}, {Delbo}, {Delgado}, {Delisle}, {Demouchy}, {Dharmawardena}, {Diakite},
  {Distefano}, {Dolding}, {Enke}, {Fabre}, {Fabrizio}, {Faigler}, {Fedorets},
  {Fernique}, {Figueras}, {Fournier}, {Fouron}, {Fragkoudi}, {Gai},
  {Garcia-Gutierrez}, {Garcia-Reinaldos}, {Garc{\'\i}a-Torres}, {Garofalo},
  {Gavel}, {Gerlach}, {Geyer}, {Giacobbe}, {Gilmore}, {Girona}, {Giuffrida},
  {Gomel}, {Gomez}, {Gonz{\'a}lez-N{\'u}{\~n}ez},
  {Gonz{\'a}lez-Santamar{\'\i}a}, {Gonz{\'a}lez-Vidal}, {Granvik}, {Guillout},
  {Guiraud}, {Guti{\'e}rrez-S{\'a}nchez}, {Guy}, {Hauser}, {Haywood}, {Helmer},
  {Helmi}, {Sarmiento}, {Hidalgo}, {H{\l}adczuk}, {Hobbs}, {Holland}, {Huckle},
  {Jardine}, {Jasniewicz}, {Jean-Antoine Piccolo}, {Jim{\'e}nez-Arranz},
  {Juaristi Campillo}, {Julbe}, {Karbevska}, {Kervella}, {Khanna}, {Kontizas},
  {Kordopatis}, {Korn}, {K{\'o}sp{\'a}l}, {Kostrzewa-Rutkowska},
  {Kruszy{\'n}ska}, {Kun}, {Laizeau}, {Lambert}, {Lanza}, {Lasne}, {Le
  Campion}, {Lebreton}, {Lebzelter}, {Leccia}, {Leclerc}, {Lecoeur-Taibi},
  {Liao}, {Licata}, {Lindstr{\o}m}, {Lister}, {Livanou}, {Lobel}, {Lorca},
  {Loup}, {Madrero Pardo}, {Magdaleno Romeo}, {Managau}, {Mann}, {Manteiga},
  {Marchant}, {Marconi}, {Marcos}, {Marcos Santos}, {Mar{\'\i}n Pina},
  {Marinoni}, {Marocco}, {Marshall}, {Polo}, {Mart{\'\i}n-Fleitas}, {Marton},
  {Mary}, {Masip}, {Massari}, {Mastrobuono-Battisti}, {Mazeh}, {McMillan},
  {Messina}, {Michalik}, {Millar}, {Mints}, {Molina}, {Molinaro}, {Moln{\'a}r},
  {Monari}, {Mongui{\'o}}, {Montegriffo}, {Montero}, {Mor}, {Mora},
  {Morbidelli}, {Morel}, {Morris}, {Muraveva}, {Murphy}, {Musella}, {Nagy},
  {Noval}, {Oca{\~n}a}, {Ogden}, {Ordenovic}, {Osinde}, {Pagani}, {Pagano},
  {Palaversa}, {Palicio}, {Pallas-Quintela}, {Panahi}, {Payne-Wardenaar},
  {Pe{\~n}alosa Esteller}, {Penttil{\"a}}, {Pichon}, {Piersimoni}, {Pineau},
  {Plachy}, {Plum}, {Poggio}, {Pr{\v{s}}a}, {Pulone}, {Racero}, {Ragaini},
  {Rainer}, {Ramos}, {Ramos-Lerate}, {Re Fiorentin}, {Regibo}, {Richards},
  {Rios Diaz}, {Ripepi}, {Riva}, {Rix}, {Rixon}, {Robichon}, {Robin}, {Robin},
  {Roelens}, {Rogues}, {Rohrbasser}, {Romero-G{\'o}mez}, {Rowell}, {Royer},
  {Ruz Mieres}, {Rybicki}, {Sadowski}, {S{\'a}ez N{\'u}{\~n}ez}, {Sagrist{\`a}
  Sell{\'e}s}, {Sahlmann}, {Salguero}, {Samaras}, {Sanchez Gimenez}, {Sanna},
  {Santove{\~n}a}, {Sarasso}, {Schultheis}, {Sciacca}, {Segol}, {Segovia},
  {S{\'e}gransan}, {Semeux}, {Shahaf}, {Siddiqui}, {Siebert}, {Siltala},
  {Silvelo}, {Slezak}, {Slezak}, {Smart}, {Snaith}, {Solano}, {Solitro},
  {Souami}, {Souchay}, {Spagna}, {Spina}, {Spoto}, {Steele},
  {Steidelm{\"u}ller}, {Stephenson}, {S{\"u}veges}, {Surdej}, {Szabados},
  {Szegedi-Elek}, {Taris}, {Taylor}, {Tolomei}, {Tonello}, {Torra}, {Torra},
  {Torralba Elipe}, {Trabucchi}, {Tsounis}, {Turon}, {Ulla}, {Unger},
  {Vaillant}, {van Dillen}, {van Reeven}, {Vanel}, {Vecchiato}, {Viala},
  {Vicente}, {Voutsinas}, {Weiler}, {Wevers}, {Wyrzykowski}, {Yoldas}, {Yvard},
  {Zhao}, {Zorec}, {Zucker}, \& {Zwitter}}]{GaiaExtra2022}
{Gaia Collaboration}, {Bailer-Jones}, C.~A.~L., {Teyssier}, D., {et~al.} 2022,
  arXiv e-prints, arXiv:2206.05681

\bibitem[{{Gaia Collaboration} {et~al.}(2021){Gaia Collaboration}, {Brown},
  {Vallenari}, {Prusti}, {de Bruijne}, {Babusiaux}, {Biermann}, {Creevey},
  {Evans}, {Eyer}, {Hutton}, {Jansen}, {Jordi}, {Klioner}, {Lammers},
  {Lindegren}, {Luri}, {Mignard}, {Panem}, {Pourbaix}, {Randich}, {Sartoretti},
  {Soubiran}, {Walton}, {Arenou}, {Bailer-Jones}, {Bastian}, {Cropper},
  {Drimmel}, {Katz}, {Lattanzi}, {van Leeuwen}, {Bakker}, {Cacciari},
  {Casta{\~n}eda}, {De Angeli}, {Ducourant}, {Fabricius}, {Fouesneau},
  {Fr{\'e}mat}, {Guerra}, {Guerrier}, {Guiraud}, {Jean-Antoine Piccolo},
  {Masana}, {Messineo}, {Mowlavi}, {Nicolas}, {Nienartowicz}, {Pailler},
  {Panuzzo}, {Riclet}, {Roux}, {Seabroke}, {Sordo}, {Tanga}, {Th{\'e}venin},
  {Gracia-Abril}, {Portell}, {Teyssier}, {Altmann}, {Andrae}, {Bellas-Velidis},
  {Benson}, {Berthier}, {Blomme}, {Brugaletta}, {Burgess}, {Busso}, {Carry},
  {Cellino}, {Cheek}, {Clementini}, {Damerdji}, {Davidson}, {Delchambre},
  {Dell'Oro}, {Fern{\'a}ndez-Hern{\'a}ndez}, {Galluccio}, {Garc{\'\i}a-Lario},
  {Garcia-Reinaldos}, {Gonz{\'a}lez-N{\'u}{\~n}ez}, {Gosset}, {Haigron},
  {Halbwachs}, {Hambly}, {Harrison}, {Hatzidimitriou}, {Heiter},
  {Hern{\'a}ndez}, {Hestroffer}, {Hodgkin}, {Holl}, {Jan{\ss}en}, {Jevardat de
  Fombelle}, {Jordan}, {Krone-Martins}, {Lanzafame}, {L{\"o}ffler}, {Lorca},
  {Manteiga}, {Marchal}, {Marrese}, {Moitinho}, {Mora}, {Muinonen}, {Osborne},
  {Pancino}, {Pauwels}, {Petit}, {Recio-Blanco}, {Richards}, {Riello},
  {Rimoldini}, {Robin}, {Roegiers}, {Rybizki}, {Sarro}, {Siopis}, {Smith},
  {Sozzetti}, {Ulla}, {Utrilla}, {van Leeuwen}, {van Reeven}, {Abbas}, {Abreu
  Aramburu}, {Accart}, {Aerts}, {Aguado}, {Ajaj}, {Altavilla}, {{\'A}lvarez},
  {{\'A}lvarez Cid-Fuentes}, {Alves}, {Anderson}, {Anglada Varela}, {Antoja},
  {Audard}, {Baines}, {Baker}, {Balaguer-N{\'u}{\~n}ez}, {Balbinot}, {Balog},
  {Barache}, {Barbato}, {Barros}, {Barstow}, {Bartolom{\'e}}, {Bassilana},
  {Bauchet}, {Baudesson-Stella}, {Becciani}, {Bellazzini}, {Bernet}, {Bertone},
  {Bianchi}, {Blanco-Cuaresma}, {Boch}, {Bombrun}, {Bossini}, {Bouquillon},
  {Bragaglia}, {Bramante}, {Breedt}, {Bressan}, {Brouillet}, {Bucciarelli},
  {Burlacu}, {Busonero}, {Butkevich}, {Buzzi}, {Caffau}, {Cancelliere},
  {C{\'a}novas}, {Cantat-Gaudin}, {Carballo}, {Carlucci}, {Carnerero},
  {Carrasco}, {Casamiquela}, {Castellani}, {Castro-Ginard}, {Castro Sampol},
  {Chaoul}, {Charlot}, {Chemin}, {Chiavassa}, {Cioni}, {Comoretto}, {Cooper},
  {Cornez}, {Cowell}, {Crifo}, {Crosta}, {Crowley}, {Dafonte}, {Dapergolas},
  {David}, {David}, {de Laverny}, {De Luise}, {De March}, {De Ridder}, {de
  Souza}, {de Teodoro}, {de Torres}, {del Peloso}, {del Pozo}, {Delbo},
  {Delgado}, {Delgado}, {Delisle}, {Di Matteo}, {Diakite}, {Diener},
  {Distefano}, {Dolding}, {Eappachen}, {Edvardsson}, {Enke}, {Esquej}, {Fabre},
  {Fabrizio}, {Faigler}, {Fedorets}, {Fernique}, {Fienga}, {Figueras},
  {Fouron}, {Fragkoudi}, {Fraile}, {Franke}, {Gai}, {Garabato},
  {Garcia-Gutierrez}, {Garc{\'\i}a-Torres}, {Garofalo}, {Gavras}, {Gerlach},
  {Geyer}, {Giacobbe}, {Gilmore}, {Girona}, {Giuffrida}, {Gomel}, {Gomez},
  {Gonzalez-Santamaria}, {Gonz{\'a}lez-Vidal}, {Granvik},
  {Guti{\'e}rrez-S{\'a}nchez}, {Guy}, {Hauser}, {Haywood}, {Helmi}, {Hidalgo},
  {Hilger}, {H{\l}adczuk}, {Hobbs}, {Holland}, {Huckle}, {Jasniewicz},
  {Jonker}, {Juaristi Campillo}, {Julbe}, {Karbevska}, {Kervella}, {Khanna},
  {Kochoska}, {Kontizas}, {Kordopatis}, {Korn}, {Kostrzewa-Rutkowska},
  {Kruszy{\'n}ska}, {Lambert}, {Lanza}, {Lasne}, {Le Campion}, {Le Fustec},
  {Lebreton}, {Lebzelter}, {Leccia}, {Leclerc}, {Lecoeur-Taibi}, {Liao},
  {Licata}, {Lindstr{\o}m}, {Lister}, {Livanou}, {Lobel}, {Madrero Pardo},
  {Managau}, {Mann}, {Marchant}, {Marconi}, {Marcos Santos}, {Marinoni},
  {Marocco}, {Marshall}, {Martin Polo}, {Mart{\'\i}n-Fleitas}, {Masip},
  {Massari}, {Mastrobuono-Battisti}, {Mazeh}, {McMillan}, {Messina},
  {Michalik}, {Millar}, {Mints}, {Molina}, {Molinaro}, {Moln{\'a}r},
  {Montegriffo}, {Mor}, {Morbidelli}, {Morel}, {Morris}, {Mulone}, {Munoz},
  {Muraveva}, {Murphy}, {Musella}, {Noval}, {Ord{\'e}novic}, {Orr{\`u}},
  {Osinde}, {Pagani}, {Pagano}, {Palaversa}, {Palicio}, {Panahi}, {Pawlak},
  {Pe{\~n}alosa Esteller}, {Penttil{\"a}}, {Piersimoni}, {Pineau}, {Plachy},
  {Plum}, {Poggio}, {Poretti}, {Poujoulet}, {Pr{\v{s}}a}, {Pulone}, {Racero},
  {Ragaini}, {Rainer}, {Raiteri}, {Rambaux}, {Ramos}, {Ramos-Lerate}, {Re
  Fiorentin}, {Regibo}, {Reyl{\'e}}, {Ripepi}, {Riva}, {Rixon}, {Robichon},
  {Robin}, {Roelens}, {Rohrbasser}, {Romero-G{\'o}mez}, {Rowell}, {Royer},
  {Rybicki}, {Sadowski}, {Sagrist{\`a} Sell{\'e}s}, {Sahlmann}, {Salgado},
  {Salguero}, {Samaras}, {Sanchez Gimenez}, {Sanna}, {Santove{\~n}a},
  {Sarasso}, {Schultheis}, {Sciacca}, {Segol}, {Segovia}, {S{\'e}gransan},
  {Semeux}, {Shahaf}, {Siddiqui}, {Siebert}, {Siltala}, {Slezak}, {Smart},
  {Solano}, {Solitro}, {Souami}, {Souchay}, {Spagna}, {Spoto}, {Steele},
  {Steidelm{\"u}ller}, {Stephenson}, {S{\"u}veges}, {Szabados}, {Szegedi-Elek},
  {Taris}, {Tauran}, {Taylor}, {Teixeira}, {Thuillot}, {Tonello}, {Torra},
  {Torra}, {Turon}, {Unger}, {Vaillant}, {van Dillen}, {Vanel}, {Vecchiato},
  {Viala}, {Vicente}, {Voutsinas}, {Weiler}, {Wevers}, {Wyrzykowski}, {Yoldas},
  {Yvard}, {Zhao}, {Zorec}, {Zucker}, {Zurbach}, \& {Zwitter}}]{gaiaedr3}
{Gaia Collaboration}, {Brown}, A.~G.~A., {Vallenari}, A., {et~al.} 2021, \aap,
  649, A1

\bibitem[{{Galliano} {et~al.}(2018){Galliano}, {Galametz}, \&
  {Jones}}]{Galliano2018}
{Galliano}, F., {Galametz}, M., \& {Jones}, A.~P. 2018, \araa, 56, 673

\bibitem[{{Georgakakis} {et~al.}(2008){Georgakakis}, {Nandra}, {Laird}, {Aird},
  \& {Trichas}}]{Georgakakis2008}
{Georgakakis}, A., {Nandra}, K., {Laird}, E.~S., {Aird}, J., \& {Trichas}, M.
  2008, \mnras, 388, 1205

\bibitem[{{Gilfanov} {et~al.}(2004){Gilfanov}, {Grimm}, \&
  {Sunyaev}}]{Gilfanov2004}
{Gilfanov}, M., {Grimm}, H.~J., \& {Sunyaev}, R. 2004, \mnras, 347, L57

\bibitem[{{Gladstone} {et~al.}(2009){Gladstone}, {Roberts}, \&
  {Done}}]{Gladstone2009}
{Gladstone}, J.~C., {Roberts}, T.~P., \& {Done}, C. 2009, \mnras, 397, 1836

\bibitem[{{Goddard} \& {Shamir}(2020)}]{PS1morph}
{Goddard}, H. \& {Shamir}, L. 2020, \apjs, 251, 28

\bibitem[{{Godet} {et~al.}(2009){Godet}, {Barret}, {Webb}, {Farrell}, \&
  {Gehrels}}]{Godet2009}
{Godet}, O., {Barret}, D., {Webb}, N.~A., {Farrell}, S.~A., \& {Gehrels}, N.
  2009, \apjl, 705, L109

\bibitem[{{Godet} {et~al.}(2014){Godet}, {Lombardi}, \& {Antonini et
  al.}}]{godet2014}
{Godet}, O., {Lombardi}, J.~C., \& {Antonini et al.}, F. 2014, APJ, 793, 105

\bibitem[{{Godet} {et~al.}(2012){Godet}, {Plazolles}, {Kawaguchi}, {Lasota},
  {Barret}, {Farrell}, {Braito}, {Servillat}, {Webb}, \& {Gehrels}}]{Godet2012}
{Godet}, O., {Plazolles}, B., {Kawaguchi}, T., {et~al.} 2012, \apj, 752, 34

\bibitem[{{Greene} \& {Ho}(2005)}]{Greene2005}
{Greene}, J.~E. \& {Ho}, L.~C. 2005, \apj, 630, 122

\bibitem[{{Greene} {et~al.}(2020){Greene}, {Strader}, \& {Ho}}]{Greene2020}
{Greene}, J.~E., {Strader}, J., \& {Ho}, L.~C. 2020, \araa, 58, 257

\bibitem[{{Grimm} {et~al.}(2002){Grimm}, {Gilfanov}, \& {Sunyaev}}]{Grimm2002}
{Grimm}, H.~J., {Gilfanov}, M., \& {Sunyaev}, R. 2002, \aap, 391, 923

\bibitem[{{G{\'u}rpide} {et~al.}(2021){G{\'u}rpide}, {Godet}, {Koliopanos},
  {Webb}, \& {Olive}}]{Gurpide2021}
{G{\'u}rpide}, A., {Godet}, O., {Koliopanos}, F., {Webb}, N., \& {Olive}, J.~F.
  2021, \aap, 649, A104

\bibitem[{{G{\'u}rpide} {et~al.}(2022){G{\'u}rpide}, {Parra}, {Godet},
  {Contini}, \& {Olive}}]{Gurpide2022}
{G{\'u}rpide}, A., {Parra}, M., {Godet}, O., {Contini}, T., \& {Olive}, J.~F.
  2022, \aap, 666, A100

\bibitem[{{Haiman}(2013)}]{Haiman2013}
{Haiman}, Z. 2013, in Astrophysics and Space Science Library, Vol. 396, The
  First Galaxies, ed. T.~{Wiklind}, B.~{Mobasher}, \& V.~{Bromm}, 293

\bibitem[{{Humphrey} \& {Buote}(2008)}]{Humphrey2008}
{Humphrey}, P.~J. \& {Buote}, D.~A. 2008, \apj, 689, 983

\bibitem[{{Iben} \& {Livio}(1993)}]{IbenLivio1993}
{Iben}, Icko, J. \& {Livio}, M. 1993, \pasp, 105, 1373

\bibitem[{{Inayoshi} {et~al.}(2016){Inayoshi}, {Haiman}, \&
  {Ostriker}}]{Inayoshi2016}
{Inayoshi}, K., {Haiman}, Z., \& {Ostriker}, J.~P. 2016, \mnras, 459, 3738

\bibitem[{{Inoue} {et~al.}(2021){Inoue}, {Yabe}, \& {Ueda}}]{Inoue2021}
{Inoue}, Y., {Yabe}, K., \& {Ueda}, Y. 2021, \pasj, 73, 1315

\bibitem[{{Israel} {et~al.}(2017){Israel}, {Belfiore}, {Stella}, {Esposito},
  {Casella}, {De Luca}, {Marelli}, {Papitto}, {Perri}, {Puccetti}, {Castillo},
  {Salvetti}, {Tiengo}, {Zampieri}, {D'Agostino}, {Greiner}, {Haberl},
  {Novara}, {Salvaterra}, {Turolla}, {Watson}, {Wilms}, \&
  {Wolter}}]{Israel2017}
{Israel}, G.~L., {Belfiore}, A., {Stella}, L., {et~al.} 2017, Science, 355, 817

\bibitem[{{Jester} {et~al.}(2005){Jester}, {Schneider}, {Richards}, {Green},
  {Schmidt}, {Hall}, {Strauss}, {Vanden Berk}, {Stoughton}, {Gunn},
  {Brinkmann}, {Kent}, {Smith}, {Tucker}, \& {Yanny}}]{Jester2005}
{Jester}, S., {Schneider}, D.~P., {Richards}, G.~T., {et~al.} 2005, \aj, 130,
  873

\bibitem[{{Jonker} {et~al.}(2012){Jonker}, {Heida}, {Torres}, {Miller-Jones},
  {Fabian}, {Ratti}, {Miniutti}, {Walton}, \& {Roberts}}]{Jonker2012}
{Jonker}, P.~G., {Heida}, M., {Torres}, M.~A.~P., {et~al.} 2012, \apj, 758, 28

\bibitem[{{Kaaret} {et~al.}(2017){Kaaret}, {Feng}, \& {Roberts}}]{Kaaret2017}
{Kaaret}, P., {Feng}, H., \& {Roberts}, T.~P. 2017, \araa, 55, 303

\bibitem[{{Karachentsev} {et~al.}(2004){Karachentsev}, {Karachentseva},
  {Huchtmeier}, \& {Makarov}}]{Karachentsev2004}
{Karachentsev}, I.~D., {Karachentseva}, V.~E., {Huchtmeier}, W.~K., \&
  {Makarov}, D.~I. 2004, \aj, 127, 2031

\bibitem[{{Kataoka} {et~al.}(2003){Kataoka}, {Leahy}, {Edwards}, {Kino},
  {Takahara}, {Serino}, {Kawai}, \& {Martel}}]{Kataoka2003}
{Kataoka}, J., {Leahy}, J.~P., {Edwards}, P.~G., {et~al.} 2003, \aap, 410, 833

\bibitem[{{Kennicutt} \& {Evans}(2012)}]{Kennicutt2012}
{Kennicutt}, R.~C. \& {Evans}, N.~J. 2012, \araa, 50, 531

\bibitem[{{Kim} \& {Fabbiano}(2004)}]{KimFabbiano2004}
{Kim}, D.-W. \& {Fabbiano}, G. 2004, \apj, 611, 846

\bibitem[{{Kim} \& {Fabbiano}(2010)}]{KimFabbiano2010}
{Kim}, D.-W. \& {Fabbiano}, G. 2010, \apj, 721, 1523

\bibitem[{{Kim} {et~al.}(2007){Kim}, {Wilkes}, {Kim}, {Green}, {Barkhouse},
  {Lee}, {Silverman}, \& {Tananbaum}}]{Kim2007}
{Kim}, M., {Wilkes}, B.~J., {Kim}, D.-W., {et~al.} 2007, \apj, 659, 29

\bibitem[{{King}(2009)}]{King2009}
{King}, A.~R. 2009, \mnras, 393, L41

\bibitem[{{King} {et~al.}(2001){King}, {Davies}, {Ward}, {Fabbiano}, \&
  {Elvis}}]{King2001}
{King}, A.~R., {Davies}, M.~B., {Ward}, M.~J., {Fabbiano}, G., \& {Elvis}, M.
  2001, \apjl, 552, L109

\bibitem[{{Koliopanos} {et~al.}(2019){Koliopanos}, {Vasilopoulos}, {Buchner},
  {Maitra}, \& {Haberl}}]{Koliopanos2019}
{Koliopanos}, F., {Vasilopoulos}, G., {Buchner}, J., {Maitra}, C., \& {Haberl},
  F. 2019, \aap, 621, A118

\bibitem[{{Koliopanos} {et~al.}(2017){Koliopanos}, {Vasilopoulos}, {Godet},
  {Bachetti}, {Webb}, \& {Barret}}]{Koliopanos2017}
{Koliopanos}, F., {Vasilopoulos}, G., {Godet}, O., {et~al.} 2017, \aap, 608,
  A47

\bibitem[{{Kovlakas} {et~al.}(2021){Kovlakas}, {Zezas}, {Andrews}, {Basu-Zych},
  {Fragos}, {Hornschemeier}, {Kouroumpatzakis}, {Lehmer}, \& {Ptak}}]{HECATE}
{Kovlakas}, K., {Zezas}, A., {Andrews}, J.~J., {et~al.} 2021, \mnras, 506, 1896

\bibitem[{{Kovlakas} {et~al.}(2020){Kovlakas}, {Zezas}, {Andrews}, {Basu-Zych},
  {Fragos}, {Hornschemeier}, {Lehmer}, \& {Ptak}}]{Kovlakas2020}
{Kovlakas}, K., {Zezas}, A., {Andrews}, J.~J., {et~al.} 2020, \mnras, 498, 4790

\bibitem[{{Kroupa}(1995)}]{Kroupa1995}
{Kroupa}, P. 1995, \apj, 453, 358

\bibitem[{{Krti{\v{c}}ka} {et~al.}(2022){Krti{\v{c}}ka}, {Kub{\'a}t}, \&
  {Krti{\v{c}}kov{\'a}}}]{Krticka2022}
{Krti{\v{c}}ka}, J., {Kub{\'a}t}, J., \& {Krti{\v{c}}kov{\'a}}, I. 2022, \aap,
  659, A117

\bibitem[{{Kuranov} {et~al.}(2020){Kuranov}, {Postnov}, \&
  {Yungelson}}]{Kuranov2020}
{Kuranov}, A.~G., {Postnov}, K.~A., \& {Yungelson}, L.~R. 2020, Astronomy
  Letters, 46, 658

\bibitem[{{Leggett} \& {Hawkins}(1988)}]{Leggett1988}
{Leggett}, S.~K. \& {Hawkins}, M.~R.~S. 1988, \mnras, 234, 1065

\bibitem[{{Lehmer} {et~al.}(2021){Lehmer}, {Eufrasio}, {Basu-Zych}, {Doore},
  {Fragos}, {Garofali}, {Kovlakas}, {Williams}, {Zezas}, \&
  {Santana-Silva}}]{Lehmer2021}
{Lehmer}, B.~D., {Eufrasio}, R.~T., {Basu-Zych}, A., {et~al.} 2021, \apj, 907,
  17

\bibitem[{{Lehmer} {et~al.}(2019){Lehmer}, {Eufrasio}, {Tzanavaris},
  {Basu-Zych}, {Fragos}, {Prestwich}, {Yukita}, {Zezas}, {Hornschemeier}, \&
  {Ptak}}]{Lehmer2019}
{Lehmer}, B.~D., {Eufrasio}, R.~T., {Tzanavaris}, P., {et~al.} 2019, \apjs,
  243, 3

\bibitem[{{Lin} {et~al.}(2017){Lin}, {Guillochon}, {Komossa}, {Ramirez-Ruiz},
  {Irwin}, {Maksym}, {Grupe}, {Godet}, {Webb}, {Barret}, {Zauderer}, {Duc},
  {Carrasco}, \& {Gwyn}}]{Lin2017}
{Lin}, D., {Guillochon}, J., {Komossa}, S., {et~al.} 2017, Nature Astronomy, 1,
  0033

\bibitem[{{Lin} {et~al.}(2018){Lin}, {Strader}, {Carrasco}, {Page},
  {Romanowsky}, {Homan}, {Irwin}, {Remillard}, {Godet}, {Webb}, {Baumgardt},
  {Wijnands}, {Barret}, {Duc}, {Brodie}, \& {Gwyn}}]{Lin2018}
{Lin}, D., {Strader}, J., {Carrasco}, E.~R., {et~al.} 2018, Nature Astronomy,
  2, 656

\bibitem[{{Lin} {et~al.}(2014){Lin}, {Webb}, \& {Barret}}]{Lin2014}
{Lin}, D., {Webb}, N.~A., \& {Barret}, D. 2014, \apj, 780, 39

\bibitem[{{Liu} \& {Bregman}(2005)}]{LiuBreg2005}
{Liu}, J.-F. \& {Bregman}, J.~N. 2005, \apjs, 157, 59

\bibitem[{{Liu} {et~al.}(2013){Liu}, {Bregman}, {Bai}, {Justham}, \&
  {Crowther}}]{Liu2013}
{Liu}, J.-F., {Bregman}, J.~N., {Bai}, Y., {Justham}, S., \& {Crowther}, P.
  2013, \nat, 503, 500

\bibitem[{{Liu} \& {Mirabel}(2005)}]{LiuMir2005}
{Liu}, Q.~Z. \& {Mirabel}, I.~F. 2005, \aap, 429, 1125

\bibitem[{{Loveday} {et~al.}(1992){Loveday}, {Peterson}, {Efstathiou}, \&
  {Maddox}}]{Loveday1992}
{Loveday}, J., {Peterson}, B.~A., {Efstathiou}, G., \& {Maddox}, S.~J. 1992,
  \apj, 390, 338

\bibitem[{{Lynden-Bell}(1971)}]{LyndenBell1971}
{Lynden-Bell}, D. 1971, \mnras, 155, 95

\bibitem[{{Maccacaro} {et~al.}(1988){Maccacaro}, {Gioia}, {Wolter}, {Zamorani},
  \& {Stocke}}]{Maccacaro1988}
{Maccacaro}, T., {Gioia}, I.~M., {Wolter}, A., {Zamorani}, G., \& {Stocke},
  J.~T. 1988, \apj, 326, 680

\bibitem[{{Marconi} {et~al.}(2004){Marconi}, {Risaliti}, {Gilli}, {Hunt},
  {Maiolino}, \& {Salvati}}]{Marconi2004}
{Marconi}, A., {Risaliti}, G., {Gilli}, R., {et~al.} 2004, \mnras, 351, 169

\bibitem[{{Martel} {et~al.}(1998){Martel}, {Sparks}, {Macchetto}, {Biretta},
  {Baum}, {Golombek}, {McCarthy}, {de Koff}, \& {Miley}}]{Martel1998}
{Martel}, A.~R., {Sparks}, W.~B., {Macchetto}, D., {et~al.} 1998, \apj, 496,
  203

\bibitem[{{Massonneau} {et~al.}(2023){Massonneau}, {Volonteri}, {Dubois}, \&
  {Beckmann}}]{Massonneau2022}
{Massonneau}, W., {Volonteri}, M., {Dubois}, Y., \& {Beckmann}, R.~S. 2023,
  \aap, 670, A180

\bibitem[{{McLean} {et~al.}(2000){McLean}, {Greene}, {Lattanzi}, \&
  {Pirenne}}]{McLean2000}
{McLean}, B.~J., {Greene}, G.~R., {Lattanzi}, M.~G., \& {Pirenne}, B. 2000, in
  Astronomical Society of the Pacific Conference Series, Vol. 216, Astronomical
  Data Analysis Software and Systems IX, ed. N.~{Manset}, C.~{Veillet}, \&
  D.~{Crabtree}, 145

\bibitem[{{Mezcua}(2017)}]{Mezcua2017}
{Mezcua}, M. 2017, International Journal of Modern Physics D, 26, 1730021

\bibitem[{{Middleton} {et~al.}(2015){Middleton}, {Heil}, {Pintore}, {Walton},
  \& {Roberts}}]{Middleton2015}
{Middleton}, M.~J., {Heil}, L., {Pintore}, F., {Walton}, D.~J., \& {Roberts},
  T.~P. 2015, \mnras, 447, 3243

\bibitem[{{Mineo} {et~al.}(2012){Mineo}, {Gilfanov}, \& {Sunyaev}}]{Mineo2012}
{Mineo}, S., {Gilfanov}, M., \& {Sunyaev}, R. 2012, \mnras, 419, 2095

\bibitem[{{Moretti} {et~al.}(2003){Moretti}, {Campana}, {Lazzati}, \&
  {Tagliaferri}}]{Moretti2003}
{Moretti}, A., {Campana}, S., {Lazzati}, D., \& {Tagliaferri}, G. 2003, \apj,
  588, 696

\bibitem[{{Mortlock} {et~al.}(2011){Mortlock}, {Warren}, {Venemans}, {Patel},
  {Hewett}, {McMahon}, {Simpson}, {Theuns}, {Gonz{\'a}les-Solares}, {Adamson},
  {Dye}, {Hambly}, {Hirst}, {Irwin}, {Kuiper}, {Lawrence}, \&
  {R{\"o}ttgering}}]{Mortlock2011}
{Mortlock}, D.~J., {Warren}, S.~J., {Venemans}, B.~P., {et~al.} 2011, \nat,
  474, 616

\bibitem[{{Mushotzky} {et~al.}(2000){Mushotzky}, {Cowie}, {Barger}, \&
  {Arnaud}}]{Mushotzky2000}
{Mushotzky}, R.~F., {Cowie}, L.~L., {Barger}, A.~J., \& {Arnaud}, K.~A. 2000,
  \nat, 404, 459

\bibitem[{{Mushtukov} {et~al.}(2017){Mushtukov}, {Suleimanov}, {Tsygankov}, \&
  {Ingram}}]{Mushtukov2017}
{Mushtukov}, A.~A., {Suleimanov}, V.~F., {Tsygankov}, S.~S., \& {Ingram}, A.
  2017, \mnras, 467, 1202

\bibitem[{{Mushtukov} {et~al.}(2015){Mushtukov}, {Suleimanov}, {Tsygankov}, \&
  {Poutanen}}]{Mushtukov2015}
{Mushtukov}, A.~A., {Suleimanov}, V.~F., {Tsygankov}, S.~S., \& {Poutanen}, J.
  2015, \mnras, 454, 2539

\bibitem[{{Narayan} {et~al.}(2017){Narayan}, {Sa{\`I}{\textsection}dowski}, \&
  {Soria}}]{Narayan2017}
{Narayan}, R., {Sa{\`I}{\textsection}dowski}, A., \& {Soria}, R. 2017, \mnras,
  469, 2997

\bibitem[{{Nelemans} {et~al.}(2000){Nelemans}, {Verbunt}, {Yungelson}, \&
  {Portegies Zwart}}]{Nelemans2000}
{Nelemans}, G., {Verbunt}, F., {Yungelson}, L.~R., \& {Portegies Zwart}, S.~F.
  2000, \aap, 360, 1011

\bibitem[{{Ochsenbein} {et~al.}(2000){Ochsenbein}, {Bauer}, \&
  {Marcout}}]{Ochsenbein2000}
{Ochsenbein}, F., {Bauer}, P., \& {Marcout}, J. 2000, \aaps, 143, 23

\bibitem[{{Pacucci} \& {Loeb}(2022)}]{Pacucci2022}
{Pacucci}, F. \& {Loeb}, A. 2022, \mnras, 509, 1885

\bibitem[{{Padovani} {et~al.}(2017){Padovani}, {Alexander}, {Assef}, {De
  Marco}, {Giommi}, {Hickox}, {Richards}, {Smol{\v{c}}i{\'c}},
  {Hatziminaoglou}, {Mainieri}, \& {Salvato}}]{Padovani2017}
{Padovani}, P., {Alexander}, D.~M., {Assef}, R.~J., {et~al.} 2017, \aapr, 25, 2

\bibitem[{{Pakull} \& {Mirioni}(2002)}]{Pakull2002}
{Pakull}, M.~W. \& {Mirioni}, L. 2002, arXiv e-prints, astro

\bibitem[{{Paturel} {et~al.}(2003){Paturel}, {Petit}, {Prugniel}, {Theureau},
  {Rousseau}, {Brouty}, {Dubois}, \& {Cambr{\'e}sy}}]{Paturel2003}
{Paturel}, G., {Petit}, C., {Prugniel}, P., {et~al.} 2003, \aap, 412, 45

\bibitem[{{Pavlovskii} {et~al.}(2017){Pavlovskii}, {Ivanova}, {Belczynski}, \&
  {Van}}]{Pavlovskii2017}
{Pavlovskii}, K., {Ivanova}, N., {Belczynski}, K., \& {Van}, K.~X. 2017,
  \mnras, 465, 2092

\bibitem[{{Poutanen} {et~al.}(2007){Poutanen}, {Lipunova}, {Fabrika},
  {Butkevich}, \& {Abolmasov}}]{Poutanen2007}
{Poutanen}, J., {Lipunova}, G., {Fabrika}, S., {Butkevich}, A.~G., \&
  {Abolmasov}, P. 2007, \mnras, 377, 1187

\bibitem[{{Quintin} {et~al.}(2021){Quintin}, {Webb}, {G{\'u}rpide}, {Bachetti},
  \& {F{\"u}rst}}]{Quintin2021}
{Quintin}, E., {Webb}, N.~A., {G{\'u}rpide}, A., {Bachetti}, M., \&
  {F{\"u}rst}, F. 2021, \mnras, 503, 5485

\bibitem[{{Reines} {et~al.}(2013){Reines}, {Greene}, \& {Geha}}]{Reines2013}
{Reines}, A.~E., {Greene}, J.~E., \& {Geha}, M. 2013, \apj, 775, 116

\bibitem[{{Reines} \& {Volonteri}(2015)}]{Reines2015}
{Reines}, A.~E. \& {Volonteri}, M. 2015, \apj, 813, 82

\bibitem[{{Remillard} \& {McClintock}(2006)}]{Remillard2006}
{Remillard}, R.~A. \& {McClintock}, J.~E. 2006, \araa, 44, 49

\bibitem[{{Repetto} {et~al.}(2017){Repetto}, {Igoshev}, \&
  {Nelemans}}]{Repetto2017}
{Repetto}, S., {Igoshev}, A.~P., \& {Nelemans}, G. 2017, \mnras, 467, 298

\bibitem[{{Salvato} {et~al.}(2018){Salvato}, {Buchner}, {Budav{\'a}ri},
  {Dwelly}, {Merloni}, {Brusa}, {Rau}, {Fotopoulou}, \& {Nandra}}]{Salvato2018}
{Salvato}, M., {Buchner}, J., {Budav{\'a}ri}, T., {et~al.} 2018, \mnras, 473,
  4937

\bibitem[{{Sambruna} {et~al.}(2006){Sambruna}, {Gliozzi}, {Tavecchio},
  {Maraschi}, \& {Foschini}}]{Sambruna2006}
{Sambruna}, R.~M., {Gliozzi}, M., {Tavecchio}, F., {Maraschi}, L., \&
  {Foschini}, L. 2006, \apj, 652, 146

\bibitem[{{Secrest} {et~al.}(2015){Secrest}, {Dudik}, {Dorland}, {Zacharias},
  {Makarov}, {Fey}, {Frouard}, \& {Finch}}]{Secrest2015}
{Secrest}, N.~J., {Dudik}, R.~P., {Dorland}, B.~N., {et~al.} 2015, \apjs, 221,
  12

\bibitem[{{Servillat} {et~al.}(2011){Servillat}, {Farrell}, \& {Lin et
  al.}}]{Servillat2011}
{Servillat}, M., {Farrell}, S.~A., \& {Lin et al.}, D. 2011, APJ, 743, 6

\bibitem[{{Shakura} \& {Sunyaev}(1973)}]{ShakuraSunyaev1973}
{Shakura}, N.~I. \& {Sunyaev}, R.~A. 1973, \aap, 24, 337

\bibitem[{{Shen} {et~al.}(2015){Shen}, {Barniol Duran}, {Nakar}, \&
  {Piran}}]{Shen2015}
{Shen}, R.~F., {Barniol Duran}, R., {Nakar}, E., \& {Piran}, T. 2015, \mnras,
  447, L60

\bibitem[{{Sidoli} \& {Paizis}(2018)}]{Sidoli2018}
{Sidoli}, L. \& {Paizis}, A. 2018, \mnras, 481, 2779

\bibitem[{{Skrutskie} {et~al.}(2006){Skrutskie}, {Cutri}, {Stiening},
  {Weinberg}, {Schneider}, {Carpenter}, {Beichman}, {Capps}, {Chester},
  {Elias}, {Huchra}, {Liebert}, {Lonsdale}, {Monet}, {Price}, {Seitzer},
  {Jarrett}, {Kirkpatrick}, {Gizis}, {Howard}, {Evans}, {Fowler}, {Fullmer},
  {Hurt}, {Light}, {Kopan}, {Marsh}, {McCallon}, {Tam}, {Van Dyk}, \&
  {Wheelock}}]{Skrutskie2006}
{Skrutskie}, M.~F., {Cutri}, R.~M., {Stiening}, R., {et~al.} 2006, \aj, 131,
  1163

\bibitem[{{Soria} {et~al.}(2022){Soria}, {Kolehmainen}, {Graham}, {Swartz},
  {Yukita}, {Motch}, {Jarrett}, {Miller-Jones}, {Plotkin}, {Maccarone},
  {Ferrarese}, {Guest}, \& {Lan{\c{c}}on}}]{Soria2022}
{Soria}, R., {Kolehmainen}, M., {Graham}, A.~W., {et~al.} 2022, \mnras, 512,
  3284

\bibitem[{{Stobie} {et~al.}(1989){Stobie}, {Ishida}, \& {Peacock}}]{Stobie1989}
{Stobie}, R.~S., {Ishida}, K., \& {Peacock}, J.~A. 1989, \mnras, 238, 709

\bibitem[{{Sutton} {et~al.}(2015){Sutton}, {Roberts}, {Gladstone}, \&
  {Walton}}]{Sutton2015}
{Sutton}, A.~D., {Roberts}, T.~P., {Gladstone}, J.~C., \& {Walton}, D.~J. 2015,
  \mnras, 450, 787

\bibitem[{{Sutton} {et~al.}(2013){Sutton}, {Roberts}, \&
  {Middleton}}]{Sutton2013}
{Sutton}, A.~D., {Roberts}, T.~P., \& {Middleton}, M.~J. 2013, \mnras, 435,
  1758

\bibitem[{{Sutton} {et~al.}(2012){Sutton}, {Roberts}, {Walton}, {Gladstone}, \&
  {Scott}}]{Sutton2012}
{Sutton}, A.~D., {Roberts}, T.~P., {Walton}, D.~J., {Gladstone}, J.~C., \&
  {Scott}, A.~E. 2012, \mnras, 423, 1154

\bibitem[{{Swartz} {et~al.}(2004){Swartz}, {Ghosh}, {Tennant}, \&
  {Wu}}]{Swartz2004}
{Swartz}, D.~A., {Ghosh}, K.~K., {Tennant}, A.~F., \& {Wu}, K. 2004, \apjs,
  154, 519

\bibitem[{{Swartz} {et~al.}(2008){Swartz}, {Soria}, \& {Tennant}}]{Swartz2008}
{Swartz}, D.~A., {Soria}, R., \& {Tennant}, A.~F. 2008, \apj, 684, 282

\bibitem[{{Swartz} {et~al.}(2011){Swartz}, {Soria}, {Tennant}, \&
  {Yukita}}]{Swartz2011}
{Swartz}, D.~A., {Soria}, R., {Tennant}, A.~F., \& {Yukita}, M. 2011, \apj,
  741, 49

\bibitem[{{Tarr{\'\i}o} \& {Zarattini}(2020)}]{Tarrio2020}
{Tarr{\'\i}o}, P. \& {Zarattini}, S. 2020, \aap, 642, A102

\bibitem[{{Taylor}(2005)}]{Taylor2005}
{Taylor}, M.~B. 2005, in Astronomical Society of the Pacific Conference Series,
  Vol. 347, Astronomical Data Analysis Software and Systems XIV, ed.
  P.~{Shopbell}, M.~{Britton}, \& R.~{Ebert}, 29

\bibitem[{{Tranin} {et~al.}(2022){Tranin}, {Godet}, {Webb}, \&
  {Primorac}}]{Tranin2022}
{Tranin}, H., {Godet}, O., {Webb}, N., \& {Primorac}, D. 2022, \aap, 657, A138

\bibitem[{{Ueda} {et~al.}(2003){Ueda}, {Akiyama}, {Ohta}, \&
  {Miyaji}}]{Ueda2003}
{Ueda}, Y., {Akiyama}, M., {Ohta}, K., \& {Miyaji}, T. 2003, \apj, 598, 886

\bibitem[{{Urquhart} \& {Soria}(2016)}]{Urquhart2016}
{Urquhart}, R. \& {Soria}, R. 2016, \mnras, 456, 1859

\bibitem[{{van den Heuvel} \& {De Loore}(1973)}]{vanDenHeuvel1973}
{van den Heuvel}, E.~P.~J. \& {De Loore}, C. 1973, \aap, 25, 387

\bibitem[{{Volonteri} {et~al.}(2008){Volonteri}, {Lodato}, \&
  {Natarajan}}]{Volonteri2008}
{Volonteri}, M., {Lodato}, G., \& {Natarajan}, P. 2008, \mnras, 383, 1079

\bibitem[{{Walton} {et~al.}(2022){Walton}, {Mackenzie}, {Gully}, {Patel},
  {Roberts}, {Earnshaw}, \& {Mateos}}]{Walton2022}
{Walton}, D.~J., {Mackenzie}, A.~D.~A., {Gully}, H., {et~al.} 2022, \mnras,
  509, 1587

\bibitem[{{Walton} {et~al.}(2011){Walton}, {Roberts}, {Mateos}, \&
  {Heard}}]{Walton2011}
{Walton}, D.~J., {Roberts}, T.~P., {Mateos}, S., \& {Heard}, V. 2011, \mnras,
  416, 1844

\bibitem[{{Wang} {et~al.}(2021){Wang}, {Yang}, {Fan}, {Hennawi}, {Barth},
  {Banados}, {Bian}, {Boutsia}, {Connor}, {Davies}, {Decarli}, {Eilers},
  {Farina}, {Green}, {Jiang}, {Li}, {Mazzucchelli}, {Nanni}, {Schindler},
  {Venemans}, {Walter}, {Wu}, \& {Yue}}]{Wang2021}
{Wang}, F., {Yang}, J., {Fan}, X., {et~al.} 2021, \apjl, 907, L1

\bibitem[{{Wang} {et~al.}(2016){Wang}, {Qiu}, {Liu}, \& {Bregman}}]{Wang2016}
{Wang}, S., {Qiu}, Y., {Liu}, J., \& {Bregman}, J.~N. 2016, \apj, 829, 20

\bibitem[{{Webb} {et~al.}(2010){Webb}, {Barret}, \& {Godet et al.}}]{Webb2010}
{Webb}, N.~A., {Barret}, D., \& {Godet et al.}, O. 2010, APJ Letters, 712, L107

\bibitem[{{Webb} {et~al.}(2020){Webb}, {Coriat}, {Traulsen}, {Ballet}, {Motch},
  {Carrera}, {Koliopanos}, {Authier}, {de la Calle}, {Ceballos}, {Colomo},
  {Chuard}, {Freyberg}, {Garcia}, {Kolehmainen}, {Lamer}, {Lin}, {Maggi},
  {Michel}, {Page}, {Page}, {Perea-Calderon}, {Pineau}, {Rodriguez}, {Rosen},
  {Santos Lleo}, {Saxton}, {Schwope}, {Tom{\'a}s}, {Watson}, \&
  {Zakardjian}}]{Webb2020}
{Webb}, N.~A., {Coriat}, M., {Traulsen}, I., {et~al.} 2020, \aap, 641, A136

\bibitem[{{Webb} {et~al.}(2017){Webb}, {Gu{\'e}rou}, {Ciambur}, {Detoeuf},
  {Coriat}, {Godet}, {Barret}, {Combes}, {Contini}, {Graham}, {Maccarone},
  {Mrkalj}, {Servillat}, {Schroetter}, \& {Wiersema}}]{Webb2017}
{Webb}, N.~A., {Gu{\'e}rou}, A., {Ciambur}, B., {et~al.} 2017, \aap, 602, A103

\bibitem[{{White} {et~al.}(2011){White}, {Daw}, \& {Dhillon}}]{GWGC}
{White}, D.~J., {Daw}, E.~J., \& {Dhillon}, V.~S. 2011, Classical and Quantum
  Gravity, 28, 085016

\bibitem[{{Wiktorowicz} {et~al.}(2019){Wiktorowicz}, {Lasota}, {Middleton}, \&
  {Belczynski}}]{Wiktorowicz2019}
{Wiktorowicz}, G., {Lasota}, J.-P., {Middleton}, M., \& {Belczynski}, K. 2019,
  \apj, 875, 53

\bibitem[{{Wiktorowicz} {et~al.}(2015){Wiktorowicz}, {Sobolewska},
  {S{\k{a}}dowski}, \& {Belczynski}}]{Wiktorowicz2015}
{Wiktorowicz}, G., {Sobolewska}, M., {S{\k{a}}dowski}, A., \& {Belczynski}, K.
  2015, \apj, 810, 20

\bibitem[{{Wolter} {et~al.}(2015){Wolter}, {Esposito}, {Mapelli}, {Pizzolato},
  \& {Ripamonti}}]{Wolter2015}
{Wolter}, A., {Esposito}, P., {Mapelli}, M., {Pizzolato}, F., \& {Ripamonti},
  E. 2015, \mnras, 448, 781

\bibitem[{{Zolotukhin} {et~al.}(2016){Zolotukhin}, {Webb}, {Godet}, {Bachetti},
  \& {Barret}}]{Zolotukhin2016}
{Zolotukhin}, I., {Webb}, N.~A., {Godet}, O., {Bachetti}, M., \& {Barret}, D.
  2016, \apj, 817, 88

\bibitem[{{Zou} {et~al.}(2022){Zou}, {Sui}, {Xue}, {Zhou}, {Ma}, {Zhou}, {Nie},
  {Zhang}, {Feng}, {Shen}, \& {Wang}}]{Zou2022}
{Zou}, H., {Sui}, J., {Xue}, S., {et~al.} 2022, Research in Astronomy and
  Astrophysics, 22, 065001

\end{thebibliography}

\begin{appendix}
\section{List of candidates}
The following tables present a partial sample of the ULXs and HLXs identified in this study. The full samples will be available at the CDS (\url{https://cdsarc.u-strasbg.fr/}) at the time of publication of this article.

\begin{table}[h!]
\caption{List of HLX candidates obtained in this study.}
    \centering
    \resizebox{0.95\columnwidth}{!}{\begin{tabular}{c|ccccccccccc}
        Name & RA & Dec & $\texttt{POSERR}$ & $F_X$ & $D$ & Gal. type & Sep. & $L_X$  & $M_{BH}$ & $M_{*,\text{Bell}}$ & Sub-sample\\
        & deg & deg & arcsec & erg cm$^{-2}$ s$^{-1}$ & Mpc & & kpc & erg s$^{-1}$ & M$_\odot$ & M$_\odot$ & \\\hline \\ [-1.5 ex]
2CXO J001143.2-285548  & 2.9303 & -28.9301 & 1.1 & $1.4 \times 10^{-14}$ & 267.9 & E      & 11.2 & $1.2 \times 10^{41}$ & $2.5 \times 10^{4}$ & $1.8 \times 10^{7}$ & robust     \\
2CXO J002234.1+001609  & 5.6424 & 0.2693 & 0.7 & $3.7 \times 10^{-15}$ & 485.3 & S0$^0$ & 15.6 & $1.0 \times 10^{41}$ & $2.3 \times 10^{4}$ &  & robust     \\
2CXO J005151.7+474019  & 12.9658 & 47.6720 & 0.8 & $1.8 \times 10^{-14}$ & 682.5 &        & 16.7 & $10.0 \times 10^{41}$ & $1.9 \times 10^{5}$ &  & weak       \\
2CXO J010739.2+541226  & 16.9134 & 54.2074 & 0.4 & $9.9 \times 10^{-15}$ & 497.8 &        & 14.3 & $2.9 \times 10^{41}$ & $4.1 \times 10^{4}$ &  & robust     \\
2CXO J011028.2-460422  & 17.6178 & -46.0729 & 0.4 & $4.9 \times 10^{-13}$ & 90.9 & S0$^+$ & 3.2 & $4.9 \times 10^{41}$ &  &  & robust     \\
2CXO J011751.9-265850  & 19.4664 & -26.9808 & 1.3 & $8.5 \times 10^{-15}$ & 463.4 &        & 20.6 & $2.2 \times 10^{41}$ & $7.1 \times 10^{4}$ &  & weak       \\
2CXO J011854.9-010551  & 19.7288 & -1.0976 & 0.5 & $2.0 \times 10^{-14}$ & 557.2 & E      & 13.4 & $7.5 \times 10^{41}$ & $1.2 \times 10^{5}$ &  & robust     \\
2CXO J023659.5-511548  & 39.2481 & -51.2636 & 2.8 & $4.7 \times 10^{-14}$ & 343.6 &        & 24.6 & $6.6 \times 10^{41}$ & $1.1 \times 10^{5}$ & $5.8 \times 10^{7}$ & robust     \\
2CXO J024106.1-081711  & 40.2757 & -8.2865 & 0.6 & $3.3 \times 10^{-15}$ & 636.8 &        & 24.6 & $1.6 \times 10^{41}$ & $2.3 \times 10^{4}$ &  & robust     \\
2CXO J024228.5-000202  & 40.6188 & -0.0340 & 0.4 & $7.1 \times 10^{-15}$ & 691.8 & E      & 27.3 & $4.1 \times 10^{41}$ & $8.8 \times 10^{4}$ &  & robust     \\
2CXO J025843.0-523930  & 44.6794 & -52.6584 & 0.5 & $4.7 \times 10^{-15}$ & 428.3 &        & 9.5 & $1.0 \times 10^{41}$ &  &  & robust     \\
2CXO J025921.5+132913  & 44.8398 & 13.4871 & 0.6 & $8.0 \times 10^{-15}$ & 487.5 &        & 8.2 & $2.3 \times 10^{41}$ & $3.0 \times 10^{4}$ &  & galaxy pair\\
2CXO J031049.4-265431  & 47.7062 & -26.9088 & 0.8 & $2.0 \times 10^{-14}$ & 597.0 &        & 14.8 & $8.6 \times 10^{41}$ & $1.6 \times 10^{5}$ &  & robust     \\
2CXO J034115.1+152335  & 55.3129 & 15.3932 & 0.4 & $2.7 \times 10^{-14}$ & 184.4 &        & 3.8 & $1.1 \times 10^{41}$ & $1.5 \times 10^{4}$ & $3.5 \times 10^{7}$ & weak       \\
2CXO J034231.8-533841  & 55.6325 & -53.6449 & 0.4 & $2.3 \times 10^{-14}$ & 270.4 & E      & 12.2 & $2.0 \times 10^{41}$ & $3.3 \times 10^{4}$ &  & robust     \\
2CXO J041338.6+102803  & 63.4112 & 10.4678 & 0.5 & $6.0 \times 10^{-15}$ & 405.5 &        & 10.7 & $1.2 \times 10^{41}$ & $1.7 \times 10^{4}$ &  & robust     \\
2CXO J043338.4-131612  & 68.4101 & -13.2702 & 0.4 & $5.2 \times 10^{-14}$ & 141.9 & S0$^-$ & 19.8 & $1.2 \times 10^{41}$ & $2.0 \times 10^{4}$ &  & robust     \\
2CXO J044921.6-485545  & 72.3403 & -48.9294 & 0.9 & $6.7 \times 10^{-15}$ & 355.9 &        & 6.7 & $1.0 \times 10^{41}$ & $1.3 \times 10^{4}$ &  & robust     \\
2CXO J055107.0-570640  & 87.7793 & -57.1111 & 0.5 & $1.5 \times 10^{-14}$ & 267.9 &        & 15.3 & $1.3 \times 10^{41}$ & $2.9 \times 10^{4}$ &  & robust     \\
2CXO J061620.8-215616  & 94.0867 & -21.9379 & 0.5 & $1.8 \times 10^{-14}$ & $1.0 \times 10^{3}$ &        & 13.2 & $2.3 \times 10^{42}$ & $2.1 \times 10^{5}$ &  & robust     \\
2CXO J062648.6-543208  & 96.7028 & -54.5357 & 0.4 & $7.5 \times 10^{-14}$ & 225.9 & E      & 27.9 & $4.6 \times 10^{41}$ & $4.3 \times 10^{4}$ &  & robust     \\
2CXO J071051.5-561009  & 107.7150 & -56.1694 & 0.8 & $1.9 \times 10^{-14}$ & 562.3 &        & 14.3 & $7.2 \times 10^{41}$ & $1.3 \times 10^{5}$ &  & robust     \\
2CXO J071644.9+372828  & 109.1875 & 37.4746 & 1.3 & $2.5 \times 10^{-14}$ & 317.9 &        & 14.0 & $3.0 \times 10^{41}$ & $4.0 \times 10^{4}$ &  & robust     \\
2CXO J074735.8+554113  & 116.8994 & 55.6870 & 0.5 & $4.0 \times 10^{-14}$ & 152.1 &        & 4.2 & $1.1 \times 10^{41}$ & $1.6 \times 10^{4}$ & $2.5 \times 10^{9}$ & weak       \\
2CXO J081654.4-073931  & 124.2271 & -7.6588 & 1.4 & $1.3 \times 10^{-14}$ & 311.2 &        & 13.7 & $1.5 \times 10^{41}$ & $2.8 \times 10^{4}$ &  & robust     \\
2CXO J082215.9+210535  & 125.5665 & 21.0933 & 0.5 & $1.2 \times 10^{-13}$ & 92.3 & S0$^-$ & 1.9 & $1.2 \times 10^{41}$ & $2.2 \times 10^{4}$ & $3.0 \times 10^{10}$ & robust     \\
2CXO J084128.8+322447  & 130.3700 & 32.4133 & 0.4 & $1.2 \times 10^{-14}$ & 295.1 & S0$^0$ & 11.4 & $1.3 \times 10^{41}$ & $2.4 \times 10^{4}$ &  & robust     \\
2CXO J084135.0+010156  & 130.3962 & 1.0323 & 0.4 & $6.6 \times 10^{-14}$ & 503.5 &        & 7.0 & $2.0 \times 10^{42}$ & $3.8 \times 10^{5}$ &  & galaxy pair\\
2CXO J091449.0+085321  & 138.7045 & 8.8892 & 0.4 & $6.8 \times 10^{-13}$ & 648.6 & E      & 7.9 & $3.4 \times 10^{43}$ & $1.7 \times 10^{7}$ &  & galaxy pair\\
2CXO J101239.8-010628  & 153.1662 & -1.1078 & 0.5 & $9.5 \times 10^{-15}$ & 438.5 & Sa     & 9.4 & $2.2 \times 10^{41}$ & $4.1 \times 10^{4}$ & $2.6 \times 10^{4}$ & robust     \\
2CXO J103619.8-273939  & 159.0827 & -27.6610 & 0.9 & $8.6 \times 10^{-15}$ & 430.5 & S0$^+$ & 27.1 & $1.9 \times 10^{41}$ & $2.9 \times 10^{4}$ &  & robust     \\
2CXO J103844.2+485220  & 159.6845 & 48.8724 & 0.4 & $9.2 \times 10^{-15}$ & 539.5 &        & 13.9 & $3.2 \times 10^{41}$ & $5.4 \times 10^{4}$ &  & robust     \\
2CXO J104142.2+400035  & 160.4262 & 40.0098 & 0.9 & $1.5 \times 10^{-14}$ & 642.7 &        & 24.7 & $7.6 \times 10^{41}$ &  &  & galaxy pair\\
2CXO J105210.4+552243  & 163.0435 & 55.3786 & 0.5 & $6.9 \times 10^{-15}$ & 669.9 & Sb     & 12.2 & $3.7 \times 10^{41}$ & $5.7 \times 10^{4}$ &  & robust     \\
2CXO J111610.0+013035  & 169.0418 & 1.5098 & 0.5 & $1.7 \times 10^{-14}$ & 500.9 & Sbc    & 11.9 & $5.2 \times 10^{41}$ & $1.1 \times 10^{5}$ &  & galaxy pair\\
2CXO J112002.6+132830  & 170.0112 & 13.4752 & 2.0 & $4.2 \times 10^{-15}$ & 586.1 & Sc     & 20.6 & $1.7 \times 10^{41}$ & $2.2 \times 10^{4}$ &  & robust     \\
2CXO J112110.9+232331  & 170.2958 & 23.3922 & 0.6 & $4.2 \times 10^{-15}$ & 513.2 & E      & 13.1 & $1.3 \times 10^{41}$ & $2.5 \times 10^{4}$ &  & weak       \\
2CXO J112255.5+010601  & 170.7313 & 1.1004 & 0.4 & $2.4 \times 10^{-14}$ & 342.0 & E      & 17.5 & $3.4 \times 10^{41}$ & $8.1 \times 10^{4}$ & $3.3 \times 10^{8}$ & weak       \\
2CXO J121756.3+280941  & 184.4849 & 28.1615 & 0.5 & $1.2 \times 10^{-14}$ & 613.8 & E      & 17.1 & $5.3 \times 10^{41}$ & $1.1 \times 10^{5}$ &  & robust     \\
2CXO J122155.1+271022  & 185.4796 & 27.1728 & 0.4 & $1.2 \times 10^{-14}$ & 363.1 & E      & 6.4 & $1.9 \times 10^{41}$ & $2.0 \times 10^{4}$ &  & robust     \\
2CXO J123307.6+091942  & 188.2820 & 9.3285 & 0.6 & $3.3 \times 10^{-14}$ & 309.0 & Sd     & 6.1 & $3.7 \times 10^{41}$ & $8.0 \times 10^{4}$ &  & robust     \\
2CXO J123605.6+163213  & 189.0235 & 16.5370 & 1.4 & $9.7 \times 10^{-15}$ & 318.0 & E      & 12.5 & $1.2 \times 10^{41}$ & $1.3 \times 10^{4}$ &  & weak       \\
2CXO J124208.4+331854  & 190.5351 & 33.3151 & 0.4 & $8.9 \times 10^{-14}$ & 701.9 & E      & 15.2 & $5.3 \times 10^{42}$ & $1.3 \times 10^{6}$ &  & galaxy pair\\
2CXO J130155.9+291815  & 195.4833 & 29.3044 & 1.2 & $1.8 \times 10^{-14}$ & 776.2 &        & 30.3 & $1.3 \times 10^{42}$ &  &  & robust     \\
2CXO J131133.3-011656  & 197.8891 & -1.2824 & 0.4 & $6.6 \times 10^{-15}$ & 895.4 &        & 22.3 & $6.3 \times 10^{41}$ & $1.2 \times 10^{5}$ &  & weak       \\
2CXO J131619.4+620610  & 199.0809 & 62.1029 & 0.9 & $1.4 \times 10^{-14}$ & 610.9 & E      & 14.4 & $6.1 \times 10^{41}$ & $9.3 \times 10^{4}$ &  & robust     \\
2CXO J132519.7-313607  & 201.3324 & -31.6021 & 0.4 & $5.5 \times 10^{-14}$ & 214.8 &        & 4.0 & $3.0 \times 10^{41}$ & $8.1 \times 10^{4}$ &  & robust     \\
2CXO J133338.8-314045  & 203.4120 & -31.6792 & 0.5 & $2.3 \times 10^{-14}$ & 214.8 & S0$^+$ & 55.0 & $1.3 \times 10^{41}$ & $2.2 \times 10^{3}$ &  & robust     \\
2CXO J134736.4+173404  & 206.9018 & 17.5680 & 0.4 & $3.2 \times 10^{-13}$ & 199.2 & Sbc    & 8.2 & $1.5 \times 10^{42}$ & $4.5 \times 10^{5}$ &  & galaxy pair\\
2CXO J140155.7-113809  & 210.4823 & -11.6360 & 0.4 & $3.2 \times 10^{-14}$ & 230.3 &        & 5.4 & $2.0 \times 10^{41}$ & $3.1 \times 10^{4}$ &  & robust     \\
2CXO J142010.7+533348  & 215.0448 & 53.5635 & 0.6 & $1.2 \times 10^{-14}$ & 293.8 & Irr    & 8.3 & $1.2 \times 10^{41}$ & $1.6 \times 10^{4}$ & $2.1 \times 10^{6}$ & robust     \\
2CXO J145409.9+183700  & 223.5415 & 18.6169 & 1.6 & $4.0 \times 10^{-15}$ & 520.0 & Sc     & 24.2 & $1.3 \times 10^{41}$ & $2.3 \times 10^{4}$ &  & robust     \\
2CXO J150930.4+333330  & 227.3769 & 33.5585 & 0.6 & $3.1 \times 10^{-14}$ & $1.0 \times 10^{3}$ &        & 24.3 & $4.0 \times 10^{42}$ & $9.3 \times 10^{5}$ & $5.5 \times 10^{9}$ & weak       \\
2CXO J150940.1+333033  & 227.4172 & 33.5094 & 0.6 & $7.6 \times 10^{-15}$ & 549.5 & E      & 13.4 & $2.8 \times 10^{41}$ & $6.9 \times 10^{4}$ &  & weak       \\
2CXO J152435.6+295302  & 231.1486 & 29.8841 & 0.7 & $3.2 \times 10^{-15}$ & 557.2 &        & 14.7 & $1.2 \times 10^{41}$ & $4.5 \times 10^{4}$ &  & robust     \\
2CXO J153143.2+240421  & 232.9304 & 24.0725 & 0.4 & $5.7 \times 10^{-14}$ & 438.5 & Sab    & 5.9 & $1.3 \times 10^{42}$ & $6.4 \times 10^{4}$ &  & galaxy pair\\
2CXO J160211.7+155437  & 240.5491 & 15.9103 & 0.4 & $3.9 \times 10^{-14}$ & 159.2 & S0$^0$ & 12.5 & $1.2 \times 10^{41}$ & $2.8 \times 10^{4}$ &  & robust     \\
2CXO J161249.8+540832  & 243.2076 & 54.1423 & 0.5 & $3.8 \times 10^{-14}$ & 711.2 & E      & 14.8 & $2.3 \times 10^{42}$ & $4.6 \times 10^{5}$ &  & robust     \\
2CXO J162953.8+394601  & 247.4743 & 39.7670 & 6.2 & $4.1 \times 10^{-14}$ & 143.2 & Sbc    & 21.2 & $1.0 \times 10^{41}$ & $1.7 \times 10^{4}$ &  & robust     \\
2CXO J163121.5-750652  & 247.8397 & -75.1145 & 0.4 & $1.8 \times 10^{-14}$ & 498.9 &        & 57.1 & $5.3 \times 10^{41}$ & $8.5 \times 10^{4}$ &  & robust     \\
2CXO J163249.0+053852  & 248.2045 & 5.6479 & 0.5 & $7.1 \times 10^{-15}$ & 376.4 &        & 13.6 & $1.2 \times 10^{41}$ & $1.1 \times 10^{4}$ &  & weak       \\
2CXO J164237.1+272634  & 250.6548 & 27.4428 & 0.4 & $3.1 \times 10^{-14}$ & 469.9 & E      & 38.7 & $8.3 \times 10^{41}$ & $1.4 \times 10^{5}$ &  & robust     \\
2CXO J170932.2+342542  & 257.3845 & 34.4286 & 0.5 & $1.7 \times 10^{-14}$ & 353.2 & Sc     & 9.5 & $2.6 \times 10^{41}$ & $4.7 \times 10^{4}$ &  & robust     \\
2CXO J171215.5+640212  & 258.0646 & 64.0368 & 0.5 & $1.1 \times 10^{-14}$ & 335.1 & E      & 7.1 & $1.5 \times 10^{41}$ & $2.5 \times 10^{4}$ &  & robust     \\
    \end{tabular}}
    \tablefoot{Column (1): name of the source. (2--4): source coordinates and position error. (5): broad-band mean X-ray flux. (6): galaxy distance. (7): galaxy morphology. (8): physical separation of the source to the galaxy centre. (9): broad-band mean X-ray luminosity. (10): black hole mass using the $r$-band luminosity and $g-r$ colour \citep{Bell2003}. (11): galaxy mass given in GLADE+ \citep{Glade2021}. (12): category assigned to the HLX in our study.}
    \label{tab:list_HLX0}
\end{table}

\begin{table}
\caption{List of ULX candidates obtained in this study.}
    \centering
    \resizebox{0.95\columnwidth}{!}{\begin{tabular}{c|cccccccccc}
         Name & RA & Dec & $\texttt{POSERR}$ & $F_X$ & $D$ & Gal. type & Sep. & $L_X$ & $P_{XRB}$ & $f_{cont}$\\
        & deg & deg & arcsec & erg cm$^{-2}$ s$^{-1}$ & Mpc & & kpc & erg s$^{-1}$ & & \\\hline \\ [-1.5 ex]
2CXO J000120.2+130641 & 0.3342 & 13.1114 & 0.4 & $5.5\times 10^{-15}$ & 76.5 & S0/E & 1.3 & $3.9\times 10^{39}$ & 0.56 & 0.01 \\ 
2CXO J000125.1+130708 & 0.3549 & 13.1189 & 0.5 & $2.0\times 10^{-15}$ & 75.0 & S & 8.8 & $1.3\times 10^{39}$ & 0.41 & 0.20 \\ 
2CXO J000126.7+130649 & 0.3616 & 13.1136 & 0.5 & $8.3\times 10^{-15}$ & 75.0 & S & 3.9 & $5.6\times 10^{39}$ & 0.82 & 0.08 \\
2CXO J000141.1+232949 & 0.4215 & 23.4972 & 0.4 & $2.8\times 10^{-15}$ & 62.5 & S & 3.4 & $1.3\times 10^{39}$ & 0.55 & 0.02 \\ 
2CXO J000142.3+232941 & 0.4263 & 23.4947 & 0.4 & $3.9\times 10^{-15}$ & 62.5 & S & 1.9 & $1.8\times 10^{39}$ & 0.43 & 0.00 \\ 
2CXO J000142.9+232935 & 0.4290 & 23.4933 & 0.4 & $3.9\times 10^{-15}$ & 62.5 & S & 5.0 & $1.8\times 10^{39}$ & 0.80 & 0.02 \\ 
2CXO J000254.5-354308 & 0.7274 & -35.7189 & 0.5 & $4.9\times 10^{-15}$ & 212.8 & S0/E & 10.6 & $2.6\times 10^{40}$ & 0.56 & 0.40 \\ 
2CXO J000619.5-413019 & 1.5814 & -41.5054 & 0.4 & $2.3\times 10^{-14}$ & 19.5 & S & 1.9 & $1.1\times 10^{39}$ & 0.58 & 0.00 \\ 
2CXO J000620.2-413005 & 1.5842 & -41.5016 & 0.4 & $8.9\times 10^{-14}$ & 19.5 & S & 0.7 & $4.0\times 10^{39}$ & 0.44 & 0.00 \\ 
2CXO J000952.9+255547 & 2.4708 & 25.9299 & 0.5 & $4.4\times 10^{-15}$ & 65.8 & S & 7.1 & $2.3\times 10^{39}$ &   &   \\ 
2CXO J000953.2+255454 & 2.4720 & 25.9153 & 0.5 & $2.1\times 10^{-15}$ & 65.8 & S & 9.5 & $1.1\times 10^{39}$ & 0.83 & 0.29 \\ 
2CXO J000955.1+255536 & 2.4800 & 25.9267 & 0.7 & $2.5\times 10^{-15}$ & 65.8 & S & 8.1 & $1.3\times 10^{39}$ & 0.75 & 0.32 \\ 
2CXO J001106.9-120631 & 2.7788 & -12.1088 & 0.4 & $8.7\times 10^{-15}$ & 83.3 & S & 2.2 & $7.2\times 10^{39}$ & 0.25 & 0.09 \\ 
2CXO J001815.4+300343 & 4.5646 & 30.0620 & 0.7 & $1.9\times 10^{-15}$ & 93.9 & S0/E & 4.3 & $2.0\times 10^{39}$ & 0.59 & 0.13 \\ 
2CXO J001827.1+300210 & 4.6131 & 30.0363 & 0.5 & $6.7\times 10^{-15}$ & 106.7 & S & 11.1 & $9.1\times 10^{39}$ & 0.63 & 0.34 \\ 
2CXO J001849.8-102134 & 4.7077 & -10.3596 & 0.4 & $9.7\times 10^{-15}$ & 116.4 & S & 4.5 & $1.6\times 10^{40}$ & 0.43 & 0.28 \\ 
2CXO J001850.4-102228 & 4.7102 & -10.3746 & 0.4 & $4.8\times 10^{-15}$ & 119.7 & S & 5.7 & $8.2\times 10^{39}$ & 0.58 & 0.05 \\ 
2CXO J001850.9-102232 & 4.7122 & -10.3757 & 0.5 & $1.1\times 10^{-14}$ & 119.7 & S & 2.2 & $1.8\times 10^{40}$ & 0.57 & 0.01 \\ 
2CXO J001850.9-102249 & 4.7125 & -10.3804 & 0.5 & $1.8\times 10^{-15}$ & 119.7 & S & 7.1 & $3.0\times 10^{39}$ & 0.76 & 0.12 \\ 
2CXO J001851.2-102244 & 4.7134 & -10.3791 & 0.4 & $7.5\times 10^{-15}$ & 119.7 & S & 5.1 & $1.3\times 10^{40}$ & 0.56 & 0.05 \\ 
2CXO J002751.6-014813 & 6.9653 & -1.8037 & 2.7 & $1.4\times 10^{-13}$ & 57.6 & S & 6.9 & $5.4\times 10^{40}$ & 0.30 & 0.24 \\ 
2CXO J003358.1-094222 & 8.4922 & -9.7062 & 0.5 & $1.2\times 10^{-14}$ & 52.9 & S & 17.2 & $4.2\times 10^{39}$ & 0.75 & 0.54 \\ 
2CXO J003404.3-094248 & 8.5182 & -9.7133 & 0.5 & $1.1\times 10^{-14}$ & 52.9 & S & 9.3 & $3.7\times 10^{39}$ & 0.79 & 0.35 \\ 
2CXO J003404.5-094239 & 8.5191 & -9.7110 & 0.6 & $7.8\times 10^{-15}$ & 52.9 & S & 8.4 & $2.6\times 10^{39}$ & 0.41 & 0.24 \\ 
2CXO J003408.4-094140 & 8.5351 & -9.6947 & 1.2 & $1.3\times 10^{-14}$ & 67.8 &  & 2.6 & $7.1\times 10^{39}$ & 0.29 & 0.02 \\ 
2CXO J003413.6-212803 & 8.5568 & -21.4676 & 0.7 & $1.5\times 10^{-15}$ & 96.3 & S & 4.7 & $1.7\times 10^{39}$ & 0.61 & 0.09 \\ 
2CXO J003414.9-212455 & 8.5622 & -21.4155 & 0.5 & $1.8\times 10^{-15}$ & 270.4 & S0/E & 12.6 & $1.6\times 10^{40}$ & 0.84 & 0.18 \\ 
2CXO J003416.0-212517 & 8.5669 & -21.4215 & 0.4 & $2.7\times 10^{-15}$ & 270.4 & S0/E & 18.9 & $2.4\times 10^{40}$ & 0.60 & 0.46 \\ 
2CXO J003654.5-010737 & 9.2271 & -1.1271 & 1.1 & $1.4\times 10^{-15}$ & 326.6 & S & 9.8 & $1.7\times 10^{40}$ & 0.26 & 0.23 \\ 
2CXO J003703.9-010904 & 9.2665 & -1.1514 & 0.4 & $1.0\times 10^{-14}$ & 328.1 & S0/E & 5.2 & $1.3\times 10^{41}$ & 0.24 & 0.07 \\ 
2CXO J003737.5-334255 & 9.4067 & -33.7153 & 0.4 & $6.3\times 10^{-15}$ & 126.5 & S & 25.5 & $1.2\times 10^{40}$ &   &   \\ 
2CXO J003738.1-334306 & 9.4088 & -33.7183 & 0.4 & $9.0\times 10^{-16}$ & 126.5 & S & 22.2 & $1.7\times 10^{39}$ & 0.67 & 0.15 \\ 
2CXO J003738.3-334309 & 9.4098 & -33.7192 & 0.4 & $1.0\times 10^{-15}$ & 126.5 & S & 20.8 & $1.9\times 10^{39}$ & 0.61 & 0.13 \\ 
2CXO J003738.7-334316 & 9.4114 & -33.7212 & 0.4 & $7.5\times 10^{-15}$ & 126.5 & S & 20.0 & $1.4\times 10^{40}$ & 0.37 & 0.13 \\ 
2CXO J003739.1-334229 & 9.4131 & -33.7083 & 0.4 & $2.9\times 10^{-15}$ & 126.5 & S & 22.0 & $5.6\times 10^{39}$ & 0.75 & 0.13 \\ 
2CXO J003739.2-334250 & 9.4134 & -33.7139 & 0.4 & $4.4\times 10^{-15}$ & 126.5 & S & 14.6 & $8.5\times 10^{39}$ & 0.83 & 0.06 \\ 
2CXO J003739.3-334323 & 9.4141 & -33.7231 & 0.4 & $2.6\times 10^{-14}$ & 126.5 & S & 18.9 & $5.0\times 10^{40}$ & 0.38 & 0.11 \\ 
2CXO J003740.2-334327 & 9.4177 & -33.7242 & 0.4 & $1.5\times 10^{-15}$ & 126.5 & S & 17.6 & $2.9\times 10^{39}$ & 0.32 & 0.10 \\ 
2CXO J003740.4-334324 & 9.4185 & -33.7236 & 0.4 & $1.0\times 10^{-15}$ & 126.5 & S & 15.9 & $2.0\times 10^{39}$ & 0.66 & 0.08 \\ 
2CXO J003740.7-334258 & 9.4198 & -33.7161 & 0.4 & $5.3\times 10^{-16}$ & 126.5 & S & 2.7 & $1.0\times 10^{39}$ & 0.53 & 0.00 \\ 
2CXO J003740.8-334331 & 9.4203 & -33.7253 & 0.4 & $4.1\times 10^{-15}$ & 126.5 & S & 18.8 & $7.8\times 10^{39}$ & 0.56 & 0.11 \\ 
2CXO J003741.2-334232 & 9.4218 & -33.7090 & 0.4 & $1.3\times 10^{-15}$ & 126.5 & S & 15.4 & $2.6\times 10^{39}$ & 0.83 & 0.07 \\ 
2CXO J003741.3-334331 & 9.4223 & -33.7254 & 0.4 & $8.9\times 10^{-16}$ & 126.5 & S & 19.0 & $1.7\times 10^{39}$ & 0.68 & 0.11 \\ 
2CXO J003742.4-334249 & 9.4267 & -33.7138 & 0.4 & $3.7\times 10^{-15}$ & 126.5 & S & 10.8 & $7.1\times 10^{39}$ & 0.70 & 0.04 \\ 
2CXO J003742.4-334304 & 9.4270 & -33.7178 & 0.4 & $9.1\times 10^{-16}$ & 126.5 & S & 10.4 & $1.7\times 10^{39}$ & 0.81 & 0.03 \\ 
2CXO J003742.7-334212 & 9.4283 & -33.7035 & 0.4 & $1.1\times 10^{-15}$ & 128.2 & S & 6.7 & $2.3\times 10^{39}$ & 0.87 & 0.03 \\ 
2CXO J003742.8-334314 & 9.4286 & -33.7207 & 0.4 & $8.7\times 10^{-16}$ & 126.5 & S & 15.6 & $1.7\times 10^{39}$ & 0.80 & 0.07 \\ 
2CXO J003742.9-334204 & 9.4289 & -33.7012 & 0.4 & $2.4\times 10^{-15}$ & 128.2 & S & 4.7 & $4.8\times 10^{39}$ & 0.63 & 0.01 \\ 
2CXO J003743.8-334209 & 9.4327 & -33.7028 & 0.4 & $2.4\times 10^{-15}$ & 128.2 & S & 3.0 & $4.7\times 10^{39}$ & 0.69 & 0.01 \\ 
2CXO J003745.3-334228 & 9.4388 & -33.7079 & 0.4 & $2.2\times 10^{-15}$ & 155.4 & S0/E & 6.3 & $6.3\times 10^{39}$ & 0.82 & 0.20 \\ 
2CXO J003747.0-333953 & 9.4459 & -33.6647 & 0.5 & $1.8\times 10^{-15}$ & 127.9 &  & 5.5 & $3.6\times 10^{39}$ & 0.70 & 0.07 \\ 
2CXO J003912.0+005153 & 9.8000 & 0.8650 & 1.2 & $2.7\times 10^{-15}$ & 59.7 & S & 5.9 & $1.1\times 10^{39}$ & 0.74 & 0.31 \\ 
2CXO J003913.2+005142 & 9.8051 & 0.8617 & 0.5 & $2.9\times 10^{-14}$ & 59.7 & S & 2.6 & $1.2\times 10^{40}$ & 0.68 & 0.03 \\ 
2CXO J003917.1+031940 & 9.8214 & 3.3279 & 0.4 & $3.4\times 10^{-15}$ & 61.6 & S0/E & 7.2 & $1.5\times 10^{39}$ & 0.70 & 0.10 \\ 
2CXO J003919.7+031945 & 9.8324 & 3.3294 & 0.4 & $4.2\times 10^{-15}$ & 61.6 & S0/E & 5.6 & $1.9\times 10^{39}$ & 0.83 & 0.10 \\

    \end{tabular}}
    \tablefoot{Column (1): name of the source. (2--4): source coordinates and position error. (5): broad-band mean X-ray flux. (6): galaxy distance. (7): galaxy morphology. (8): physical separation of the source to the galaxy centre. (9): broad-band mean X-ray luminosity. (10): probability that the source is an XRB, given by the classifier. (11) fraction of ULX candidates being background contaminants within the separation of the ULX, given by the $\log N-\log S$ method.}
    \label{tab:list_ULX1}
\end{table}


\end{appendix}

\end{document}